\documentclass[twocolumn,pra,aps,showpacs,longbibliography]{revtex4-1}

\usepackage{lineno}
\usepackage{mathptmx}
\usepackage{subfigure}
\usepackage{dcolumn}
\usepackage{amsmath,amssymb}
\usepackage{bm}
\usepackage{color}
\usepackage{overpic}
\usepackage{latexsym}
\usepackage{epstopdf}
\usepackage{color}
\usepackage[english]{babel}
\usepackage{latexsym}
\usepackage{stmaryrd}

\usepackage{psfrag,graphicx}
\usepackage{epsf}
\usepackage{subfigure}
\usepackage{amsmath}
\usepackage{amssymb}
\usepackage{amsfonts}
\usepackage{bm}
\usepackage{natbib}
\usepackage{epstopdf}
\usepackage{appendix}

\definecolor{mygrey}{gray}{0.35}
\definecolor{myblue}{rgb}{0.2,0.2,0.8}
\definecolor{myzard}{cmyk}{0,0,0.05,0}
\definecolor{mywhite}{rgb}{1,1,1}
\definecolor{myred}{rgb}{1,0.,0.3}
\definecolor{darkred}{rgb}{0.9,0.1,0.1}
\definecolor{darkblue}{rgb}{0.1,0.1,0.7}

\usepackage[colorlinks=true,citecolor=myblue,linkcolor=myred]{hyperref}
\def\be{\begin{equation}}
\def\ee{\end{equation}}
\def\ba{\begin{align}}
\def\enda{\end{align}}
\def\bi{\begin{itemize}}
\def\ei{\end{itemize}}

 \def\ee{\mathord{\rm e}}
 
 \def\ii{\mathord{\rm i}}

\def\half{\textstyle\frac{1}{2}}

\def\fourth{\textstyle\frac{1}{4}}

 \def\ee{\mathord{\rm e}}
 
 \def\ii{\mathord{\rm i}}

\def\half{\textstyle\frac{1}{2}}

\def\fourth{\textstyle\frac{1}{4}}

\renewcommand{\ii}{{\rm i}}
\renewcommand{\ee}{{\rm e}}

\def\beq{\begin{equation}}
\def\beq{\begin{equation}}
\def\eeq{\end{equation}}

\newcommand{\ket}[1]{|#1\rangle}
\newcommand{\bra}[1]{\langle #1|}

\newcommand{\ketbradif}[2]{\ket{#1}\bra{#2}}
\newcommand{\ketbra}[1]{\ketbradif {#1}{#1}}

\begin{document}


\title[Short Title]{Controlling and measuring quantum transport of heat in trapped-ion crystals}

\author{A. Bermudez}
\affiliation{Institut f\"ur Theoretische Physik, Albert-Einstein-Allee 11, Universit\"at Ulm, 89069 Ulm, Germany\\ Center for Integrated Quantum Science and Technology,
Albert-Einstein-Allee 11, Universit\"at Ulm, 89069 Ulm}

\author{M. Bruderer}
\affiliation{Institut f\"ur Theoretische Physik, Albert-Einstein-Allee 11, Universit\"at Ulm, 89069 Ulm, Germany\\ Center for Integrated Quantum Science and Technology,
Albert-Einstein-Allee 11, Universit\"at Ulm, 89069 Ulm}

\author{M. B. Plenio}
\affiliation{Institut f\"ur Theoretische Physik, Albert-Einstein-Allee 11, Universit\"at Ulm, 89069 Ulm, Germany\\ Center for Integrated Quantum Science and Technology,
Albert-Einstein-Allee 11, Universit\"at Ulm, 89069 Ulm}





\begin{abstract}
Measuring heat flow through nanoscale devices poses formidable practical
difficulties as there is no `ampere meter' for heat. We propose to overcome
this problem in a chain of trapped ions, where
laser cooling the chain edges to different temperatures induces a heat current of
local vibrations (vibrons). We show how to efficiently control and measure
this current, including fluctuations, by coupling vibrons to internal ion
states. This demonstrates that ion crystals provide an ideal platform for
studying quantum transport, e.g., through thermal analogues of quantum wires
and quantum dots. Notably, ion crystals may give access to measurements of
the elusive bosonic fluctuations in heat currents and the onset of Fourier's
law. Our results are strongly supported by numerical simulations for a
realistic implementation with specific ions and system parameters.
\end{abstract}



\maketitle

In view of the rapid development of nanoscale
technologies~\cite{Cahill-JAP-2003}, understanding charge and heat transport
at the microscopic level has become a central topic of current research. As
already shown for fermions~\cite{Zimbovskaya-PR-2011}, charge transport at
the nanoscale is typically governed by quantum effects. Transport of heat by
bosons, e.g, phonons, is expected to have analogous
properties~\cite{Dubi-RMP-2011}. Thermal experiments, however, are considerably more
challenging as there is no device capable of measuring local
heat currents~\cite{Dubi-RMP-2011}. Moreover, heat reservoirs and
temperature probes required to study heat transport usually entail spurious
interface effects. Within these restrictions, most experimental efforts have
focused on detecting temperature profiles~\cite{Majumdar-ARMS-1999} in
different devices~\cite{Cahill-JAP-2003,Kim-PRL-2001,Schwab-NAT-2000}.

\begin{figure}
\centering
\includegraphics[width=.9\columnwidth]{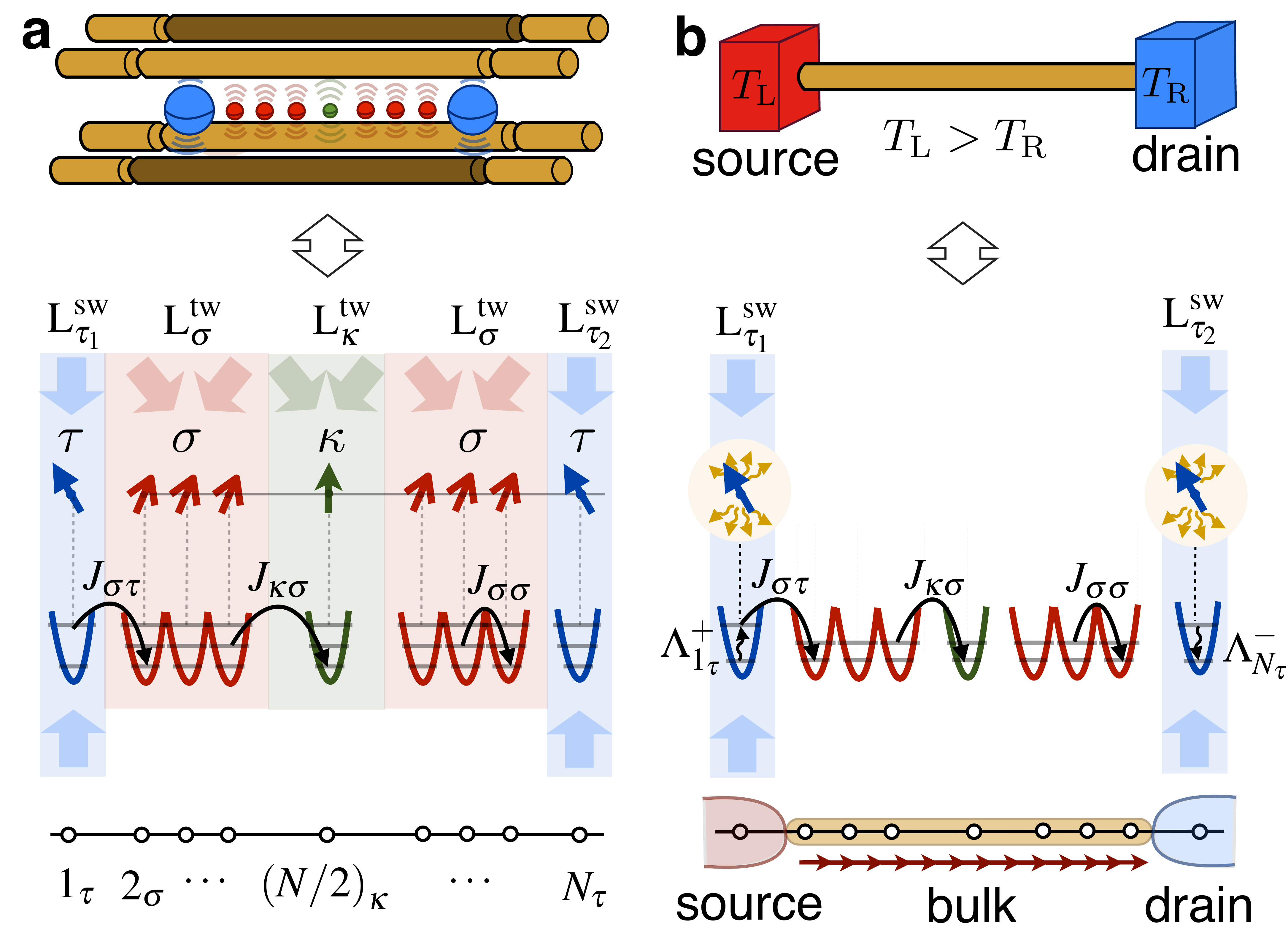}
\caption{ {\bf Heat transport toolbox: }{\bf (a)} (top) A mixed-species ion crystal in a linear Paul trap (similarly for surface trap arrays). (bottom) Spins
and vibrons are indicated by arrows and wells, respectively. Laser arrangements ${\rm L}^{\rm sw}_{\tau}$ (${\rm L}^{\rm tw}_{\sigma},{\rm L}^{\rm tw}_{\kappa}$) control the
incoherent (coherent) vibrational dynamics of the ions, with $J_{\alpha\beta}$ the vibron tunneling. {\bf (b)} (top) A thermal quantum wire (TQW) connected to two reservoirs at different
temperatures. (bottom) Strong laser cooling with strengths $\Lambda_{\ell_{\tau}}^->\Lambda_{\ell_{\tau}}^+$ allows us to treat $\tau$-ions as heat reservoirs, whereas bulk ions $\sigma,\kappa$ act as
the TQW.}
\label{fig_scheme}
\end{figure}

{In this Letter, we show that trapped-ion crystals are  promising platforms
for thermal experiments  overcoming these limitations. We introduce a
{\it quantum transport toolbox} containing all functionalities required for
treating heat currents on the same footing as electrical currents. By
exploiting laser-induced couplings between transverse quantized vibrations
(vibrons) and internal degrees of freedom (spins) of the ions, we show how
to control and measure heat currents across ion
chains~[Fig.~\ref{fig_scheme}{\bf (a)}]. Specifically, ions at the edges of
the crystal are Doppler-cooled to {different vibron numbers, equivalent to different temperatures.
The edge ions act as unbalanced
thermal reservoirs sustaining a heat flow in the form of vibron hopping through the
bulk~[Fig.~\ref{fig_scheme}{\bf (b)}].} For probing vibron numbers and heat
currents (including fluctuations) we map their values onto the spins, which
can be measured via spin-dependent fluorescence~\cite{wineland_review}.
We note that thermal experiments with ions is a topic of increasing interest:
{Propagation} of vibrational excitations  has been assessed
in~\cite{dynamic_transport}, while the use of single  ions as heat
engines has been proposed in~\cite{heat_engine}. More relevant to the topic of this work, the
thermalization of sympathetically-cooled  chains has been studied
in~\cite{steady_state_thermalization} by Langevin dynamics~\cite{talkner}.
Our  toolbox, which is  based on thorough first-principle
derivations, will be  useful for the development of 
experiments about  non-equilibrium statistical mechanics in the quantum regime  with trapped ions.}

We demonstrate the versatility of this toolbox by the examples of a {\it
thermal quantum wire} (TQW) and a {\it thermal quantum dot} (TQD). We first
study the onset of temperature gradients across the TQW according to
Fourier's law~\cite{Fourier}. This requires the transition from ballistic to
diffusive transport, which we induce by {\it i)} dephasing through noisy
modulations of the trap frequencies~\cite{dephasing_noise} or {\it ii)}
disorder {in the ion crystal} due to engineered spin-vibron couplings~\cite{Bermudez-NJP-2010}.
The TQD highlights the differences between bosonic and fermionic
transport~\cite{Esposito-RMP-2009}, captured by the statistics of the
fluctuations {in the heat current}~\cite{Harbola-PRB-2007}. Building on laser-assisted
tunneling~\cite{assisted_tunneling,photon_assisted_tunneling_ions} we show
how to measure current fluctuations. Moreover, the TQD can be operated as a
switch for heat currents, a first step towards a single-spin heat
transistor.

{\it Model.--} We consider a linear Coulomb crystal with three types of
ions~[Fig.~\ref{fig_scheme}{\bf (a)}]. Unlike in phonon-mediated quantum computing~\cite{qip}, we focus on {\it vibrons}: the quanta of individual transverse
oscillations responsible for a local
electric dipole. As demonstrated experimentally~\cite{tunneling_exp}, the
interaction between these dipoles leads to a  tight-binding model
($\hbar=1$)
\begin{equation}
\label{tbm}
H_{\rm tb}\!\!=\!\!\sum_{\alpha,i_{\alpha}}\!\!\omega_{i_\alpha}a_{i_{\alpha}}^{\dagger}a_{i_{\alpha}}^{\phantom{\dagger}}\!+\!\sum_{\alpha,\beta}\!\sum_{i_{\alpha}\neq
j_{\beta}}\!\!\big(J_{i_{\alpha}j_{\beta}}a_{i_{\alpha}}^{\dagger}a_{j_{\beta}}^{\phantom{\dagger}}+\text{H.c.}\big),
\end{equation}
where the bosonic operators
$a_{i_{\alpha}}^{\dagger}(a_{i_{\alpha}}^{\phantom{\dagger}})$ create
(annihilate) local vibrons; latin indices label lattice sites
$i,j\in\{1\cdots N\}$ and greek sub-indices label species
$\alpha,\beta\in\{\sigma,\tau,\kappa\}$~[Fig.~\ref{fig_scheme}{\bf (a)}].
The trapping and dipole-dipole couplings yield the on-site energies
$\omega_{i_\alpha}$ and long-range tunnelings
$J_{i_{\alpha}j_{\beta}}$. The ion crystal is a natural playground
for bosonic lattice models~\cite{porras_hubbard_model}, where vibrons
correspond to bosonic particles{, hopping between different lattice sites,} and the lattice is determined by the
underlying crystal structure. Additionally, we exploit two atomic levels of
each ion, denoted spins
$\ket{s_{i_{\alpha}}}\in\{\ket{{\uparrow_{i_\alpha}}},\ket{{\downarrow_{i_\alpha}}}\}$,
with the Hamiltonian $H_{\rm
s}^{\alpha}=\half\sum_{i_{\alpha}}\omega_0^{\alpha}\sigma_{i_{\alpha}}^z$
and
$\sigma_{i_{\alpha}}^z=\ket{{\uparrow_{i_\alpha}}}\bra{{\uparrow_{i_\alpha}}}-\ket{{\downarrow_{i_\alpha}}}\bra{{\downarrow_{i_\alpha}}}$.
The atomic transitions are characterized by their frequency
$\omega_0^{\alpha}$ and linewidth $\Gamma_{\alpha}^{\rm eff}$.

{We supplement  the dynamics of the vibrons by incoherent
and coherent laser-induced processes which are necessary to develop the
tools for studying quantum transport:}

{\it (i)} For \emph{incoherent} dynamics, we employ a laser forming a standing wave along the vibron direction. This drives
dipole-allowed transitions of the $\alpha$-spins and simultaneously {increases or decreases}
the corresponding vibron number. For fast decaying spins,
the two processes yield an effective vibron dissipation
\begin{equation}
\label{dissipation}
\mathcal{D}_{\rm v}^{i_\alpha}(\mu)= {\mathcal{D}}[\Lambda^+_{i_\alpha},a^{{\dagger}}_{i_\alpha},a^{\phantom{\dagger}}_{i_\alpha}](\mu)
+{\mathcal{D}}[\Lambda^-_{i_\alpha},a^{\phantom{\dagger}}_{i_\alpha},a^{{\dagger}}_{i_\alpha}](\mu),
\end{equation}
where  ${\mathcal{D}}[\Lambda,O_1, O_2](\bullet)=\Lambda (O_1 \bullet
O_2-O_2O_1 \bullet)+\text{H.c.}$ is a super-operator {acting on the density
matrix $\mu$.} The local heating
(cooling) strength $\Lambda^{+}_{i_\alpha}$~$(\Lambda^{-}_{i_\alpha})$
depends on the spectral functions of the couplings~\cite{laser_cooling_ions}
and is controlled by the laser parameters~\cite{comment_cooling}.

{\it (ii)} For \emph{coherent} dynamics, we apply a spin-dependent traveling
wave  consisting of two {non-copropagating} laser
beams. The spin-vibron couplings originates from two-photon processes~\cite{comment_drivings}, whereby the spin is
virtually excited by absorption/emission of a photon from/into
a different laser beam
\begin{equation}
\label{driving}
H_{\rm
sv}^{i_\alpha}(t)=\half(\Delta\omega^+_{\alpha}+\Delta\omega^-_{\alpha}\sigma_{i_{\alpha}}^z)\cos(\nu_{\alpha}t-\varphi_{{\alpha}})a_{i_\alpha}^{{\dagger}}a_{i_\alpha}^{\phantom{\dagger}},
\end{equation}
where $\Delta\omega^\pm_{\alpha},\nu_\alpha,\varphi_\alpha$ are fully
controllable.

Equations~\eqref{tbm} to \eqref{driving} form our Liouvillian heat transport
toolbox, the {\it driven dissipative spin-vibron model}
\begin{equation}
\label{ddsp}
\mathcal{L}_{\rm ddsv}(\mu)=-\ii\bigg[H_{\rm tb}+\hspace{-1ex}\sum_{\alpha,i_\alpha\in \mathfrak{C}}\hspace{-1ex}H_{\rm sv}^{i_\alpha}(t), \mu\bigg]+\hspace{-1ex}\sum_{\alpha, i_\alpha\in
\mathfrak{I}}\hspace{-1ex}\mathcal{D}_{\rm v}^{i_\alpha}(\mu),
\end{equation}
where the sets $\mathfrak{C},\mathfrak{I}$ comprise ions subjected to
coherent/incoherent effects. {We avoid single-ion laser
addressing by employing different species for each functionality, such as the implementation of thermal reservoirs. Ideally, these  are
capable of supplying/absorbing vibrons without changing their state.
This is achieved by using a red-detuned laser,
such that the cooling~\eqref{dissipation}
 with a  rate $\gamma_{i_\alpha}={\rm
Re}\{(\Lambda^-_{i_\alpha})^*-\Lambda^+_{i_\alpha}\}$  dominates over the tunneling
$\gamma_{i_\alpha}\gg J_{i_\alpha,j_\beta}$ (i.e. strong-cooling limit)~\cite{sm}. Thus, the ions
remain in a thermal state,  providing an accurate
implementation of  vibronic reservoirs.}

{\it Thermal quantum wire.--} For designing a TQW we choose ion
species with $\Gamma_{\tau}^{\rm eff}\gg\Gamma_{\sigma}^{\rm eff},\Gamma_{\kappa}^{\rm eff}$, and
implement dissipation only for the $\tau$-ions (i.e.
$\sigma,\kappa\in\mathfrak{C}$, $\tau\in\mathfrak{I}$). The
$\tau$-ions, placed at the edges of the chain, are cooled to mean vibron numbers $\bar{n}_{1_{\tau}}>\bar{n}_{N_{\tau}}$, such that they act as vibronic batteries,
realising the starting point for many transport
studies~\cite{meso_reservoirs}. The left (right) reservoir constantly
supplies (absorbs) vibrons in the attempt to equilibrate with the TQW. If
combined, the reservoirs sustain a flow of heat along the
TQW~[Fig.~\ref{fig_scheme}{\bf (b)}].

We assess how the TQW thermalizes in contact with the reservoirs. In the strong-cooling regime, the edge vibrons can be
integrated out to obtain a
dissipative spin-vibron model for the reduced density matrix of the bulk
$\partial_t{\mu}_{{\rm bulk}}=\mathcal{L}_{\rm ddsv}^{\rm bulk}(\mu_{{\rm
bulk}})$,
\begin{equation}
\label{bulk_liouvillian}
\mathcal{L}_{\rm ddsv}^{\rm bulk}(\bullet)=-\ii\bigg[{H}_{\rm
rtb}+\hspace{-1ex}\sum_{\alpha,i_\alpha\in \mathfrak{C}}\hspace{-1ex}H_{\rm sv}^{i_\alpha}(t), \bullet \bigg]+\hspace{-1ex}\sum_{\alpha i_\alpha,\beta j_\beta\in\mathfrak{C}}\hspace{-1ex}\mathcal{D}_{i_\alpha j_\beta}(\bullet).
\end{equation}
Here, ${H}_{\rm rtb}$ is identical to~\eqref{tbm} with renormalized
parameters. { The
dissipator is similar to~\eqref{dissipation}, but extended to {all} bulk ions
$
\mathcal{D}_{i_\alpha j_\beta}=\mathcal{D}[\tilde{\Lambda}^+_{i_\alpha,j_\beta},
a^{{\dagger}}_{i_\alpha},a^{\phantom{\dagger}}_{j_\beta}]+\mathcal{D}[\tilde{\Lambda}^-_{i_\alpha,j_\beta},a^{\phantom{\dagger}}_{i_\alpha},a^{{\dagger}}_{j_\beta}],
$
where $\tilde{\Lambda}^{\pm}_{i_\alpha,j_\beta}$  depend on the  tunneling via {the spectral densities}
$\Gamma^{\ell_\tau}_{i_\alpha,j_\beta} = 2\pi
J_{i_\alpha,\ell_\tau}\rho_{\ell_\tau}(\omega_{i_\alpha})J_{\ell_\tau,j_\beta}$, {including the reservoir  density of
states $\rho_{\ell_\tau}$}~\cite{bulk_parameters}.
Hence, the bulk-reservoir-bulk tunneling of vibrons introduces an effective
dissipation responsible for the thermalization of the TQW.}

The dipolar decay
of tunneling with distance suggests that vibron exchange with bulk ions
adjacent to the reservoirs dominates thermalization. In the strong-cooling regime, we thus predict a
homogeneous steady-state vibron occupation
\begin{equation}
\label{occupation}
\langle n_{i_{\alpha}}\rangle_{\rm ss} =\frac{\Gamma_{\rm L} \bar{n}_{\rm L}+\Gamma_{\rm R} \bar{n}_{\rm R}}{\Gamma_{\rm L} + \Gamma_{\rm R}},\hspace{3ex}\alpha,i_\alpha\in\mathfrak{C},
\end{equation}
with the {local} couplings $\Gamma_{\rm L}=\Gamma^{1_\tau}_{2_{\sigma},2_{\sigma}}$,
$\Gamma_{\rm R}=\Gamma^{N_\tau}_{(N-1)_{\sigma},(N-1)_{\sigma}}$ and the
reservoir mean occupations $\bar{n}_{\rm L}=\bar{n}_{1_\tau}$, $\bar{n}_{\rm
R}=\bar{n}_{N_\tau}$. Similar arguments apply to the vibron current, defined
through {$\partial_t{n}_{i_\alpha}=I_{ \rightarrow i_{\alpha}}^{\rm vib}-
I_{ i_{\alpha}\rightarrow}^{\rm vib}$}, which is independent of the TQW
length
\begin{equation}
\label{current}
\langle I_{ i_{\alpha}\rightarrow}^{\rm vib}\rangle_{\rm ss} =\langle I_{\rightarrow i_{\alpha}}^{\rm vib}\rangle_{\rm ss} = \frac{\Gamma_{\rm L} \Gamma_{\rm R}}{\Gamma_{\rm L} + \Gamma_{\rm R}}( \bar{n}_{\rm L}-\bar{n}_{\rm
R}),\hspace{2ex}\alpha,i_\alpha\in\mathfrak{C}.
\end{equation}
Numerical solutions of the complete dissipative dynamics in
Eq.~\eqref{bulk_liouvillian} fully confirm these predictions~\cite{sm}. {Our results are
 different from   {\it Fourier's law} of thermal conduction~\cite{Fourier}}
which predicts: {\it (i)} a linear temperature gradient, i.e., $\langle
n_{i_{\alpha}}\rangle_{\rm FL}=\bar{n}_{\rm L}+(\bar{n}_{\rm R}-\bar{n}_{\rm
L}) i_{\alpha}/N$. {\it (ii)} a heat current inversely proportional to the
length of the wire $\langle I_{ i_{\alpha}\rightarrow}^{\rm vib}\rangle_{\rm
FL}\propto(\bar{n}_{\rm L}-\bar{n}_{\rm R})/N$. The disagreement is expected since Fourier's law applies to diffusive processes;
in contrast, Eqs.~\eqref{occupation}-\eqref{current} describe ballistic
transport of vibrons, analogous to ballistic electronic
transport~\cite{cuevas}.

\begin{figure}[t]
\centering
\includegraphics[width=.9\columnwidth]{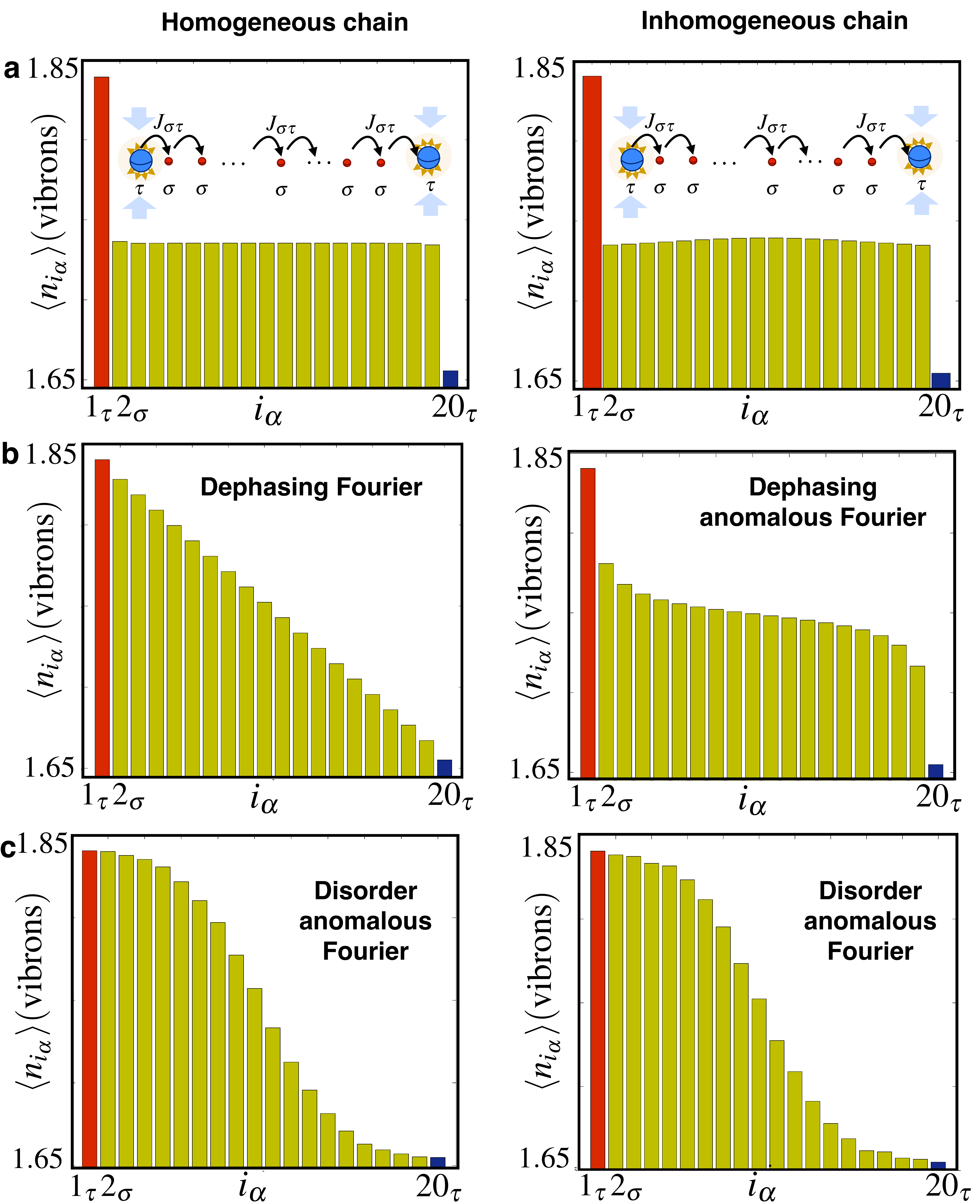}
\caption{ {\bf Fourier's law: } Vibron
distribution in the steady-state of a chain with $N=20$ ions
(left: microtrap array, right: linear Paul trap).  {\bf (a)}  Ballistic regime (agreeing with~\cite{steady_state_thermalization})
{\bf (b)} the dephasing-induced diffusive regime, and {\bf (c)} the disorder-induced
diffusive regime.} \label{fig_vibron_distribution}
\end{figure}

We now consider two phase-breaking processes resulting in a ballistic-diffusive
crossover: {\it dephasing}~\cite{dephasing_fl} and {\it
disorder}~\cite{disorder_fl}. Dephasing can be engineered by modulating trap
frequencies with a noisy voltage~\cite{dephasing_noise}. We model such noise
as dynamic fluctuations of the on-site energies in Eq.~\eqref{tbm},
$\omega_{i_\alpha}\to\omega_{i_\alpha}+\delta\omega_{i_\alpha}(t)$, with
$\delta\omega_{i_\alpha}(t)$ a random process. In a Born-Markov
approximation, this leads to an additional term in Eq.~\eqref{ddsp},
$\mathcal{L}_{\rm ddsv}\to\mathcal{L}_{\rm ddsv}+\mathcal{D}_{\rm d}$, where
\beq \nonumber \mathcal{D}_{\rm
d}(\bullet)=\sum_{\alpha,\beta}\sum_{i_\alpha, j_\beta}\Gamma_{\rm
d}\ee^{-\frac{|{\bf r}_{i_\alpha}^0-{\bf r}_{j_\beta}^0|}{\xi_{\rm
c}}}(n_{i_\alpha}\bullet n_{j_\beta}-n_{j_\beta}n_{i_\alpha}\bullet)+{\rm
H.c.}, \eeq
with the dephasing rate $\Gamma_{\rm d}$ and the noise correlation length
$\xi_{\rm c}$~\cite{comment}. Fig.~\ref{fig_vibron_distribution}{\bf (a)}
shows homogeneous vibron distributions along the TQW without dephasing. For
dephasing with $\xi_{\rm c}\ll L$, we observe the onset of a linear gradient
along the microtrap array {[Fig.~\ref{fig_vibron_distribution}{\bf (b)}
left]}, pinpointing diffusive transport. For the linear Paul trap, the
inhomogeneous crystal modifies the  gradient yielding an anomalous
Fourier's law {[Fig.~\ref{fig_vibron_distribution}{\bf (b)} right]}.

Disorder can be modeled by modifying the on-site energies of
Eq.~\eqref{tbm},
$\omega_{i_\alpha}\to\omega_{i_\alpha}+\Delta\omega_{\alpha}$, with
$\Delta\omega_{\alpha}$ a static random variable. To obtain such disorder,
we apply a strong static spin-vibron coupling~\eqref{driving} with
parameters $\nu_{\alpha}=0,\varphi_\alpha=0,\Delta\omega_\alpha^+=0$,
$\Delta\omega_{\alpha}^-\neq0$, such that the vibrons experience a
spin-dependent inhomogeneous landscape of on-site energies, resulting in
vibron scattering~\cite{Bermudez-NJP-2010}. {With each bulk spin initialised
in
$\ket{+_{i_{\alpha}}}=(\ket{{\uparrow_{i_{\alpha}}}}+\ket{{\downarrow_{i_{\alpha}}}})/\sqrt{2}$,
the tight-binding model becomes
stochastic $H_{\rm tb}\rightarrow H_{\rm stb}\!\!=\!\!\sum_{\alpha,i_{\alpha}}\!\epsilon_{i_{\alpha}}a_{i_{\alpha}}^{\dagger}a_{i_{\alpha}}^{\phantom{\dagger}}\!+\!\sum_{\alpha,\beta}\sum_{ i_{\alpha}\neq
j_{\beta}}\!\!{J}_{i_{\alpha}j_{\beta}}a_{i_{\alpha}}^{\dagger}a_{j_{\beta}}^{\phantom{\dagger}}\!+\!\text{H.c.}$. }Here, the on-site energies are binary random
variables sampling $\epsilon_{i_{\alpha}}\in\{\omega_ \alpha
-\half\Delta\omega^-_{{\alpha}},\omega_\alpha
+\half\Delta\omega^-_{{\alpha}}\}$ with probabilities
$p(\epsilon_{i_{\alpha}})=\half$ inherited from the quantum parallelism.
This randomness leads to Anderson localization, whereby normal modes display
a finite localization length $\xi_{\rm loc}$~\cite{and_loc}. For the small
ion crystals of length $L$, the modes with $\xi_{\rm loc} \gg L$ contribute
ballistically, those with $\xi_{\rm loc} \lesssim L$ introduce diffusion,
and with $\xi_{\rm loc} \ll L$ do not contribute to transport. We thus
expect that the heat transport is much richer in the disordered case.
Figure~\ref{fig_vibron_distribution}{\bf (c)} shows the disorder-averaged
distribution of vibron occupations along the TQW, where we find clear  anomalies in Fourier's law, measurable in  experiments.

To distinguish  ballistic from diffusive transport, we suggest a
measuring scheme inspired by~\cite{ramsey1,ramsey2}. We map the
mean {value} of  any vibron operator $\langle O_{i_\alpha}\rangle_{\rm ss}$,
and its fluctuation spectrum
\beq
\nonumber
S_{O_{i_\alpha}O_{i_\alpha}}\hspace{-0.5ex}(\omega)=\hspace{-0.5ex}\int_0^{\infty}\hspace{-1.5ex}{\rm d}t\langle \tilde{O}_{i_\alpha}(t)\tilde{O}_{i_\alpha}(0)\rangle_{\rm ss}\ee^{-\ii\omega t},\hspace{1ex}
\tilde{O}_{i_\alpha}={O}_{i_\alpha}-\langle\tilde{O}_{i_\alpha}\rangle_{\rm ss},
\eeq
onto the spin coherences, while disturbing the vibron states minimally. This is
achieved through Ramsey-type interferometry based on engineered spin-vibron
interactions, $ \tilde{H}_{\rm sv}^{O}=\sum_{i_{\kappa}}\half\lambda_OO_{i_
\kappa}\sigma_{i_ \kappa}^z$, with weak coupling $\lambda_O$~\cite{sm}.

A single $\kappa$-ion~\cite{comment_2} initialised in the state
$\ket{+_{i_{\kappa}}}$ by a $\pi/2$-pulse acquires phase information about
the steady-state vibron observable. We perform another $\pi/2$-pulse and
measure the probability of observing the state
$\ket{{\downarrow_{i_{\kappa}}}}$, which is equivalent to measuring the spin
coherences
\begin{equation}
\label{ramsey}
\langle\tilde{\sigma}_{i_{\kappa}}^x(t)\rangle=\cos(\lambda_{O}\langle O_{i_{\kappa}}\rangle_{\rm ss}t)\ee^{-\lambda_{O}^2{\rm Re}\{S_{{O}_{i_\kappa}{O}_{i_\kappa}}\hspace{-0.5ex}(0)\}t}.
\end{equation}
Therefore, the period (decay) of the spin oscillations yields the mean value
(zero-frequency fluctuations) of the vibron operator $\langle
O_{i_\kappa}\rangle_{\rm ss}$ ($S_{O_{i_\kappa}O_{i_\kappa}}(0)$).
Considering the excellent accuracies achieved in projective spin
measurements~\cite{haeffner_review}, probing steady-state vibrons with this
method promises to be very efficient. For measuring the mean vibron number
$O_{i_\kappa}=n_{i_\kappa}$ we choose a weak static spin-vibron
coupling~\eqref{driving} with
$\nu_{\kappa}=0,\varphi_\kappa=0,\Delta\omega_\sigma^+=0$, and
$\Delta\omega_\kappa^-\neq0$. Similarly, vibron density-density correlators
can be probed by using several $\kappa$-ions.

{\it Thermal quantum dot and single-spin heat switch.--} {The TQD is formed by a single $\kappa$-ion at position $p_{\kappa}$
in the center of the bulk.} We use the remaining $\sigma$-ions as thermal contacts
by employing a strong static spin-phonon coupling~\eqref{driving} with
parameters $\nu_{\sigma}=0,\varphi_\sigma=0,\Delta\omega_ \sigma^+=0$, and
$\Delta\omega_\sigma^-\neq0$. If the spins are initialised in
$\ket{\psi_0}=\ket{{\downarrow_{\sigma}\cdots\downarrow_{\sigma}}}\ket{{\phi_{{\kappa}}}}\ket{{\uparrow_{\sigma}\cdots\uparrow_{\sigma}}}$
there is a large shift of the on-site energies across $p_\kappa$, inhibiting
tunneling through the TQD. The two halves of the chain thus
thermalize independently, i.e., $ \langle n_{i_\sigma}\rangle_{\rm
ss}=\bar{n}_{\rm L}$ for $i_{\sigma}<p_{\kappa}$ and $\langle
n_{i_\sigma}\rangle_{\rm ss}=\bar{n}_{\rm R}$ for $i_{\sigma}>p_{\kappa}$,
{functioning as thermal leads connected to the quantum dot.} The
Liouvillian is $\mathcal{L}_{\rm ddtb}^{\rm bulk}=\mathcal{L}_{\rm
L}+\mathcal{L}_{{\rm L}\kappa{\rm R}}+\mathcal{L}_{\rm R}$, where
$\mathcal{L}_{\rm L/R}$ describe  the uncoupled
halves~\eqref{bulk_liouvillian} and $\mathcal{L}_{\rm L\kappa R}$ describes
the TQD. Transport through the TQD is achieved by using a dynamical
spin-vibron coupling~\eqref{driving} for the $\kappa$-ion. For
spin-independent drivings $\Delta\omega^-_{\kappa}=0$, the periodic
modulation of the on-site energies results in photon-assisted tunneling
overcoming the on-site energy
gradient {between the dot and the leads}~\cite{photon_assisted_tunneling_ions}. We exploit the
spin-dependence of this driving to build a {\it single-spin heat switch} and
a {\it current probe}.

For the single-spin heat switch, the parameters of the spin-vibron
coupling~\eqref{driving} are
$\nu_{\kappa}=\half\Delta\omega^-_{\sigma},\hspace{1ex}
\varphi_{\kappa}=0,\hspace{1ex}\Delta\omega_{\kappa}^-=\Delta\omega_{\kappa}^+$,
which lead to $\mathcal{L}_{{\rm L}\kappa{\rm R}}(\bullet)=-\ii[H_{{\rm
L}\kappa{\rm R}}^{\rm PAT},\bullet]+\mathcal{D}_{p_\kappa
p_\kappa}(\bullet)$, where
\begin{equation}
\nonumber
H_{{\rm L}\kappa{\rm R}}^{\rm PAT}\!=-\!\!\!\!\!\sum_{i_{\sigma}<p_{\kappa}}\!\!\!J_{p_{\kappa}i_{\sigma}}^{\rm
PAT}(\sigma_{p_\kappa}^z)a_{p_{\kappa}}^{\dagger}a_{i_{\sigma}}^{\phantom{\dagger}}+\!\!\!\!\!\sum_{i_{\sigma}>p_{\kappa}}\!\!\!\!J_{p_{\kappa}i_{\sigma}}^{\rm
PAT}(\sigma_{p_\kappa}^z)a_{p_{\kappa}}^{\dagger}a_{i_{\sigma}}^{\phantom{\dagger}}+\text{H.c.}.
 \end{equation}
The tunneling is spin-dependent, i.e., $J_{p_{\kappa}i_{\sigma}}^{\rm
PAT}(\sigma_{p_\kappa}^z)\ket{{\downarrow_{p_\kappa}}}=0$ and
$J_{p_{\kappa}i_{\sigma}}^{\rm
PAT}(\sigma_{p_\kappa}^z)\ket{{\uparrow_{p_\kappa}}}\neq 0$, because of the
operator argument in the first Bessel function
$J_{p_{\kappa}i_{\sigma}}^{\rm
PAT}(\sigma_{p_\kappa}^z)=\tilde{J}_{p_{\kappa}i_{\sigma}}\mathfrak{J}_1\big[\zeta_\kappa(1+\sigma_{p_\kappa}^z)\big]$,
with $\zeta_{{\kappa}}=\Delta\omega_{\kappa}^+/2\nu_{\kappa} $~\cite{sm}.
Therefore, by controlling the $\kappa$-spin state via microwave
$\pi$-pulses, we can switch on/off the heat current through the
TQD. Different switches have  been studied
in~\cite{entanglement_harmonic_lattices} to control the entanglement
 in harmonic chains.

Probing vibron currents requires a minimally-perturbing mapping of the
current onto the $\kappa$-spin. This requires a bichromatic spin-vibron
coupling~\eqref{driving} with specific parameters~\cite{comment_parameters}.
The first frequency {induces photon-assisted tunneling}
\beq \nonumber H_{\rm L\kappa
R}^{\rm PAT}=\sum_{i_\sigma}(\tilde{J}^{\rm
PAT}_{i_\sigma,p_\kappa}a_{i_\sigma}^{\dagger}a_{p_\kappa}^{\phantom{\dagger}}+{\rm
H.c.}),\hspace{2ex}\tilde{J}^{\rm
PAT}_{i_\sigma,p_\kappa}=-\ii2\tilde{J}_{i_\sigma
p_\kappa}\mathfrak{J}_1(\pi), \eeq
such that the  tunneling amplitude becomes
purely imaginary. This is crucial to devise the probe since the second
frequency leads to {the necessary spin-current interactions} { $ H_{\rm sv}^{I}=\half\tilde{\lambda}_I
I_{p_\kappa}^{\rm vib}\sigma_{p_\kappa}^z$, where
$I_{p_\kappa}=\half(I_{p_\kappa\rightarrow}^{\rm vib}+I_{\rightarrow
p_\kappa}^{\rm vib})$. In the limit  $\zeta_{\kappa,2}\to 0$,
$\tilde{\lambda}_I\approx4\zeta_{\kappa,2}/\pi $ we get a Ramsey
probe~\eqref{ramsey} for the current mean value $\langle I^{\rm
vib}_{p_\kappa}\rangle$ and fluctuations $S_{I^{\rm vib}_{p_\kappa}I^{\rm
vib}_{p_\kappa}}(0)$.

Measuring fluctuations is essential for comparing fermionic and bosonic
currents via the Fano factor $\mathcal{F}=S_{I^{\rm vib}_{p_\kappa}I^{\rm
vib}_{p_\kappa}}(0)/2\langle I^{\rm vib}_{p_\kappa}\rangle$. For heat
currents through a symmetrically coupled TQD, we expect strong
super-Poissonian fluctuations {$\mathcal{F}\gg 1$}, which increase linearly with $\bar{n}_{L}$ in
the regime $\bar{n}_{L}\gg\bar{n}_{R}$~\cite{Harbola-PRB-2007,kindermann}.
Unlike the sub-Poissonian fluctuations {$\mathcal{F}< 1$} in electrical currents,
super-Poissonian fluctuations in heat currents have not been observed yet.

\vspace{1ex} {\it Conclusions.--} We have outlined the implementation of an
ion-trap toolbox for quantum heat transport, which provides {\it (i)}
thermal reservoirs, quantum dots and wires; {\it (ii)} engineered on-site
disorder and dephasing, and {\it (iii)} noninvasive probes for vibron
occupations and currents.  It would be of the utmost interest to assess  the validity of the proposed probes  for capturing the full counting statistics of  heat transport. All these functionalities significantly extend the
possible range of experiments on heat transport. Laser-cooled edge ions in
coherent or squeezed vibron states~\cite{squeezed} may constitute valuable
supplementary gadgets. We expect, moreover, interesting effects in the
presence of non-linearities, e.g., the interplay with Mott
insulators~\cite{porras_hubbard_model}, competition between dephasing and
interactions~\cite{clark}, thermal rectification~\cite{nitzan}, and
structural phase transitions~\cite{zigzag}. In a non-equilibrium version of
the spin-Peierls instability~\cite{peierls} correlations between structural
change and heat currents may be explored. 

\vspace{5pt}

A.B., M.B. and M.B.P are supported by PICC and the Alexander von Humboldt
Foundation. A.B. thanks FIS2009-10061, QUITEMAD.




\vspace*{50ex}
\newpage

\hypertarget{sm}{\section{Supplemental Material:\\ Controlling and Measuring the Quantum Transport of Heat in Trapped-ion Crystals}}

\appendix

We present a detailed derivation of the expressions used in the main text, and test their validity  by  comparing the analytical expressions to numerical results for ion-trap setups with
realistic parameters. Therefore, this SM will also be useful to guide an experimental realisation of quantum heat transport.

\begingroup
\hypersetup{linkcolor=black}
\tableofcontents
\endgroup

\section{Trapped-ion toolbox for quantum transport}

We present a detailed derivation, supported by numerics, of  our toolbox gadgets: the vibronic tight-binding model~\eqref{tbm}, the controlled
dissipation~\eqref{dissipation}, and the spin-vibron coupling~\eqref{driving}.

\subsection{Tight-binding model for the vibrons} Let us start by introducing the notation. We consider  an ensemble of
$N=N_{\sigma}+N_{\tau}+N_{\kappa}$ ions of three different species/isotopes with mass $m_{\alpha}$ and charge $e$. These ions are   labelled by latin indexes $i,j\in\{1\cdots N\}$, and by greek sub-indexes $\alpha,\beta\in\{\sigma,\tau,\kappa\}$ specifying the particular   ion species. The dynamics of the ions is controlled by the Hamiltonian
\begin{equation}
\label{coulomb}
H=\!\sum_{\alpha,i_{\alpha}}\!\!\bigg(\frac{{\bf p}^2_{i_{\alpha}}}{2m_{\alpha}}+\frac{1}{2}m_{\alpha}{\bf r}_{i_{\alpha}}\cdot\boldsymbol{\omega}_{\alpha}^2\cdot{\bf r}_{i_{\alpha}}\bigg)+\frac{e_0^2}{2}\!\sum_{\alpha,\beta}\sum_{i_{\alpha}\neq j_{\beta}}\!\frac{1}{|{\bf r}_{i_{\alpha}}-{\bf r}_{j_{\beta}}|},
\end{equation}
where the matrix
${\bf \omega}_{\alpha}={\rm diag}(\omega_{\alpha x},\omega_{\alpha y},\omega_{\alpha z})$ contains the trap frequencies along the different axes, and $e_0^2=e^2/4\pi\epsilon_0$ is expressed in terms of the vacuum permittivity $\epsilon_0$. At low-enough temperatures, the ions from a Wigner-type crystal with a geometry that depends on the trapping potential. We shall focus on  linear ion chains with equilibrium positions ${\bf
r}_{i_{\alpha}}^0=z_{i_{\alpha}}^0{\bf e}_{z}$. For linear Paul traps, one obtains an inhomogeneous
chain with ions   closer at the centre than  at the boundaries~\cite{james}. By segmenting the electrodes, it is possible to make
the crystal  more homogeneous~\cite{quartic}. Moreover, with the advent of the so-called micro-fabricated surface traps, the ion lattice can be designed at will~\cite{surface_traps}. Therefore, we will also investigate homogeneous  ion chains for quantum heat transport.

As customary, a Taylor expansion to second order in the small displacements around the equilibrium positions leads to a quadratic model: the  harmonic crystal~\cite{feynman}. For the ion chain, the
vibrations along each direction  decouple~\cite{james}, and we can focus  on the transversal direction $\delta{x}_{i_{\alpha}}{\bf e}_x$. The harmonic crystal contains a coupling between the vibrations of distant ions, which can  be understood as the result of a dipole-dipole interaction $V_{\rm dd}\propto { d}_{i_{\alpha}}{ d }_{j_{\beta}}/|z^0_{i_{\alpha}}-z^0_{j_{\beta}}|^3$  between the
 effective dipoles ${ d}_{i_{\alpha}}=e \delta x_{i_{\alpha}} $ induced by the vibrating  charges~\cite{blatt_tunneling}. Quantising the vibrations via  the  creation-annihilation operators
\begin{equation}
\label{ca}
\textstyle{p_{i_{\alpha}}=\ii\sqrt{\frac{m_{\alpha}\omega_{\alpha}}{2}}\big(a_{i_{\alpha}}^{\dagger}-a_{i_{\alpha}}^{\phantom{\dagger}}\big), \hspace{1.ex}
\delta x_{i_{\alpha}}=\sqrt{\frac{1}{2m_{\alpha}\omega_{\alpha}}}\big(a_{i_{\alpha}}^{\dagger}+a_{i_{\alpha}}^{\phantom{\dagger}}\big),}
\end{equation}
 we get  the announced quadratic model for lattice vibrons
\begin{equation}
\nonumber
\begin{split}
H&=\sum_{\alpha,
i_{\alpha}}\omega_{\alpha}a_{i_{\alpha}}^{\dagger}a_{i_{\alpha}}^{\phantom{\dagger}}+\frac{1}{2}\sum_{\alpha,\beta}\sum_{i_{\alpha},j_{\beta}}J_{i_{\alpha}j_{\beta}}\big(a_{i_{\alpha}}^{\dagger}+a_{i_{\alpha}}^{\phantom{\dagger}}\big)\big(a_{j_{\beta}}^{\dagger}+a_{j_{\beta}}^{\phantom{\dagger}}\big),
\end{split}
\end{equation}
where we have defined the tunneling strengths for $i_\alpha\neq j_\beta$
\beq
\nonumber
J_{i_{\alpha}j_{\beta}}=\frac{e_0^2}{2\sqrt{m_{\alpha}\omega_{\alpha}m_{\beta}\omega_{\beta}}}\frac{1}{|z^0_{i_{\alpha}}-z^0_{j_{\beta}}|^3},
\eeq
and the renormalization of  the trap frequencies of an ion due to its surrounding ions $J_{i_{\alpha}i_{\alpha}}=-\sum_{\beta}\sum_{j_{\beta}\neq i_\alpha}J_{i_\alpha,j_\beta}$.

To obtain the desired tight-binding
model~\eqref{tbm}, we need  to neglect  terms in the Hamiltonian that do not
conserve the number of vibrons. This is justified if the trap frequencies
are much stronger than the tunneling~\cite{porras_hubbard}, namely
$J_{i_{\alpha}j_{\beta}}\ll\omega_{\alpha}+\omega_{\beta}$. Using a rotating
wave approximation (RWA), the Hamiltonian becomes
the sum of the vibronic on-site energy  $H_{\rm vo}$, and the
tunneling $H_{\rm vt}$. This gives rise to  the  tight-binding model
$H_{\rm tb}=H_{\rm vo}+H_{\rm vt}$ of Eq.~\eqref{tbm} in the main text,
namely
\begin{equation}
\label{vibron_tbm}
\begin{split}
H_{\rm vo}&\!\!=\!\!\sum_{\alpha, i_{\alpha}}\!\!\omega_{i_\alpha}a_{i_{\alpha}}^{\dagger}a_{i_{\alpha}}^{\phantom{\dagger}}, \hspace{1 ex}
H_{\rm vt}\!=\!\!\sum_{\alpha,\beta}\sum_{i_{\alpha}\neq j_{\beta}}\!\!\!\!\!\big(J_{i_{\alpha}j_{\beta}}a_{i_{\alpha}}^{\dagger}a_{j_{\beta}}^{\phantom{\dagger}}\!+\text{H.c.}\big),
\end{split}
\end{equation}
where $\omega_{i_\alpha}=\omega_\alpha+J_{i_\alpha i_\alpha}$.
As   demonstrated in recent experiments~\cite{blatt_tunneling,wineland_tunneling,paul_trap_tunneling}, this model allows for a controlled tunneling of vibrons between different ions.

\begin{table}
  \centering
   \caption{{\bf Vibrational parameters for each ion species} }
  \begin{tabular}{  c c c c }
\hline
\hline
 $\omega_{\alpha}/2\pi$$\,$ &   $\,$$|z_{i_{\alpha}}^0-z_{(i+1)_{\beta}}^0|$$\,$ & $\,$$J_{i_{\alpha}{{(i+1)}_\beta}}/2\pi$$\,$ & $\,$$\mu_{\alpha}=\frac{m_{\alpha}}{m_{\sigma}}$ \\
\hline
\hline
$1$-$10$\,${\rm MHz}$ & $1$-$10$\,$\mu{\rm m}$&1-100$\,${\rm kHz} &1-10 \\
\hline
\hline
\end{tabular}
  \label{table_vib}
\end{table}

In order to show that  the approximations leading to the  tight-binding model are satisfied, we perform a numerical comparison of the dynamics under the original~\eqref{coulomb}
and the effective~\eqref{vibron_tbm} Hamiltonians. The typical orders of magnitude for the vibronic parameters are summarised in Table~\ref{table_vib}. Let us consider a particular example of a two-ion crystal with the species $\sigma=$$^{25}{\rm Mg}^{+}$,
and $\tau=$$^{24}{\rm Mg}^+$.  The trap frequencies  are $(\omega_{\alpha x},\omega_{\alpha y},\omega_{\alpha z})/2\pi=(5,5,0.5)\hspace{0.2ex}$MHz, which lead to an  inter-ion distance of
$|z^0_{1_\tau}-z^0_{2_{\sigma}}|\approx10\hspace{0.2ex}\mu$m, and to a vibron tunneling strength of $J_{1_{\tau}2_{\sigma}}/2\pi\approx12\hspace{0.2ex}$kHz. We consider an initial pure state with a single vibronic excitation  in the $^{25}{\rm Mg}^{+}$ ion, which should be periodically interchanged with the  neighbouring $^{24}{\rm Mg}^{+}$ ion. In Fig.~\ref{fig1_app}, we observe the agreement of both descriptions through the
predicted periodic  tunneling. This simulation shows that the approximations leading to the tight-binding model are very accurate for realistic  parameters.

\begin{figure}
\centering
\includegraphics[width=.85\columnwidth]{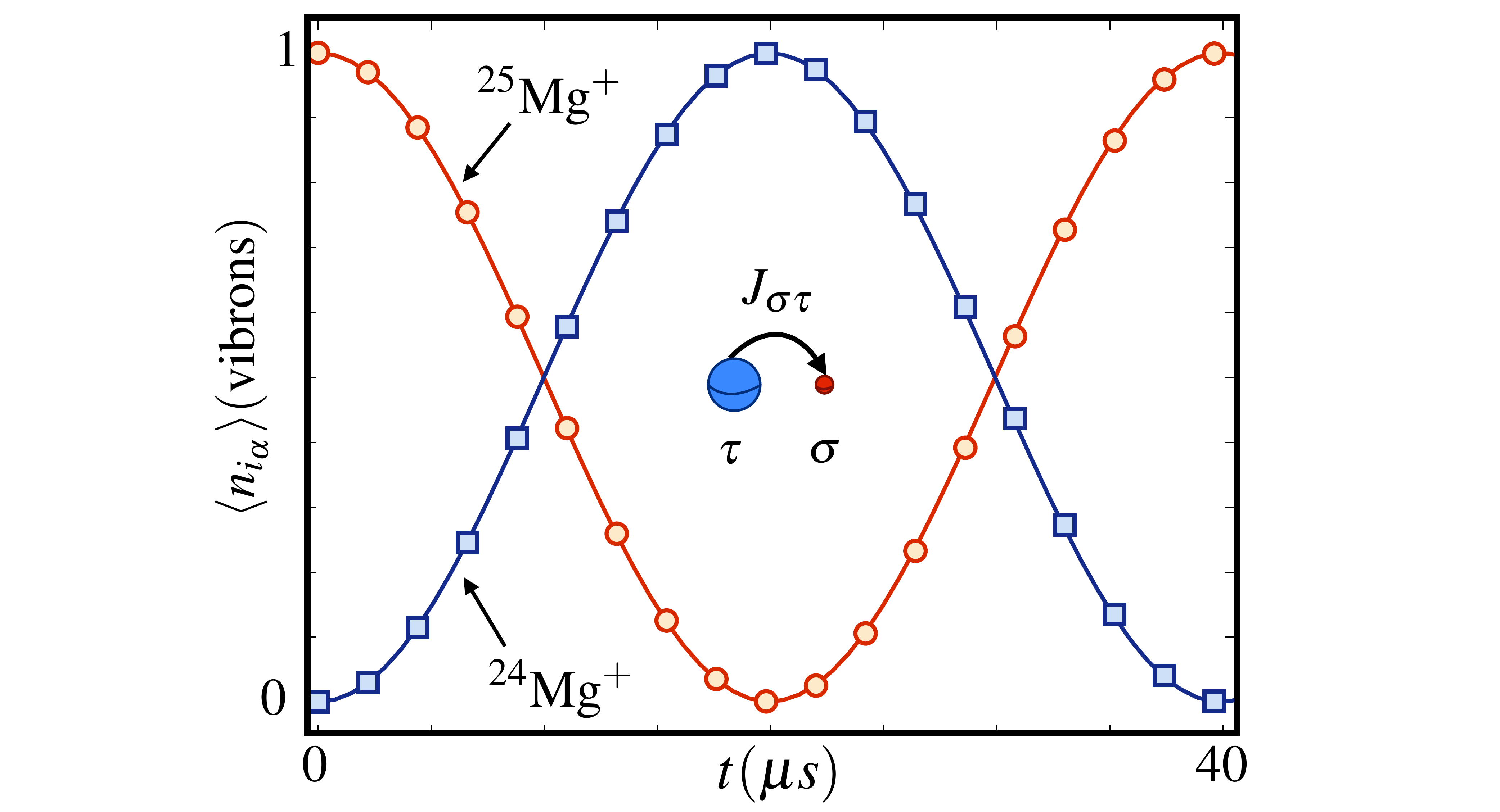}
\caption{ {\bf Vibronic quantum dynamics:} Exchange of a vibrational quantum (i.e. vibron) between two distant  $\sigma=$$^{25}{\rm Mg}^{+}$, and $\tau=$$^{24}{\rm Mg}^+$ ions. The solid lines represent the vibronic numbers  given by the original Coulomb Hamiltonian~\eqref{coulomb} (red: $\langle n_{1_\sigma}\rangle$ for $^{25}{\rm Mg}^{+}$, blue: $\langle n_{2_\tau}\rangle$ for
$^{24}{\rm Mg}^{+}$), whereas the symbols stand for the vibronic numbers given  the effective tight-binding model~\eqref{vibron_tbm} (red squares: $\langle n_{1_\sigma}\rangle$ for $^{25}{\rm Mg}^{+}$, blue circles: $\langle
n_{2_\tau}\rangle$ for $^{24}{\rm Mg}^{+}$). To
obtain the dynamics numerically, we truncate the vibron Hilbert space to $n_{\rm max}=2$, and consider the three vibrational axes (i.e. 6 vibronic modes) with Coulomb non-linearities taken up to $8$-th order (e.g.
$a_{i_{\alpha}}^8$)}
\label{fig1_app}
\end{figure}

\subsection{Atomic degrees of freedom}
To derive the  controlled-dissipation gadget~\eqref{dissipation}, we need to describe first  the atomic degrees of freedom. The different ion species are divided into two groups, depending on wether we
  exploit their coherent $\mathfrak{C}$ or incoherent $\mathfrak{I}$ (i.e. dissipative) dynamics. We select   $\tau\in\mathfrak{I}$, and $\sigma,\kappa\in\mathfrak{C}$. To ease  notation, we will focus on a single ion, and keep in mind that we have to summed over all the ions in the crystal in the next sections.

 {\it Dipole-allowed transition.--} Let us start by selecting a dipole-allowed transition $\ket{{\downarrow_{\tau}}}\leftrightarrow\ket{{\uparrow_{\tau}}}$ for the $\tau$-ions [Fig.~\ref{level_scheme}{\bf (a)}]. In the absence of  laser beams, the dynamics of the atomic density matrix is given by $\dot{\rho}=-\ii[H_{\rm s}^{\tau},\rho]+\mathcal{D}_\tau(\rho)$. This master equation contains a Hamiltonian part  $H^\tau_{\rm s}=\half\omega_0^{\tau}\sigma^z$, where $\omega_0^{\tau}$ is the transition frequency and $\sigma^z=\ketbra{{\uparrow_\tau}}-\ketbra{{\downarrow_\tau}}$, and a dissipative part
characterised by a spontaneous decay  rate $\Gamma_{\tau}$. Considering
the recoil  by the emitted photons~\cite{recoil_master_equation}, the
dissipation is described by
\begin{equation}
\begin{split}
\label{recoil_dissipator}
\mathcal{D}_{\tau}(\bullet)\!=\!\int\!\!{\rm d}\xi\big(\sigma^-\ee^{-\ii k_{\tau}{\bf u}_{k}\cdot{\bf r}} \bullet \ee^{\ii k_{\tau}{\bf
u}_{k}\cdot{\bf r}}\sigma^+-\sigma^+\sigma^-\bullet+\text{H.c.}\big).
\end{split}
\end{equation}
 Here, we have defined  the raising-lowering operators  $\sigma^{+}=\ket{{\uparrow_\tau}}\bra{{\downarrow_\tau}}=(\sigma^{-})^{\dagger}$, and integrated
 (summed)
 over all    different  directions (polarisations) of the emitted photon $\int{\rm d}\xi= \frac{3\Gamma_{\tau}}{16\pi}\int_0^{4\pi}{\rm d}\Omega_{{\bf u}_{k}}\sum_{{\boldsymbol{\epsilon}}}|\boldsymbol{\epsilon}\cdot{\bf
 u}_{\tau}|^2$. In this expression,  $k_{\tau}{\bf u}_{k}$ is the wavevector of the emitted photon, whose modulus is determined by energy conservation
 $k_{\tau}=\frac{\omega_\tau}{c}$, and whose  direction is specified by the unit vector ${\bf u}_{k}$. Additionally, we have introduced the unit vectors of the atomic dipole operator ${\bf u}_{\tau}$, which depend on the
 angular-momentum selection rules, and thus on the polarisation of the emitted photon $q\in\{0,\pm1\}$~\cite{recoil_master_equation}.

\begin{figure}
\centering
\includegraphics[width=.85\columnwidth]{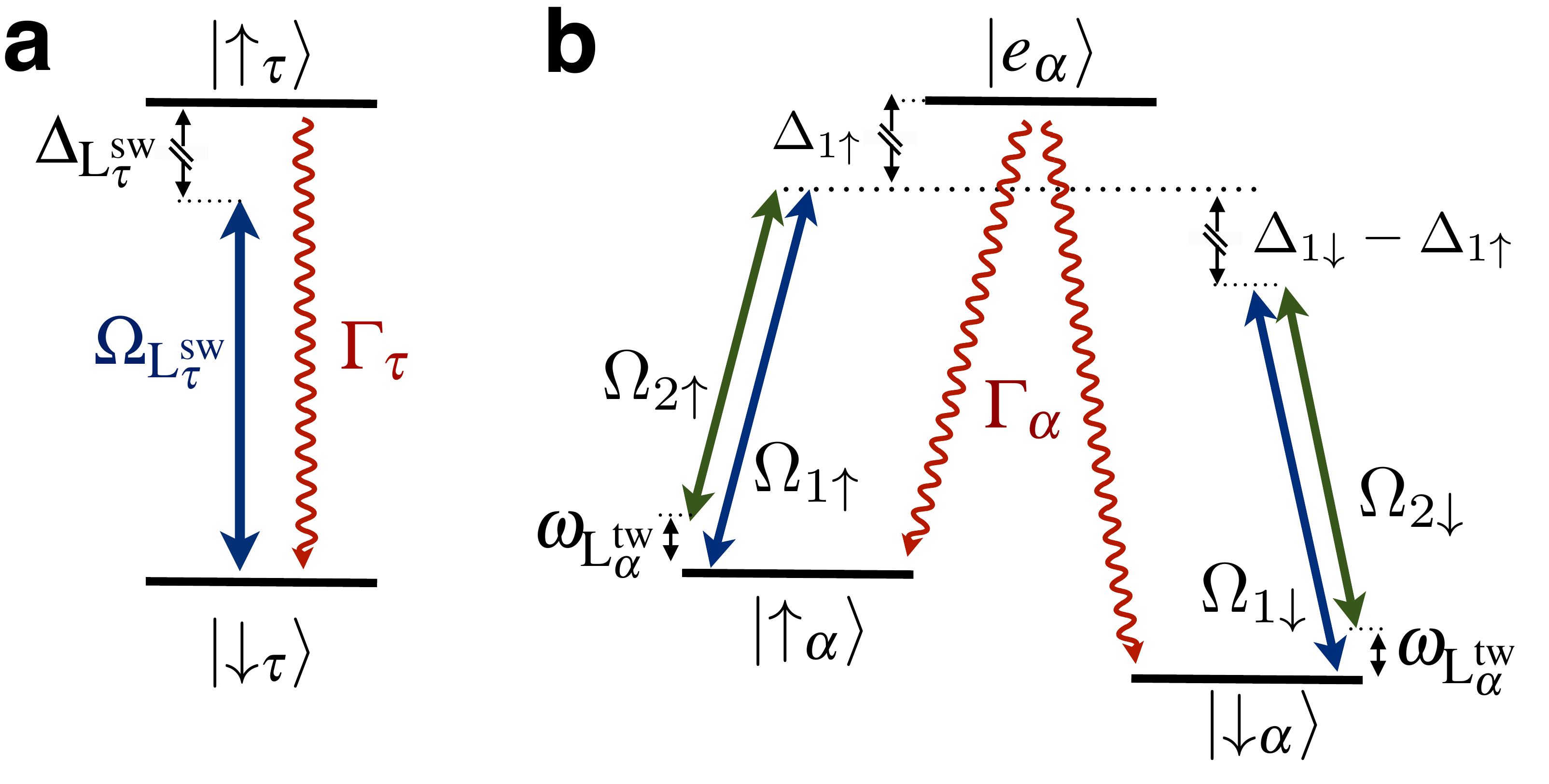}
\caption{ {\bf Atomic level scheme for the different ions:} {\bf (a)} Two-level scheme $\{{\ket{\uparrow_\tau}},\ket{{\downarrow_\tau}}\}$ for a dipole-allowed transition with decay rate $\Gamma_{\tau}$ of the $\tau$-ions, which is
driven by a laser in a standing-wave configuration ${\rm L}_{\tau}^{\rm sw}$, such that the Rabi frequency is $\Omega_{{\rm L}^{\rm sw}_{\tau}}$. The standing wave is red-detuned $\Delta_{{\rm L}_{\tau}^{\rm sw}}<0$ from the
atomic
transition, such that we can use it for laser cooling. {\bf (b)} Three-level $\Lambda$-scheme $\{\ket{{\uparrow_\alpha}},\ket{{\downarrow_\alpha}},\ket{{e_{{\alpha}}}}\}$  for a dipole-allowed transition with decay rate
$\Gamma_{\alpha}$ of the $\alpha =\{\sigma,\kappa\}$-ions. We use a couple of laser beams in a travelling-wave configuration, such that their Rabi frequencies for each of the optical transitions are
$\Omega_{ls}$,
where $s=\uparrow,\downarrow$ stands for the two possible ground-states, and  $l=1, 2$ stands for the two laser beams. In this case, the corresponding detunings $\Delta_{ls}$ will be much
larger than any other scale of the problem, such that we can manipulate the state of the ion in this ground-state manifold by tuning the
effective laser frequency  $\omega_{{\rm L}_{\alpha}^{\rm tw}}=\omega_{1}-\omega_{2}$ of the two beams.}
\label{level_scheme}
\end{figure}

This expression can be simplified further if the vibrations are
much smaller than the wavelength of the emitted light, namely ${\bf
r}={\bf r}^0+\delta{\bf r}$, such
that $k_\tau|{\bf u}_{k}\cdot \delta {\bf r}|\ll1$
(i.e. Lamb-Dicke limit). By Taylor expanding the
dissipator~\eqref{recoil_dissipator}, we find that in analogy with the Coulomb couplings~\eqref{vibron_tbm}, the recoil events to second order do not couple the vibrations along different directions. Therefore, we can focus on the transverse
vibrations along the $x$-axis directly~\eqref{ca}, and rewrite the dissipator~\eqref{recoil_dissipator} as  $\mathcal{D}_{\tau}(\rho)=\mathcal{D}_{\tau}^0(\rho)+\mathcal{D}_{\tau}^1(\rho)$, where
\begin{equation}
\label{do}
\mathcal{D}_{\tau}^0(\bullet)=\frac{\Gamma_{\tau}}{2}\big(\sigma^-\bullet
\sigma^+-\sigma^+\sigma^-\bullet+\text{H.c.}\big),
\end{equation}
describes the spontaneous emission of a collection of  atoms with mutual distances much larger than the wavelength of the emitted light. In addition, the recoil effects are contained in
\begin{equation}
\label{d1}
\begin{split}
\mathcal{D}_{\tau}^1(\bullet)\!=\!\!\frac{\tilde{\Gamma}_{{\tau}}}{2}\sigma^-\!&\big(
\!(a^{\dagger}+a)\bullet(a^{\dagger}+a)-(a^{\dagger}+a)^2\bullet\big)\sigma^++\text{H.c.},
\end{split}
\end{equation}
where  $\tilde{\Gamma}_{{\tau}}=\Gamma_{\tau}\eta_{\tau}^2(2+q^2)/(5(1+q^2))$  is smaller than the bare dissipation~\eqref{do} since
$\eta_{\tau}=k_{\tau}/\sqrt{2m_{\tau}\omega_{\tau}}\ll1$. According to this expression, the photon recoil leads to dissipative events where the number of vibrons is modified.

To have further control over the vibrons, we include a laser beam tuned close to the resonance of the dipole-allowed transition (Fig.~\ref{level_scheme}{\bf (a)}). The master equation   is
\begin{equation}
\label{incoherent_master_equation}
\dot{\rho}=-\ii[H_{\rm s}^{\tau}+H_{{\rm L}^{\rm sw}_{\tau}},\rho]+\mathcal{D}_{\tau}(\rho)
\end{equation}
where  the laser-ion interaction is given by
\begin{equation}
\label{laser_ion}
H_{{\rm L}^{\rm sw}_{\tau}}=-({\bf d}_{{\tau}}\sigma^++{\bf d}_{{\tau}}^*\sigma^-)\cdot {\bf E}_{{\rm L}^{\rm sw}_{\tau}}({\bf r},t),
\end{equation}
and we have introduced the laser electric field  ${\bf E}_{{\rm L}^{\rm sw}_{\tau}}({\bf r},t)$, and the atomic dipole ${\bf d}_{{\tau}}$. For reasons that will become clear later, we need cooling rates that are
much stronger than the vibron tunnelings~\eqref{vibron_tbm}. Therefore, the laser beam is arranged in a standing-wave configuration~\cite{laser_cooling}, ${\bf E}_{{\rm L}^{\rm sw}_{\tau}}({\bf
r},t)=\boldsymbol{\epsilon}_{{\tau}}E_{\tau}\cos(\omega_{{\rm L}^{\rm sw}_{\tau}}t)\cos({\bf k}^{\rm sw}_{{\rm L}_{\tau}}\cdot{\bf r})$, where $\boldsymbol{\epsilon}_{\tau},E_{\tau},\omega_{{\rm L}^{\rm
sw}_{\tau}}$ are the polarisation, amplitude, and frequency of the laser, and ${\bf k}_{{\rm L}^{\rm sw}_{\tau}}$ is the laser wavevector directed along the $x$-axis (i.e. direction of the vibrons). Let us also introduce the  Rabi frequency $\Omega_{{\rm L}^{\rm
sw}_{\tau}}=-E_{\tau}{\bf d}_{\tau}\cdot\boldsymbol{\epsilon}_{{\tau}}$. Besides, we consider that the
axis of the  ion-chain  lies at the node of the standing wave.
\begin{table}
  \centering
   \caption{{\bf Atomic and laser parameters for each ion species}  }
  \begin{tabular}{ c  c c c c c }
\hline
\hline
$\alpha$&$\omega^{\alpha}_0/2\pi$ & $\Gamma^{\rm eff}_{\alpha}/2\pi$ &${\rm L}_{\alpha}$ & $\omega_{{\rm L}_{\alpha}}$& $\Omega_{{\rm L}_{\alpha}}$\\
\hline
\hline
$\tau$ &$10^{2}$-$10^3\,{\rm THz}$ & 10\,${\rm MHz}$& {\rm 1-photon}& $\omega_{{\rm L}^{\rm sw}_{\tau}}\!\!=\!\omega^{\tau}_0\!-\!\frac{\Gamma_{\tau}}{2}$ &1-10\,MHz\\
\hline
$\sigma$ & $1$-$10\,{\rm GHz}$ & $1$-$10\,{\rm Hz}$ & {\rm 2-photon} &$\omega_{{\rm L}^{\rm tw}_{\sigma}}=0$ &0.1-10\,kHz\\
\hline
$\kappa$ & $1$-$10\,{\rm GHz}$  & $1$-$10\,{\rm Hz}$ & {\rm 2-photon} &$\omega_{{\rm L}^{\rm tw}_{\kappa}}\!\!\!\ll\!\omega_{\alpha}\!\ll\!\omega^{\kappa}_0$ &0.1-10\,kHz\\
\hline
\hline
\end{tabular}
  \label{table_spins}
\end{table}

 {\it Three-level $\Lambda$ scheme.--} We now focus on the remaining species $\alpha\in\{\sigma,\kappa\}$, where two dipole-allowed transitions  $\{\ket{{\uparrow_\alpha}},\ket{{\downarrow_\alpha}}\}\leftrightarrow\ket{{e_{{\alpha}}}}$ define a so-called $\Lambda$-scheme [Fig.~\ref{level_scheme}{\bf (b)}], and can be described by dissipators analogous to those of the $\tau$-ions~\eqref{do}-\eqref{d1}. As we want to exploit the coherent dynamics, $\alpha\in\{\sigma,\kappa\}\in\mathfrak{C}$,  we
   use
   laser beams ${\rm L}^{\rm tw}_{\alpha}$ that are far from the resonance of the corresponding dipole-allowed transitions.
Here, the electric field for each laser arrangement  ${\bf E}_{{\rm L}^{\rm tw}_{\alpha}}({\bf r},t)=\sum_{l}\boldsymbol{\epsilon}_{l}E_{l}\cos({\bf k}_{l}\cdot{\bf
r}-\omega_{l}t)$ consists of two $l=1,2$ travelling waves with  polarisation, amplitude,  frequency, and wavevector
$\boldsymbol{\epsilon}_{l},E_{l},\omega_{l},{\bf k}_{l}$ respectively.

Let us define the detunings $\Delta_{ls}$, and Rabi frequencies for each transition $ \Omega_{ls}=-E_{l}{\bf
d}_{\alpha,s}\cdot\boldsymbol{\epsilon}_{l}$, where $s\in\{{\uparrow},{\downarrow}\}$,
as depicted in Fig.~\ref{level_scheme}{\bf (b)}. In   weak-coupling regime $|\Omega_{ls}|\ll|\Delta_{ls}|$, the auxiliary state is seldom populated,  and the dynamics is due to two-photon processes that connect the ground-states via the excited state (see e.g.~\cite{wineland_review_sm}). Moreover, if $\Gamma_{\alpha}\ll|\Delta_{ls}|$,
the spontaneous decay due to the finite lifetime of the excited state is negligible in comparison to the coherent dynamics. In addition to the free evolution $H_{\rm s}^{\alpha}=\half\omega_0^\alpha\sigma^z$, where $\omega_0^{\alpha}$ is the transition frequency and $\sigma^z=\ketbra{{\uparrow_\alpha}}-\ketbra{{\downarrow_\alpha}}$,  the coherent evolution is given by the effective Hamiltonian
\begin{equation}
\label{coherent_master_equation}
\dot{\rho}=-\ii[H_{\rm s}^{\alpha}+\Delta H_{\alpha}(t),\rho], \hspace{1ex}\Delta {H}_{\alpha}(t)=\sum_{s,s'}\mathcal{G}_{ss'}({\bf
r},t)\ket{s}\bra{s'},
\end{equation}
where we have defined the two-photon amplitudes
\begin{equation}
\nonumber
\mathcal{G}_{ss'}({\bf r},t)=-\sum_{l,l'}\frac{\Omega_{ls}^*\Omega_{l's'}}{4\Delta_{l's'}}\ee^{-\ii\big(({\bf
k}_{l}-{\bf k}_{l'})\cdot{\bf r}-({\omega}_{l}-{\omega}_{l'})t\big)}.
\end{equation}
 Let us remark that the effective decay rates within the ground-state manifold scale as $\Gamma_\alpha^{\rm eff}=\Gamma_\alpha(|\Omega_{ls}|/\Delta_{ls})^2\ll\Gamma_{\alpha}$, and can be thus neglected for large-enough
detunings. This is precisely the regime  $\Gamma_{\tau}\gg \Gamma_{\sigma}^{\rm eff},\Gamma^{\rm eff}_{\kappa}$ considered in this work.

{\it Typical orders of magnitude.--} Let us discuss the orders of magnitude of   the parameters  appearing in the master equation for the $\tau$~\eqref{incoherent_master_equation} and $\{\sigma,\kappa\}$~\eqref{coherent_master_equation} ions (see Table~\ref{table_spins}). In order to be more precise, let us consider a particular mixed ion crystal with  species $\sigma=$$^{25}{\rm Mg}^{+}$,
$\kappa=$$^{9}{\rm Be}^+$, and $\tau=$$^{24}{\rm Mg}^{+}$. The internal states corresponding to the level structure in Fig.~\ref{level_scheme} can be expressed in terms of the hyperfine atomic levels $\ket{nL_J,F,M}$, where $n$ is
the principal quantum number, $L,J$ are the orbital and total  electronic angular momenta, and $F,M$ are the total angular momentum and its Zeeman component along a quantising magnetic field.
The $\tau=$$^{24}{\rm Mg}^{+}$ ions have no nuclear spin, and thus no hyperfine structure. For the two levels in Fig.~\ref{level_scheme}{\bf (a)}, we choose
$\ket{{\uparrow_\tau}}=\ket{3P_{1/2}},\ket{{\downarrow_\tau}}=\ket{2S_{1/2}}$, such that the transition frequency is $\omega_0^{\tau}/2\pi\approx10^3\hspace{0.2ex}$THz, and the
natural linewidth $\Gamma_{\tau}/2\pi=41.4\hspace{0.2ex}$MHz. Conversely, the $\sigma=$$^{25}{\rm Mg}^{+}$ and $\kappa=$$^{9}{\rm Be}^+$ ions display a hyperfine structure, which allows us to select two states from the hyperfine
ground-state manifold and a single excited state to form the desired $\Lambda$-scheme of Fig.~\ref{level_scheme}{\bf (b)}. For $\sigma=$$^{25}{\rm Mg}^{+}$, we take
$\ket{{\uparrow_\sigma}}=\ket{3S_{1/2},2,2},\ket{{\downarrow_\sigma}}=\ket{3S_{1/2},3,3}$, and the excited state in the $\ket{{e_{{\sigma}}}}=\ket{3P_{3/2}}$ manifold. The corresponding transition frequency between the
ground-states lies in the microwave regime  $\omega_0^{\sigma}/2\pi=1.8\hspace{0.2ex}$GHz,  and there is a negligible decay rate  (i.e.
$\Gamma_{\sigma}/2\pi\approx10^{-14}\hspace{0.2ex}$Hz). Therefore, all the spontaneous emission occurs via transitions to the excited state, which has a natural linewidth of  $\Gamma_{\sigma}/2\pi=41.4\hspace{0.2ex}$MHz. Finally,
for $\kappa=$$^{9}{\rm Be}^+$, the ground-states would be $\ket{{\uparrow_\kappa}}=\ket{2S_{1/2},1,-1},\ket{{\downarrow_\kappa}}=\ket{2S_{1/2},2,-2}$ with a transition frequency
$\omega_0^{\kappa}/2\pi=1.25\hspace{0.2ex}$GHz, and also a negligible linewidth. In this case, the excited state is in the manifold $\ket{{e_{{\kappa}}}}=\ket{2P_{1/2}}$, such
that    $\Gamma_{\kappa}/2\pi=19.4$ MHz. The detunings of the $\Lambda$-scheme are $|\Delta_{ls}|/2\pi\sim10$-$100\,$GHz.

\subsection{Edge dissipation by Doppler cooling}
\label{edge_dissipation_sec}

 We move onto the derivation of the effective dissipation~\eqref{dissipation} of the $\tau$-vibrons. We will be interested in positioning these ions $\ell_\tau\in\{1\cdots N_\tau\}$ at the edges of the chain, such that they can act as reservoirs for quantum
 transport [Fig.~\ref{fig_scheme}{\bf (a)}].

  We will show how the master equation~\eqref{incoherent_master_equation} allows for the control of the dissipation of the edge vibrons.
Let us introduce  the
Lamb-Dicke parameter $\eta_{{\rm L}^{\rm sw}_{\tau}}=k_{{\rm L}^{\rm
sw}_{\tau}}/\sqrt{2m_{\tau}\omega_{\tau}}$, and the detuning $\Delta_{{\rm
L}^{\rm sw}_\tau}=\omega_{{\rm L}^{\rm
sw}_{\tau}}-\omega_0^\tau$. If $|\Omega_{{\rm L}^{\rm
sw}_{\tau}}|,|\Delta_{{\rm L}^{\rm
sw}_{\tau}}|\ll\omega_0^{\tau}$,
and $\eta_{{\rm L}^{\rm sw}_{\tau}}\ll1$, we can approximate  the laser-ion
coupling~\eqref{laser_ion} by
 \begin{equation}
 \nonumber
{H}_{{\rm L}^{\rm sw}_{\tau}}\!=\!\!\sum_{\ell_{\tau}}F_{\ell_{\tau}}(a^{\phantom{\dagger}}_{\ell_{\tau}}+a_{\ell_{\tau}}^{\dagger}),\hspace{1.5ex}F_{\ell_{\tau}}=-\half\Omega_{{\rm L}^{\rm sw}_{\tau}}\eta_{{\rm L}^{\rm
sw}_{\tau}}\sigma_{\ell_{\tau}}^+\ee^{-\ii\omega_{{\rm L}^{\rm sw}_{\tau}}t}+{\rm H.c.}
\end{equation}
Since we are working at the node of the standing wave, let us note that the component of the laser-ion interaction that would drive the carrier is exactly cancelled. Therefore, the only fundamental constraint over the
standing-wave
Rabi frequency will be  $|\Omega_{{\rm L}^{\rm sw}_{\tau}}|\ll\omega_0^{\tau}$, still allowing for high driving strengths.

To derive the effective dissipation~\eqref{dissipation} of the $\tau$-vibrons,  the crucial point is to appreciate the   separation of time-scales
\begin{equation}
\label{conditions_cooling}
|J_{i_{\alpha}j_{\beta}}|,|\half\Omega_{{\rm L}^{\rm sw}_{\tau}}\eta_{{\rm L}^{\rm sw}_{\tau}}|, |\Gamma_{\tau}\eta_{{\rm L}^{\rm sw}_\tau}^2|\ll\Gamma_{\tau},
\end{equation}
 which implies that the spontaneous decay of the atomic states of the $\tau$-ions is  faster than any other dynamics. This allows us to partition the Liouvillian~\eqref{incoherent_master_equation} as follows
 $\mathcal{\tilde{L}}=\mathcal{\tilde{L}}_0+\mathcal{\tilde{L}}_1$
 \begin{equation}
 \nonumber
 \begin{split}
 \mathcal{\tilde{L}}_0(\tilde{\rho})&=\mathcal{\tilde{D}}^0_{\tau}(\tilde{\rho}), \hspace{2ex}
 \mathcal{\tilde{L}}_1(\tilde{\rho})=-\ii[\tilde{H}_{\rm vt}+\tilde{H}_{{\rm L}^{\rm
 sw}_{\tau}},\tilde{\rho}]+\mathcal{\tilde{D}}^1_{\tau}(\tilde{\rho}), \end{split}
 \end{equation}
 where the "tildes" refer to the interaction picture with respect to $ H_0=H_{\rm s}^{\tau}+ {H}_{\rm vo}$.
 We can  eliminate  the fast degrees  of freedom of the $\tau$ atomic states by projecting onto the steady-state of $ \mathcal{{L}}_0 (\mu^{\tau}_{\rm ss})=0$. This can be accomplished  by  projector-operator
 techniques~\cite{ad_elim}, which  to   second-order lead to
 \begin{equation}
\label{ad_el}
\dot{\tilde{\rho}}=\bigg\{\mathcal{P}\mathcal{\tilde{L}}_1(t)\mathcal{P}+\!\!\int_0^{\infty}\!\!{\rm
d}s\mathcal{P}\mathcal{\tilde{L}}_{1}(t)\mathcal{Q}\ee^{\mathcal{\tilde{L}}_0s}\mathcal{Q}\mathcal{\tilde{L}}_{1}(t-s)\mathcal{P}\bigg\}\tilde{\rho}.
\end{equation}
Here, $\mathcal{P}$ and $\mathcal{Q}=1-\mathcal{P}$ are the projectors of interest, which  correspond to $\mathcal{P}_{\tau}\{\bullet\}=\mu^{\tau}_{\rm ss}\otimes{\rm Tr}_{\tau,{\rm at}}\{\bullet\}$ in this case. Since  the ion
chain lies at the node of the standing wave,
 the atomic steady state   is $\mu^{\tau}_{\rm ss}=\otimes_{\ell_{\tau}}\ket{{\downarrow_{\ell_{\tau}}}}\bra{{\downarrow_{\ell_{\tau}}}}$. After  tracing over the atomic degrees of freedom of the $\tau$-species $\mu={\rm Tr}_{\tau,{\rm at}}\{\rho\}$, and moving  back in the Schr\"{o}dinger picture, we obtain the master equation   $
\dot{\mu}=-\ii[H_{\rm tb},\mu]+\sum_{\ell_\tau}\mathcal{D}^{\ell_\tau}_{\rm v}(\mu).$
 Here, we have introduced a dissipation super-operator that only acts on the vibrons of the ion chain, namely
\begin{equation}
\label{edge_dissipation}
\begin{split}
\mathcal{D}^{\ell_\tau}_{\rm v}(\bullet)&=\Lambda_{\ell_{\tau}}^{+}(a_{\ell_{\tau}}^{{\dagger}}\bullet a_{\ell_{\tau}}^{\phantom{\dagger}}-a_{\ell_{\tau}}^{\phantom{\dagger}}a_{\ell_{\tau}}^{{\dagger}}\bullet )+\\
&+\Lambda_{\ell_{\tau}}^{-}(a_{\ell_{\tau}}^{\phantom{\dagger}}\bullet a_{\ell_{\tau}}^{{\dagger}}-a_{\ell_{\tau}}^{{\dagger}}a_{\ell_{\tau}}^{\phantom{\dagger}}\bullet )+\text{H.c.}
\end{split}
\end{equation}
 The heating-cooling coefficients can be expressed in terms of the power spectrum of the laser-induced couplings
$F_{\ell_\tau}$
\beq
\nonumber
S_{F_{\ell_\tau},F_{\ell_\tau}}(\omega)=\int_0^{\infty}{\rm d}t\langle \tilde{F}_{\ell_{\tau}}(t)\tilde{F}_{\ell_{\tau}}(0)\rangle_{\rm ss}\ee^{\ii\omega t}.
\eeq
In particular, the cooling depends on the power spectrum at positive frequencies $\Lambda_{\ell_{\tau}}^{-}=S_{F_{\ell_\tau},F_{\ell_\tau}}(+\omega_{\ell_\tau})$, while the heating depends on the negative
frequencies $\Lambda_{\ell_{\tau}}^{+}=S_{F_{\ell_\tau},F_{\ell_\tau}}(-\omega_{\ell_\tau})$. By the quantum regression theorem~\cite{breuer}, they become
\begin{equation}
\nonumber
\Lambda_{\ell_{\tau}}^{\pm}=\frac{(\half\Omega_{{\rm L}^{\rm sw}_{\tau}}\eta_{{\rm L}^{\rm sw}_{\tau}})^2}{\half\Gamma_{\tau}+\ii(-\Delta_{{\rm L}^{\rm sw}_\tau}\pm\omega_{\ell_\tau})}.
\end{equation}
Such coefficients coincide with those of a single trapped
ion~\cite{laser_cooling}, which is not a surprise as the vibron tunneling between different ions is perturbative~\eqref{conditions_cooling}. The possibility of controlling experimentally  the frequency asymmetry of the  power spectrum which allows for an effective laser cooling of the vibrational modes, i.e.
$S_{F_{\ell_\tau},F_{\ell_\tau}}(+\omega_{\ell_\tau})>S_{F_{\ell_\tau},F_{\ell_\tau}}(-\omega_{\ell_\tau})$. Finally, by using the generic super-operator
\begin{equation}
\label{generic_so}
{\mathcal{D}}[\Lambda,O_1,
O_2](\bullet)=\Lambda (O_1 \bullet O_2-O_2O_1 \bullet)+\text{H.c.},
\end{equation}
 the  dissipator~\eqref{edge_dissipation} corresponds to Eq.~\eqref{dissipation} in the main text.

\begin{figure}
\centering
\includegraphics[width=.9\columnwidth]{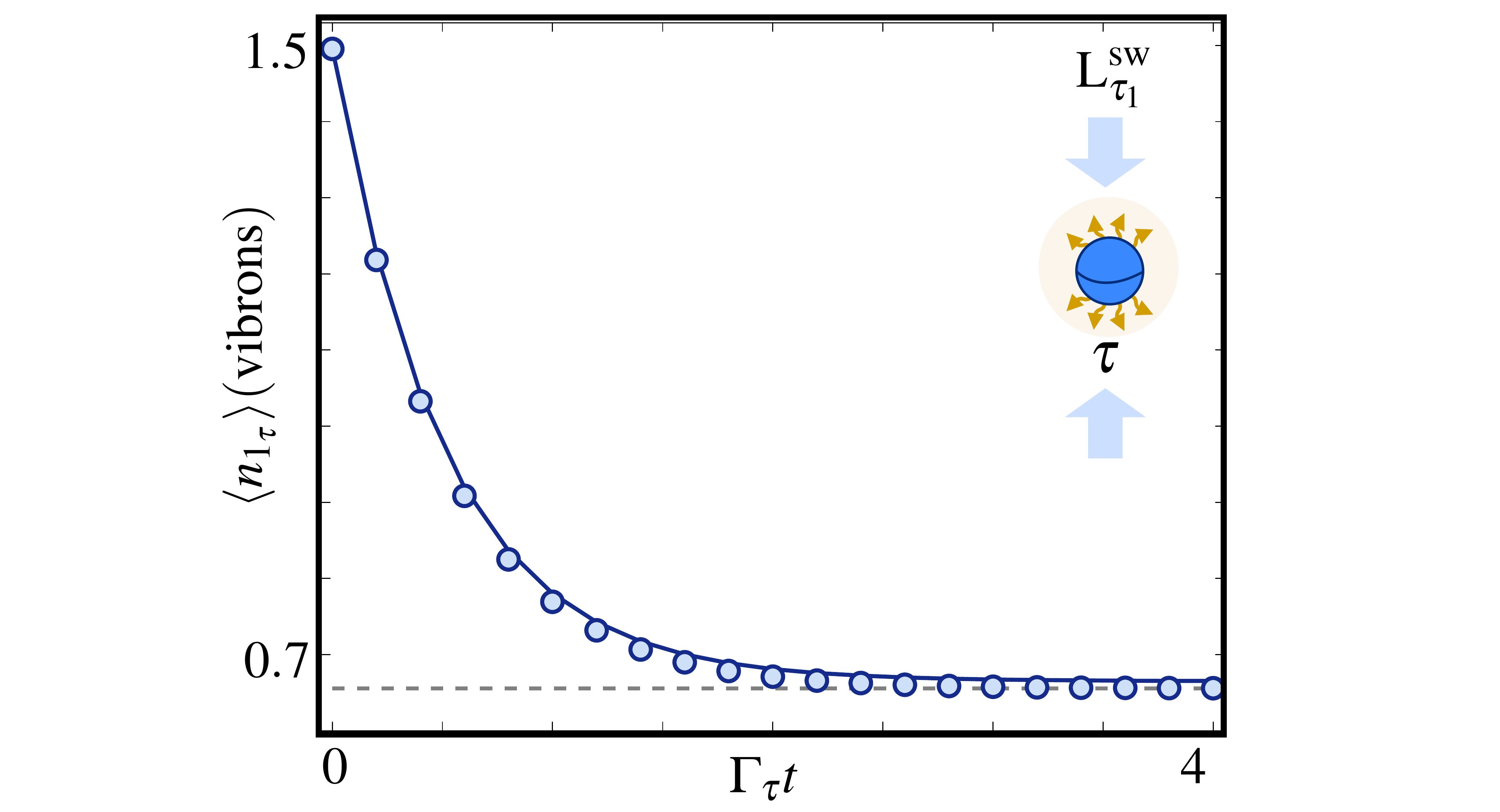}
\caption{  {\bf Damping of the vibrons by laser cooling:} Decay of the average number of vibrons $\langle n_{1_\tau}\rangle$ for a single laser-cooled  $\tau=$$^{24}{\rm Mg}^{+}$ ion. The solid line corresponds to the predictions
of the original master equation~\eqref{incoherent_master_equation}, whereas the circles are given by the effective dissipation~\eqref{edge_dissipation}. We also display in a dashed straight line, the steady-state vibron number. We   truncate the
vibron Hilbert space to $n_{\rm max}=15$ to account for the thermal effects accurately }
\label{cooling_app}
\end{figure}

 In Fig.~\ref{cooling_app}, we compare the dynamics given by the effective edge dissipator~\eqref{edge_dissipation} with that given by the original master equation~\eqref{incoherent_master_equation} restricted to a single
 $\tau=$$^{24}{\rm Mg}^{+}$ ion.
We consider an initial state $\rho_{\tau}(0)=\ket{{\uparrow}_{\tau}}\bra{{\uparrow}_{\tau}}\otimes\rho_{\tau}^{\rm th}$, where $\rho_{\tau}^{\rm th}$ is a thermal state for the $\tau$-vibrons with an average vibron number of $\bar{n}_{1_\tau}=1.5$.
In addition to the and atomic parameters for the $\tau=$$^{24}{\rm Mg}^{+}$ ions introduced above, we consider a trap frequency of $\omega_{\tau}/2\pi=10\,$MHz, and a standing-wave laser that is red-detuned $\Delta_{{\rm L}^{\rm
sw}_{\tau}}=-\half\Gamma_{\tau}$, such that its Rabi frequency is $\Omega_{{\rm L}^{\rm sw}_{\tau}}=0.1|\Delta_{{\rm L}^{\rm sw}_{\tau}}|$. As follows from the  agreement, the effective description~\eqref{edge_dissipation}   is
very accurate.

\begin{figure}
\centering
\includegraphics[width=0.7\columnwidth]{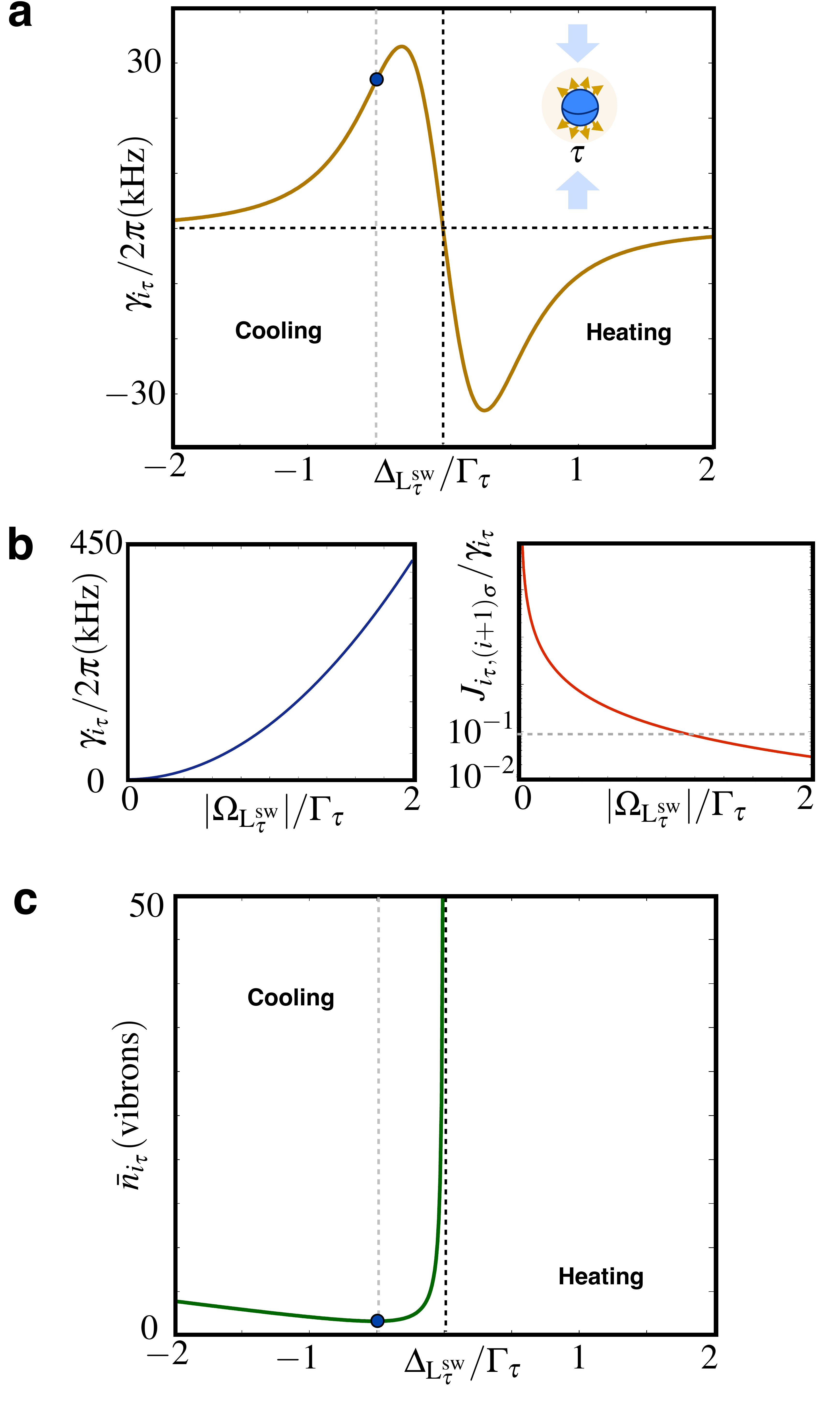}
\caption{  {\bf Doppler cooling parameters:} {\bf (a)}  Effective laser cooling strength $\gamma_{i_{\tau}}$ for $\tau=$$^{24}{\rm Mg}^{+}$ as a function of the standing-wave detuning $\Delta_{{\rm L}^{\rm
sw}_{\tau}}$ for a Rabi frequency  $\Omega_{{\rm L}^{\rm sw}_\tau}=\half\Gamma_{\tau}$. For red detunings, we obtain cooling rates that can be as high as tens of kHz. {\bf (b)} (left panel) Quadratic increase of the cooling rate
as
a function of the Rabi frequency. (right panel) Ratio of the nearest-neighbour vibron tunneling and the effective cooling rate $J_{i_{\tau},(i+1)_{\sigma}}/\gamma_{i_{\tau}}$ as a function of the standing-wave Rabi frequency. {\bf
(c)} Steady-state mean vibron number $\bar{n}_{\ell_{\tau}}$ as a function of the standing-wave detuning. }
\label{fig2_app}
\end{figure}
Let us now consider the  parameter-dependence of the cooling rate $\gamma_{\ell_{\tau}}={\rm Re}\{(\Lambda^-_{\ell_{\tau}})^*-\Lambda^+_{\ell_{\tau}}\}$, and the mean number of vibrons in the steady state $\bar{n}_{\ell_{\tau}}={\rm
Re}\{\Lambda^+_{\ell_{\tau}}\}/\gamma_{\ell_{\tau}}$. In Fig.~\ref{fig2_app}{\bf (a)},  this rate is represented as a function of the detuning in the so-called Doppler-cooling regime
$\Gamma_{\tau}\gg\omega_{\tau}$. For red detunings $\Delta_{{\rm L}^{\rm sw}_\tau}<0$, we get an effective cooling of the $\tau$-vibrons, whereas heating is obtained for blue detunings $\Delta_{{\rm
L}^{\rm sw}_\tau}>0$. Another important property is that the cooling rate increases quadratically with the laser Rabi frequency without saturation  (left panel of Fig.~\ref{fig2_app}{\bf (b)}). This  will allow us to attain  regimes where the cooling is much stronger than the vibron tunnelings $\gamma_{\ell_{\tau}}\gg |J_{i_{\alpha}j_{\beta}}|$, and the
$\tau$-ions act as vibronic reservoirs for the heat transport along the ion chain (right panel of Fig.~\ref{fig2_app}{\bf (b)}). Finally, let us also note that the mean number of vibrons in the steady state is
independent of the Rabi frequency. Therefore, increasing the laser power such that the desired regime  $\gamma_{\ell_{\tau}}\gg |J_{i_{\alpha}j_{\beta}}|$ is attained, does not limit the tunability over the vibronic reservoirs
(see
Fig.~\ref{fig2_app}{\bf (c)}), a property that will be important to study the consequences of heat transport.

\subsection{ Tailoring the spin-vibron coupling}
\label{sv_app}

 The final ingredient of our toolbox is the coherent spin-vibron coupling~\eqref{driving} for the ions $\alpha\in\{\sigma,\tau\}\in\mathfrak{C}$. In particular, these ions $i_\alpha\in\{1\cdots N_\alpha\}$ will be positioned at the
   the bulk of the chain, such that the spin-vibron coupling can be used to control and measure the quantum heat transport [Fig.~\ref{fig_scheme}{\bf (a)}]. We will show how the master equation~\eqref{coherent_master_equation} allows for the control of the spin-vibron coupling of the bulk ions.

 By tuning the two-photon frequencies $\omega_{{\rm L}^{\rm tw}_{\alpha}}=\omega_{1}-\omega_{2}$, such that $|\omega_{{\rm L}^{\rm
 tw}_{\alpha}}|\ll\omega_0^{\alpha}$,  the lasers do not provide enough energy to drive a two-photon Raman transition.
Hence, the sum in the Hamiltonian~\eqref{coherent_master_equation} should only comprise  $s=s'$.
There are two terms in this expression $\Delta \tilde{H}_{\alpha}=\Delta \tilde{H}^{\alpha}_{\rm ss}+\Delta \tilde{H}^{\alpha}_{\rm sv}$. The  processes whereby a photon is absorbed from and re-emitted into the
same laser beam (i.e. $l=l'$) contribute with an ac-Stark shift
\begin{equation}
\label{ac}
\Delta \tilde{H}^{\alpha}_{\rm
ss}=\sum_{s}\frac{1}{2}\Delta\epsilon_{s}\ket{s}\bra{s},\hspace{1.ex}\Delta\epsilon_{s}=\sum_{l}\frac{-|\Omega_{ls}|^2}{2\Delta_{l_{\alpha}s}}.
\end{equation}
 If the photon is absorbed from and re-emitted into different beams (i.e. $l\neq l'$), the corresponding term leads to a   coupling between  internal and vibrational
 degrees of freedom
\begin{equation}
\label{crossed_ss}
\Delta \tilde{H}^{\alpha}_{\rm sv}=\sum_{s}\half\Omega_{{\rm L}^{\rm tw}_{\alpha},s}\ee^{-\ii({\bf k}_{{\rm L}^{\rm tw}_{\alpha}}\cdot{\bf r}-\omega_{{\rm L}^{\rm
tw}_{\alpha}}t)}\ket{s}\bra{s}+\text{H.c.},
\end{equation}
where we have introduced the crossed-beam Rabi frequencies
$
\Omega_{{\rm L}^{\rm tw}_{\alpha},s}=-\Omega_{1s}^*\Omega_{2s}/2\Delta_{2s},
$
 the effective wavevectors ${\bf k}_{{\rm L}^{\rm tw}_{\alpha}}={\bf k}_{1}-{\bf k}_{2}$, and used the fact that  $\omega_0^{\alpha}\ll|\Delta_{ls}|$.

 This  crossed-beam Stark shift~\eqref{crossed_ss} can lead to a
 variety of spin-vibron couplings. We discuss now how to produce the desired  spin-vibron couplings~\eqref{driving}. Let us extend it to all the bulk ions $\alpha\in\{\sigma,\kappa\}$, and substitute  ${\bf r}_{i_{\alpha}}={\bf r}_{i_{\alpha}}^0+\delta{\bf r}_{i_{\alpha}}$ in  Eq.~\eqref{crossed_ss}, such that the small vibrations are expressed in terms of the creation-annihilation operators~\eqref{ca}. We
 Taylor expand in  the Lamb-Dicke parameter
$
\eta_{{\rm L}^{\rm tw}_{\alpha}}=k_{{\rm L}^{\rm tw}_{\alpha}}/\sqrt{2m_{\alpha}\omega_{\alpha}}\ll1$. When setting $|\Omega_{{\rm L}^{\rm tw}_{\alpha}}\eta_{{\rm L}^{\rm tw}_{\alpha}}|\ll\omega_{\alpha}$, and imposing that the effective laser frequency is much smaller than the trap frequency $\omega_{{\rm L}^{\rm tw}_{\alpha}}\ll \omega_{\alpha}$, we obtain
\begin{equation}
\label{spin_vibron_complete}
\Delta \tilde{H}^{\alpha}_{\rm sv}\approx\sum_{i_ \alpha,s_{i_{\alpha}}}\!\!\!\!\!\!\frac{\Omega_{{\rm L}_{\alpha},s_{i_{\alpha}}}}{2}\big(1-\eta_{{\rm L}^{\rm
tw}_{\alpha}}^2(a_{i_{\alpha}}^{\dagger}a_{i_{\alpha}}+\half)\big)\ee^{\ii\omega_{{\rm L}^{\rm tw}_{\alpha}}t}\ket{s_{i_{\alpha}}}\bra{s_{i_{\alpha}}}+\text{H.c.}
\end{equation}
The first term of this expression contributes with a periodic modulation of the Stark shift~\eqref{ac}, namely
\begin{equation}
\nonumber
\Delta\epsilon_{\alpha,s}\to\Delta\epsilon_{\alpha,s}+|\Omega_{{\rm L}^{\rm tw}_{\alpha},s}|(1-\half\eta_{{\rm L}^{\rm
tw}_{\alpha}}^2)\cos(\omega_{{\rm L}^{\rm tw}_{\alpha}}t-\phi_{{\rm L}^{\rm tw}_{\alpha}}),
\end{equation}
where we have introduced the phase of the  Rabi frequencies $\Omega_{{\rm L}^{\rm tw}_{\alpha},s}=|\Omega_{{\rm L}^{\rm tw}_{\alpha},s}|{\rm exp}\{-\ii\phi_{{\rm L}^{\rm tw}_{\alpha}}\}$. The second term leads to
\begin{equation}
\label{spin_vibron_app}
H_{\rm sv}^{\alpha}(t)=-\sum_{i_{\alpha}}\sum_{s_{i_{\alpha}}}|\Omega_{{\rm L}^{\rm tw}_{\alpha},s_{i_{\alpha}}}|\eta_{{\rm L}^{\rm tw}_{\alpha}}^2\cos(\omega_{{\rm L}^{\rm tw}_{\alpha}}t-\phi_{{\rm L}^{\rm
tw}_{\alpha}})\ket{s_{i_{\alpha}}}\bra{s_{i_{\alpha}}}a_{i_{\alpha}}^{\dagger}a_{i_{\alpha}}.
\end{equation}
We are now ready to derive the expression~\eqref{driving} used throughout this work. Let us make the following definitions
\beq
\label{driving_equations_1}
\Delta\omega_{s_{i_{\alpha}}}\!\!\!=-|\Omega_{{\rm L}^{\rm tw}_{\alpha},s_{i_{\alpha}}}|\eta_{{\rm L}^{\rm tw}_ \alpha}^2,
\hspace{2ex}\Delta\omega^{\pm}_{\alpha}=\Delta\omega_{{\uparrow}_{{\alpha}}}\pm\Delta\omega_{{\downarrow}_{{\alpha}}}
\eeq
together with the frequency and phase of the lasers
\beq
\label{driving_equations_2}
\nu_{\alpha }=\omega_{{\rm L}^{\rm tw}_{\alpha}}, \hspace{1.5ex}\varphi_{\alpha}=\phi_{{\rm L}^{\rm tw}_{\alpha}}.
\eeq
Then, the crossed-beam Stark shift~\eqref{spin_vibron_app}
 becomes exactly   the desired spin-vibron coupling in Eq.~\eqref{driving} of the main text. Let us note that the above drivings in the  spin-independent regime, $\Delta\omega_\alpha^-=0$, were used
 in~\cite{photon_assisted_tunneling_ions_sm} to mimic the effects of an external gauge field in the dynamics of the vibrons.

\begin{figure}
\centering
\includegraphics[width=1\columnwidth]{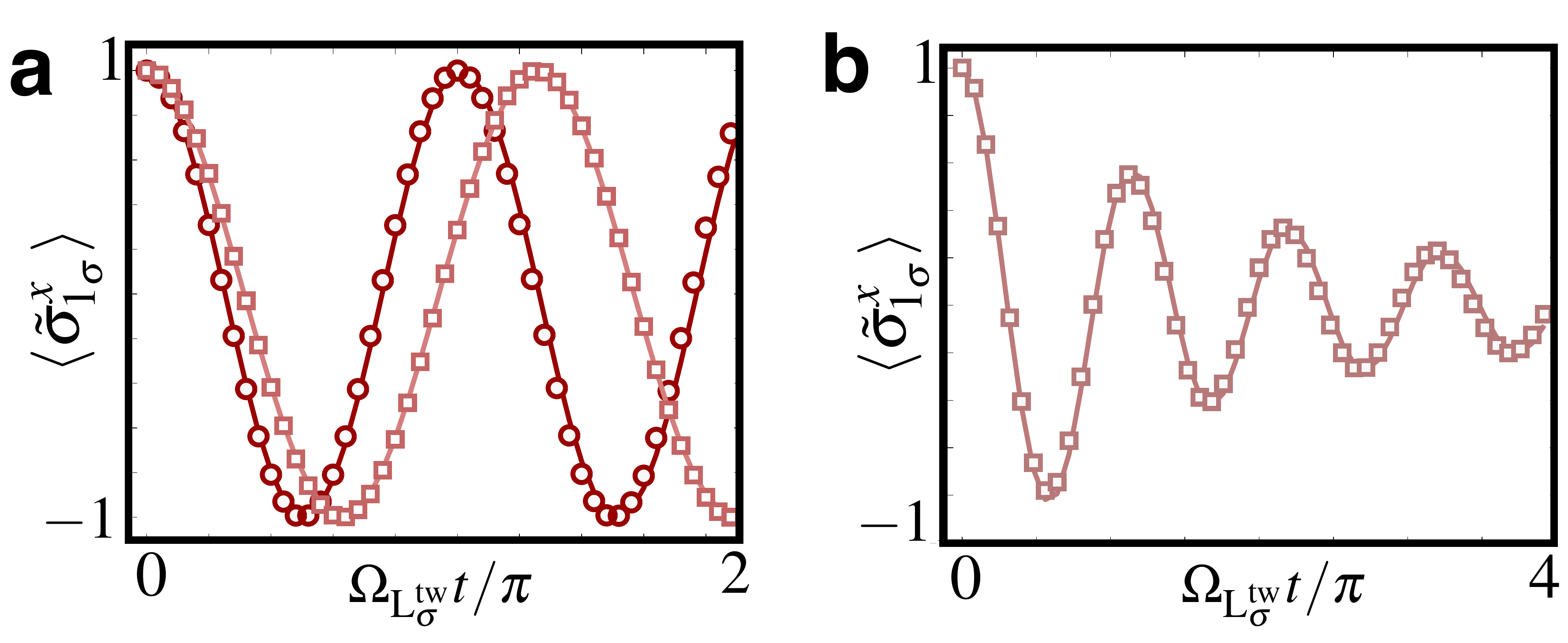}
\caption{  {\bf Spin-vibron coupling:} {\bf (a)} Dynamics of the spin coherence of a single ion initially prepared in $\ket{\Psi_{\sigma}(0)}=(\ket{{\uparrow_\sigma}}+\ket{{\downarrow_\sigma}})/\sqrt{2}$ in the $\Lambda$-scheme
[Fig.~\ref{level_scheme}{\bf (a)}] (see text for the particular parameters). For an initial vibrational Fock state  with $\bar{n}_\sigma=0 $ (red solid line describes the
coherences
$\langle \tilde{\sigma}_1^x\rangle$ given by~\eqref{crossed_ss_single}; red circles correspond to~\eqref{spin_vibron_complete}), we obtain a periodic oscillation of the coherences. For $\bar{n}_\sigma=10$
(pink solid line describes the coherences $\langle \tilde{\sigma}_1^x\rangle$ given by~\eqref{crossed_ss_single}; pink squares correspond to~\eqref{spin_vibron_complete}),  one observes a shift of the oscillation period due
to the vibronic state. In {\bf (b)}, we consider an initial thermal state with $\bar{n}_\sigma=10 $ (pink solid line describes the coherences $\langle \tilde{\sigma}_1^x\rangle$ given by~\eqref{crossed_ss_single}; pink squares
correspond to~\eqref{spin_vibron_complete}). In addition to the frequency shift,   damping of the coherences is caused by the fluctuations of the vibron number in the thermal state. We truncate the vibron Hilbert space to $n_{\rm max}=60$.}
\label{spin_vibron_fig}
\end{figure}

We now support numerically this derivation for a single $\sigma=$$^{25}{\rm Mg}^+$ ion. We will consider that the standard ac-Stark shift~\eqref{ac} is compensated, such that the dynamics is given by the crossed-beam ac-Stark shift~\eqref{crossed_ss}. Moreover, we choose the Rabi frequencies such that it becomes
\begin{equation}
\label{crossed_ss_single}
\Delta \tilde{H}^{\sigma}_{\rm sv}=\half\Omega_{{\rm L}^{\rm tw}_{\sigma}}\sigma^z\ee^{-\ii({\bf k}_{{\rm L}^{\rm tw}_{\sigma}}\cdot{\bf r}-\omega_{{\rm L}^{\rm tw}_{\sigma}}t)}+\text{H.c.}
\end{equation}
We will  align the laser wavevectors such that ${\bf k}_{1}=-{\bf k}_{2}=\half{\bf k}_{{\rm L}^{\rm tw}_\sigma}\parallel{\bf e}_x$. In this case, the crossed-beam Stark shift~\eqref{crossed_ss_single}  introduces a coupling between
the spin and vibrational degrees of freedom that will affect the coherences. We want to assess numerically the validity of the leading spin-vibron coupling derived in Eq.~\eqref{spin_vibron_app}. Hence, we consider a slowly
oscillating $\omega_{{\rm L}^{\rm tw}_{\sigma}}/2\pi=10^{-3}\omega_{\sigma}/2\pi=5\,$kHz laser arrangement with $\Omega_{{\rm L}^{\rm tw}_{\sigma}}=0.1\omega_{\sigma}/\eta_{{\rm L}^{\rm tw}_{\sigma}}$, where we recall that the transverse trap
frequency for $\sigma=$$^{25}{\rm Mg}^+$ ion is $\omega_{\sigma}/2\pi=5\,$MHz, and the Lamb-Dicke parameter is $\eta_{{\rm L}^{\rm tw}_{\sigma}}\approx0.15$.
In Fig.~\ref{spin_vibron_fig}{\bf (a)}, we represent the spin coherences. We consider two initial Fock states with $\bar{n}_{\sigma}\in\{0,10\}$. First, we observe that the effective spin-vibron coupling ~\eqref{spin_vibron_app}
is
an accurate description. Second, we see that the period of the coherence oscillations depends on the number of vibrons, a feature that will be crucial to use this coupling as a measurement device. Finally, in
Fig.~\ref{spin_vibron_fig}{\bf (b)}, we initialise the vibrons in a thermal state with $\bar{n}_{\sigma}=10$. We observe that, for thermal states, the intrinsic fluctuations in the number of vibrons lead to a decoherence of the
spin states. This  feature will be crucial for heat transport measurements.

\section{Thermalization: vibron number and current }

The objective of this section is to present a detailed derivation, supported by numerical simulations, of the effective dissipation of the bulk vibrons~\eqref{bulk_liouvillian}, which forms the basis to understand the ballistic
heat transport across an ion chain~\eqref{occupation}-\eqref{current}. Additionally, we describe how to introduce dephasing and disorder in the ion chain, and how they affect the transport.

\subsection{Effective dissipation of the bulk vibrons}
\label{sec_diss_bulk}

Let us  derive the effective thermalization of the bulk vibrons~\eqref{bulk_liouvillian} starting from the driven dissipative spin-vibron model~\eqref{ddsp}. In Fig.~\ref{fig2_app}, we showed that the Doppler cooling by a standing
wave leads to cooling rates
that can be much stronger than the vibron tunnelings $2\gamma_{\ell_{\tau}}\gg J_{i_{\alpha}j_{\beta}}$ . In this
limit, there is again a separation of time-scales: the thermalization of the edge $\tau$-vibrons is much faster than any other term in the
Liouvillian~\eqref{ddsp}. This allows us to  regroup the Liouvillian~\eqref{ddsp}
\begin{equation}
\nonumber
\begin{split}
\mathcal{\tilde{L}}_0(\tilde{\mu})&=\sum_{\ell_\tau}\mathcal{\tilde{D}}^{\ell_\tau}_{\rm v}(\tilde{\mu}),\\
\mathcal{\tilde{L}}_{1}(\tilde{\mu})&=-\ii[\tilde{H}_{\rm vt}+\tilde{H}_{\rm sv}^{\sigma}(t)+\tilde{H}_{\rm sv}^{\kappa}(t),\tilde{\mu}],
\end{split}
\end{equation}
where the "tildes" refer to the interaction picture with respect to $H_0=H^{\sigma}_{\rm s}+H^{\kappa}_{\rm s}+ {H}_{\rm vo}$. Let us start by switching off the spin-vibron couplings. To integrate out the edge vibrons, we use the
projection-operator techniques~\eqref{ad_el} for a projector $\mathcal{P}_{{\rm edge}}\{\bullet\}= \mu_{\rm ss}^{\tau}\otimes {\rm Tr}_{\tau,{\rm vib}}\{\bullet\}$. Here, $\mu_{\rm ss}^{\tau}$ is the steady state of the
laser-cooled $\tau$-vibrons at each edge  $\mu_{\rm ss}^{\tau}= \mu_{1_{\tau}}^{\rm th}\otimes \mu_{N_{\tau}}^{\rm th}$. In particular, it corresponds the
thermal states
\begin{equation}
\nonumber
\mu_{\ell_{\tau}}^{\rm th}=\sum_{n_{\ell_{\tau}}=0}^{\infty}(\bar{n}_{\ell_{\tau}})^{n_{\ell_{\tau}}}(1+\bar{n}_{\ell_{\tau}})^{-(1+n_{\ell_{\tau}})}\ket{n_{\ell_{\tau}}}\bra{n_{\ell_{\tau}}},
\end{equation}
 with different mean vibron numbers  $\bar{n}_{\ell_{\tau}}={\rm Re}\{\Lambda_{\ell_{\tau}}^+\}/\gamma_{\ell_{\tau}}$. As discussed in the main text, as long as the laser-cooling is switched on, the edge ions remain in a
 vibrational thermal state that can be controlled by the laser parameters. These edge $\tau$-ions act as a reservoir of vibrons for the  bulk of the ion chain. We now derive the effective bulk Liouvillian.

By making use of the quantum regression theorem, we obtain the two-time correlation functions of the vibrons
\begin{equation}
\nonumber
\begin{split}
\langle a_{\ell_{\tau}}^{\dagger}(s)a_{\ell'_{\tau}}(0)\rangle_{\rm ss}&=\delta_{\ell_{\tau},\ell'_{\tau}}\bar{n}_{\ell_{\tau}}\ee^{-(\gamma_{\ell_\tau}-\ii\delta_{\ell_\tau})s},\\
\langle a_{\ell_{\tau}}(s)a_{\ell'_{\tau}}^{\dagger}(0)\rangle_{\rm ss}&=\delta_{\ell_{\tau},\ell'_{\tau}}(\bar{n}_{\ell_{\tau}}+1)\ee^{-(\gamma_{\ell_\tau}-\ii\delta_{\ell_\tau})s},
\end{split}
\end{equation}
where we have introduced  $\delta_{\ell_{\tau}}=-{\rm Im}\{(\Lambda^-_{\ell_{\tau}})^*-\Lambda^+_{\ell_{\tau}}\}>0$, and
$\delta_{\ell_{\tau},\ell'_{\tau}}$ is the Kronecker delta. Using these expressions, together with the projection-operator formula~\eqref{ad_el},
we arrive at a master equation that only involves the bulk ions
\begin{equation}
\label{bulk_meq}
\dot{\mu}_{{\rm bulk}}=-\ii\big[\sum_{\alpha}H_{\rm s}^{\alpha}+H_{\rm tb}^{\rm bulk},\mu_{{\rm bulk}}\big]+\Delta\mathcal{L}(\mu_{{\rm bulk}}),
\end{equation}
where $\mu_{{\rm bulk}}={\rm Tr}_{\tau,{\rm vib}}\{\mu\}$, and $H_{\rm tb}^{\rm bulk}$ is the vibron tight-binding model restricted to the bulk ion species $\alpha,\beta\in\{\sigma,\kappa\}$. In the expression above, we have
introduced the super-operator
\begin{equation}
\nonumber
\begin{split}
\Delta\mathcal{L}(\bullet)\!=\!\sum_{\alpha,\beta}\sum_{i_{\alpha},j_{\beta},\ell_{\tau}}\!\!\Upsilon^{\ell_{\tau}}_{i_{\alpha}j_{\beta}}&\bigg\{(\bar{n}_{\ell_{\tau}}+1)(a_{j_{\beta}}\bullet
a_{i_{\alpha}}^{\dagger}-a_{i_{\alpha}}^{\dagger}a_{j_{\beta}}\bullet)\\
&+\bar{n}_{\ell_{\tau}}(a_{i_{\alpha}}^{\dagger}\bullet a_{j_{\beta}}-a_{j_{\beta}}a_{i_{\alpha}}^{\dagger}\bullet)\bigg\}+\text{H.c.},
\end{split}
\end{equation}
which is expressed in terms of the couplings
\begin{equation}
\nonumber
\Upsilon^{\ell_{\tau}}_{i_{\alpha}j_{\beta}}=\frac{J_{i_{\alpha}\ell_{\tau}}J_{\ell_{\tau}j_{\beta}}}{\gamma_{\ell_{\tau}}-\ii((\omega_{i_\alpha}-\delta_{\ell_{\tau}})-\omega_{\ell_\tau})}.
\end{equation}
The imaginary part of the $\Upsilon$-coefficients can be rewritten as a Hamiltonian term, which yields a renormalization of the vibron tunnelings and the on-site energies
\begin{equation}
\nonumber
\tilde{J}_{i_{\alpha}j_{\beta}}=J_{i_{\alpha}j_{\beta}}+\sum_{\ell_{\tau}}{\rm Im}\{\Upsilon^{\ell_{\tau}}_{i_{\alpha}j_{\beta}}\},\hspace{2ex}\tilde{\omega}_{i_\alpha}=\omega_{\alpha}+\tilde{J}_{i_{\alpha}i_{\alpha}}.
\end{equation}
This leads to the renormalized tight-binding model
\beq
\nonumber
H_{\rm rtb}=\sum_{\alpha,i_\alpha}\tilde{\omega}_{i_\alpha}a_{i_{\alpha}}^{\dagger}a_{i_{\alpha}}+\sum_{\alpha,\beta}\sum_{i_\alpha\neq j_\beta}\tilde{J}_{i_\alpha j_\beta}a_{i_{\alpha}}^{\dagger}a_{j_{\beta}},
\eeq
 introduced in Eq.~\eqref{bulk_liouvillian} of the main text. In addition, the real part of the $\Upsilon$-coefficients leads to a dissipative super-operator
\begin{equation}
\nonumber
\begin{split}
\mathcal{D}_{\rm bulk}(\bullet)=\sum_{\alpha, \beta}\sum_{i_{\alpha}j_{\beta}}&\bigg\{\tilde{\Lambda}^+_{i_{\alpha}j_{\beta}}\big(a^{{\dagger}}_{j_{\beta}}\bullet
a^{\phantom{\dagger}}_{i_{\alpha}}-a^{\phantom{\dagger}}_{i_{\alpha}}a^{{\dagger}}_{j_{\beta}}\bullet\big)\\
&+\tilde{\Lambda}^-_{i_{\alpha}j_{\beta}}\big(a^{\phantom{\dagger}}_{j_{\beta}}\bullet a^{{\dagger}}_{i_{\alpha}}-a^{{\dagger}}_{i_{\alpha}}a^{\phantom{\dagger}}_{j_{\beta}}\bullet\big)\bigg\}+\text{H.c.},
\end{split}
\end{equation}
where the dissipation rates are the following
\begin{equation}
\label{eff_gamma_app}
\tilde{\Lambda}^+_{i_{\alpha}j_{\beta}}=\sum_{\ell_{\tau}}{\rm Re}\{\Upsilon^{\ell_{\tau}}_{i_{\alpha}j_{\beta}}\}\bar{n}_{\ell_{\tau}}, \hspace{2ex} \tilde{\Lambda}^-_{i_{\alpha}j_{\beta}}=\sum_{\ell_{\tau}}{\rm
Re}\{\Upsilon^{\ell_{\tau}}_{i_{\alpha}j_{\beta}}\}(\bar{n}_{\ell_{\tau}}+1).
\end{equation}
Using the  super-operator~\eqref{generic_so}, the above dissipator can be written as the bulk dissipator below Eq.~\eqref{bulk_liouvillian} of the main text.

\subsection{Mesoscopic transport in ion chains}
\label{mes_transport_chains}

The objective of this section is to provide numerical evidence supporting Eq.~\eqref{bulk_liouvillian}. Additionally, we will also check the accuracy of the predictions derived
thereof, namely Eqs.~\eqref{occupation} and~\eqref{current} for the vibronic number and current through the ion chain.
Following the philosophy of "$one$, $two$, $many$", we first consider the smallest setup,  a single-ion channel that will play the role of a thermal quantum dot (TQD), and allow us to test the validity of
Eqs.~\eqref{bulk_liouvillian}, and~\eqref{occupation} . Then, we will move to a two-ion channel that will act as a double thermal quantum dot (DTQD), which will allow us to test the validity of
Eq.~\eqref{current}. Finally, we will explore a thermal quantum wire (TQW)  formed by a longer ion chain, or a TQD connected to two thermal leads, where  the leads are formed by a large
number of ions.

To test these predictions, we integrate the dynamics given by the bulk~\eqref{bulk_liouvillian} and edge~\eqref{ddsp} master equations. Since both
Liouvillians are quadratic in creation-annihilation operators, it is possible to obtain a closed system of $(N-2)^2$ or $N^2$ differential equations for the two-point correlators $C_{i_{\alpha}j_{\beta}}=\langle
a_{i_{\alpha}}^{\dagger}a_{j_{\beta}}\rangle$, respectively. Both theories can be recast into
\begin{equation}
\label{ode}
\frac{{\rm d}C}{{\rm d}t}=\ii[\mathbb{J},C]-(\mathbb{W}C+C\mathbb{W}^*)+\mathbb{K},
\end{equation}
where the matrices $\mathbb{J},\mathbb{W},\mathbb{K}$ depend on the particular master equation.  For the edge dissipation~\eqref{ddsp}, we find
\begin{equation}
\label{edge_matrices}
\begin{split}
\mathbb{J}^{\rm edge}_{i_{\alpha}j_{\beta}}&=\omega_{\alpha}\delta_{i_{\alpha}j_{\beta}}+J_{i_{\alpha}j_{\beta}},\\
\mathbb{W}^{\rm edge}_{i_{\alpha}j_{\beta}}&=((\Lambda^-_{\ell_{\tau}})^*-\Lambda^+_{\ell_{\tau}})\delta_{i_{\alpha},\ell_{\tau}}\delta_{j_{\beta},\ell_{\tau}},\hspace{3ex} \alpha,\beta\in\{\sigma,\kappa,\tau\}\\
\mathbb{K}^{\rm edge}_{i_{\alpha}j_{\beta}}&=2{\rm Re}\{\Lambda^+_{\ell_{\tau}}\}\delta_{i_{\alpha}\ell_{\tau}}\delta_{j_{\beta},\ell_{\tau}},
\end{split}
\end{equation}
whereas for the effective bulk dissipation~\eqref{bulk_liouvillian}, we get
\begin{equation}
\label{bulk_matrices}
\begin{split}
\mathbb{J}^{\rm bulk}_{i_{\alpha}j_{\beta}}&=\omega_{\alpha}\delta_{i_{\alpha}j_{\beta}}+\tilde{J}_{i_{\alpha}j_{\beta}},\\
\mathbb{W}^{\rm bulk}_{i_{\alpha}j_{\beta}}&=(\tilde{\Lambda}^-_{i_{\alpha}j_{\beta}})^*-\tilde{\Lambda}^+_{i_{\alpha}j_{\beta}},\hspace{8ex} \alpha,\beta\in\{\sigma,\kappa\}\\
\mathbb{K}^{\rm bulk}_{i_{\alpha}j_{\beta}}&=2{\rm Re}\{\tilde{\Lambda}^+_{i_{\alpha}j_{\beta}}\}.
\end{split}
\end{equation}
The possibility of expressing the dissipative dynamics as a closed set of differential equations~\eqref{ode} allows us to circumvent  numerical limitations, which would arise  due to the large truncation of the vibron Hilbert
space  required for some of the simulations of the  dissipative vibron model.

\subsubsection{ Thermal Quantum dot: vibron number}
\label{single_ion_channel}

Let us consider the minimal scenario: the {\it thermal quantum dot}. In this case, the chain is composed of three ions
\begin{equation}
\nonumber
\tau-\sigma-\tau,
\end{equation}
such that  heat transport takes place along the minimal channel: a single-ion connecting the two $\tau$-reservoirs. In this limit, Eq.~\eqref{bulk_liouvillian} corresponds to single-oscillator master equation that can be solved exactly, and  yields a  steady-state mean vibron number of ${\bar n}_{2_{\sigma}}={\rm Re}\{\tilde{\Lambda}^+_{2_{\sigma}2_{\sigma}}\}/{\rm
Re}\{\tilde{\Lambda}^-_{2_{\sigma}2_{\sigma}}-\tilde{\Lambda}^+_{2_{\sigma}2_{\sigma}}\}$. According to Eq.~\eqref {eff_gamma_app},  this mean vibron number can be written
$
\bar{n}_{2_{\sigma}}=({\Gamma_{\rm L} \bar{n}_{\rm L}+\Gamma_{\rm R} \bar{n}_{\rm R}})/({\Gamma_{\rm L} + \Gamma_{\rm R}}),
$
where we have introduced the mean vibron numbers $\bar{n}_{\rm L}=\bar{n}_{1_\tau}, \bar{n}_{\rm R}=\bar{n}_{3_\tau}$. We thus obtain the couplings
 $\Gamma_{\rm L}=\Gamma^{1_\tau}_{2_{\sigma},2_{\sigma}}$ and
$\Gamma_{\rm R}=\Gamma^{3_\tau}_{2_{\sigma},2_{\sigma}}$, which correspond exactly to those introduced below Eq.~\eqref{bulk_liouvillian} in the main text, namely
$
    \Gamma^{\ell_\tau}_{2_\sigma,2_\sigma} = 2\pi
J_{2_\sigma,\ell_\tau}\rho_{\ell_\tau}(\omega_{2_\sigma})J_{\ell_\tau,2_\sigma},
$
where $\rho_{\ell_\tau}(\epsilon)$ is the Lorentzian density of states for the laser-cooled ions.

\begin{figure}
\centering
\includegraphics[width=0.75\columnwidth]{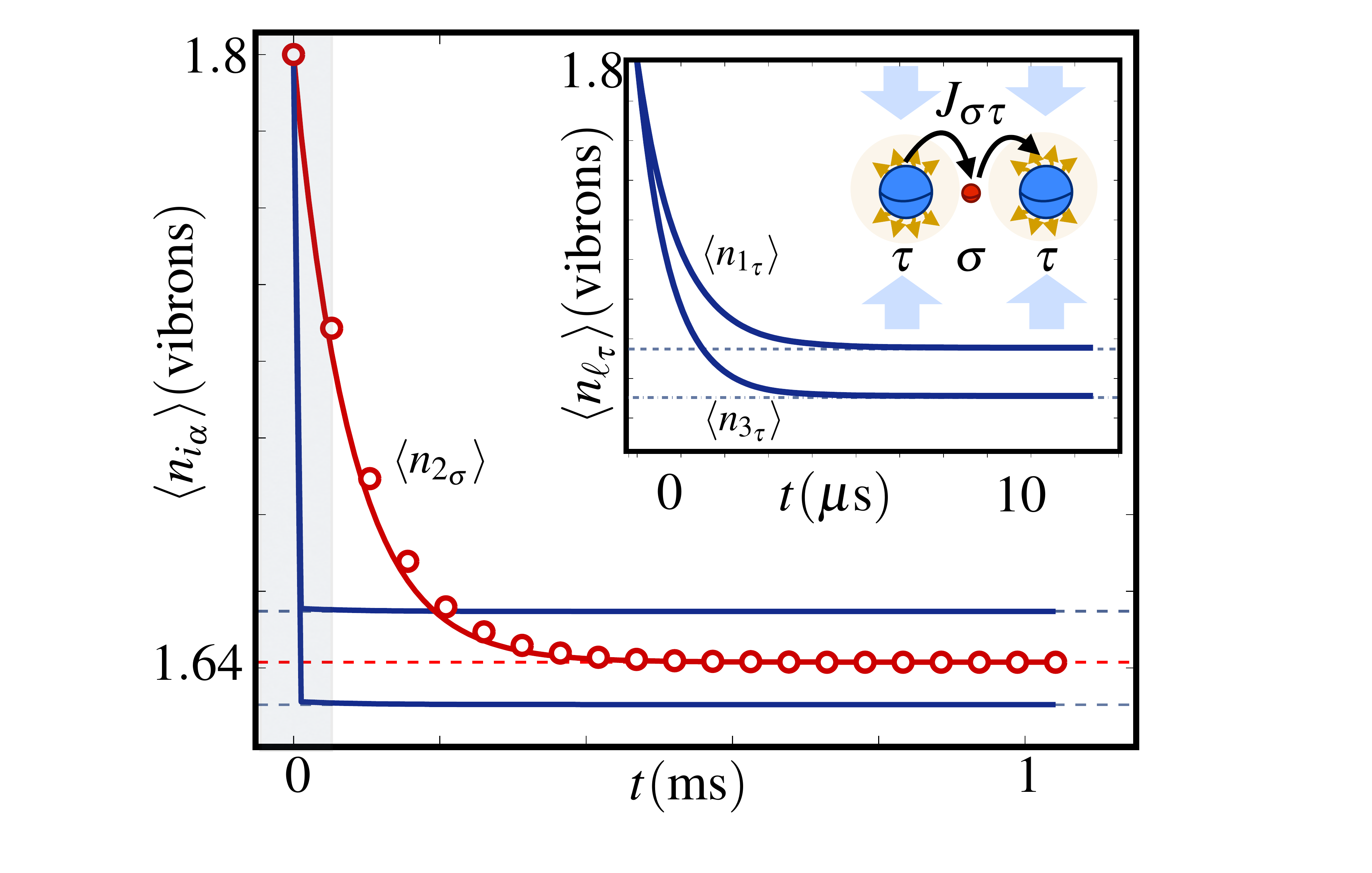}
\caption{ {\bf Thermalization of a thermal quantum dot:}  Dynamics of the vibronic numbers $\langle n_{i_{\alpha}\rangle}$ for a $^{24}{\rm Mg}^+$-$^{25}{\rm Mg}^+$-$^{24}{\rm Mg}^+$ ion chain (see the text for the
particular parameters). The solid lines represent the numerical solution of Eqs.~\eqref{ode}-\eqref{edge_matrices}, showing that the edge ions thermalize much faster (see also the inset for $\langle n_{1_{\tau}\rangle}$, $\langle
n_{3_{\tau}\rangle}$). For the bulk ion, the numerical solution for the vibron number $\langle n_{2_{\sigma}\rangle}$ given by Eqs.~\eqref{ode}-\eqref{bulk_matrices} is displayed with red circles, and shows a good agreement with
the previous dynamics.
}
\label{bulk_cooling_app}
\end{figure}

Let us now consider the realistic parameters for a $\tau$-$\sigma$-$\tau$ chain, where $\tau=$$^{24}{\rm Mg}^+$ and  $\sigma=$$^{25}{\rm Mg}^+$ as usual. In addition to the parameters introduced in previous sections, we consider the
detunings $\Delta_{1_\tau}=-0.6\Gamma_{\tau}$, $\Delta_{3_\tau}=-0.5\Gamma_{\tau}$, and the Rabi frequencies $\Omega_{{\rm L}^{\rm sw}_{1_\tau}}=\Omega_{{\rm L}^{\rm sw}_{3_\tau}}=\Gamma_{\tau}$ for the laser cooling of the
$\tau$-ions, where we recall that $\Gamma_{\tau}/2\pi=41.4\,$MHz. With these parameters, the effective cooling rates of the $\tau$-ions would be $\gamma_{1_{\tau}}/2\pi\approx 86\,$kHz, and
$\gamma_{3_{\tau}}/2\pi\approx 106\,$kHz. Additionally, the mean number of vibrons for each reservoir would be $\bar{n}_{1_{\tau}}=1.65$, and $\bar{n}_{3_{\tau}}=1.63$. The trap frequencies  are $(\omega_{\alpha
x},\omega_{\alpha y},\omega_{\alpha z})/2\pi=(5,5,0.5)\hspace{0.2ex}$MHz, which lead to an  inter-ion distance of $|z^0_{1_\tau}-z^0_{2_{\sigma}}|=|z^0_{3_\tau}-z^0_{2_{\sigma}}|\approx9\hspace{0.2ex}\mu$m, and to a vibron
tunneling strength of $J_{1_{\tau}2_{\sigma}}/2\pi=J_{3_{\tau}2_{\sigma}}/2\pi\approx30\hspace{0.2ex}$kHz. The constraint $2\gamma_{\ell_{\tau}}\gg J_{i_{\alpha}j_{\beta}}$
is thus fulfilled, such that  the $\tau$-ions thermalize  fast and act as a reservoir for the bulk $\sigma$-ion.

In  Fig.~\ref{bulk_cooling_app}, we confirm this behaviour numerically. As displayed in the figure, the edge vibrons thermalize on a $\mu$s-scale (see also the inset), whereas the bulk  vibron number  reaches the steady state on a longer millisecond-scale.
Moreover, the agreement of the numerical results shows that the effective bulk Liouvillian~\eqref{bulk_liouvillian} is a good description of the problem. Moreover, the red dashed line
represents our prediction for the stationary bulk vibrons~\eqref{occupation}, which also displays a good agreement with the numerical results. Finally, the blue dashed lines represent the laser-cooling vibron numbers
$\bar{n}_{1_{\tau}},\bar{n}_{3_{\tau}}$, which perfectly match the edge steady state.

\subsubsection{ Double thermal quantum dot: vibron current}

We turn into  the double thermal quantum dot: a two-oscillator channel  connected to the two laser-cooled reservoirs
\beq
\nonumber
\tau-\sigma-\sigma-\tau,
\eeq
 where $\tau=$$^{24}{\rm Mg}^+$ and  $\sigma=$$^{25}{\rm Mg}^+$. We choose same parameters as above, except for  the detunings $\Delta_{1_\tau}=-0.8\Gamma_{\tau}$, $\Delta_{4_\tau}=-0.6\Gamma_{\tau}$,  the Rabi frequencies $\Omega_{{\rm L}^{\rm
 sw}_{1_\tau}}=1.4\Gamma_{\tau}$, $\Omega_{{\rm L}^{\rm sw}_{4_\tau}}=\Gamma_{\tau}$, and the trap frequencies  $(\omega_{\alpha x},\omega_{\alpha y},\omega_{\alpha
 z})/2\pi=(5,5,0.2)\hspace{0.2ex}$MHz.
 We consider an initial thermal state, where the two $\sigma$-ions have a different vibronic number $\bar{n}_{2_\sigma}(0)=2.4$, and $\bar{n}_{3_\sigma}(0)=1$. Accordingly, we expect to observe a periodic exchange of vibrons
 between the bulk ions, which is additionally damped due to their contact with the reservoirs. In Fig.~\ref{bulk_cooling_2_ions}{\bf (a)}, we show the thermalization dynamics of such a two-oscillator channel.  The clear agreement between the bulk~\eqref{bulk_liouvillian} and edge~\eqref{ddsp} master equations supports once more the validity of
our  derivations.

\begin{figure}
\centering
\includegraphics[width=1.\columnwidth]{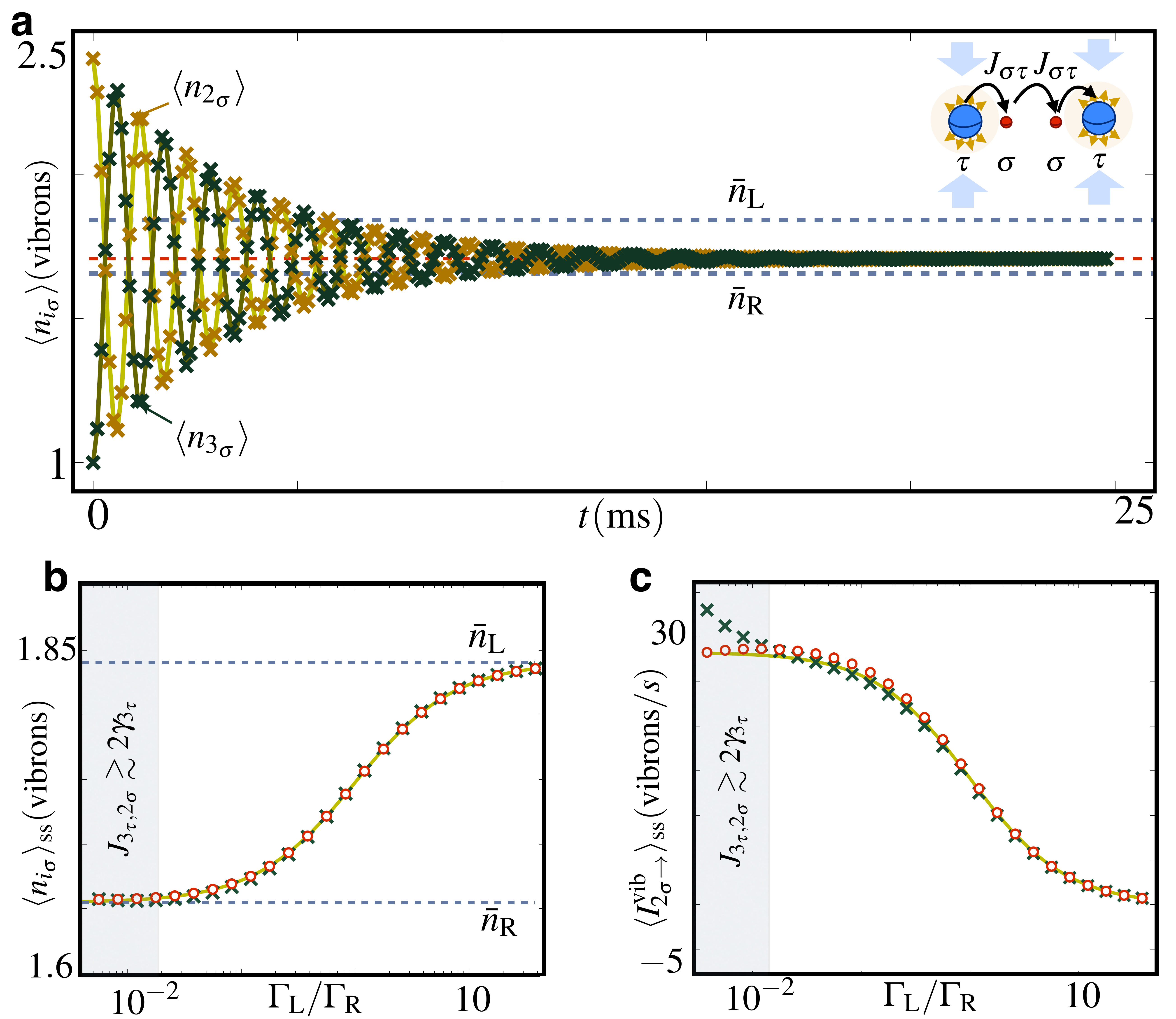}
\caption{ {\bf Vibron number and current in a double  thermal quantum dot:}  {\bf (a)} Thermalization dynamics for the number of bulk vibrons $\langle n_{2_\sigma}\rangle, \langle n_{3_\sigma}\rangle$ (see the text for the
particular parameters). The solid lines correspond to the numerical solution of Eqs.~\eqref{ode}-\eqref{edge_matrices}, and the symbols to the numerical solution of Eqs.~\eqref{ode}-\eqref{bulk_matrices}. The grey dashed lines
represent the reservoir mean vibron numbers $\bar{n}_{\rm L},\bar{n}_{\rm R}$, while the red dashed line stands for the theoretical prediction for the bulk vibron number~\eqref{occupation}. {\bf (b)} Steady-state bulk vibron
number $\langle n_{2_\sigma}\rangle=\langle n_{3_\sigma}\rangle$ as a function of the system-reservoir effective couplings $\Gamma_{\rm L},\Gamma_{\rm R}$. The green crosses represent the numerical solution of
Eqs.~\eqref{ode}-\eqref{edge_matrices}, the red circles that of Eqs.~\eqref{ode}-\eqref{bulk_matrices}, and the yellow solid line corresponds to the theoretical prediction in Eq.~\eqref{occupation}. {\bf (c)} Same as above,
but
displaying the steady-state vibron currents $\langle I_{ 2_{\sigma}\rightarrow}^{\rm vib}\rangle_{\rm ss}$ according to Eqs.~\eqref{ode}-\eqref{edge_matrices} (crosses), Eqs.~\eqref{ode}-\eqref{bulk_matrices} (circles), and the
prediction~\eqref{current} (solid line).
}
\label{bulk_cooling_2_ions}
\end{figure}

We now address the validity of the predictions for the steady-state mean vibron number~\eqref{occupation} and heat current~\eqref{current}. To calculate the vibron current,  note that the current operator can be defined through a continuity equation ${\rm d}n_{i_\alpha}/{\rm
 d}t=I^{\rm vib}_{\rightarrow i_\alpha}-I^{\rm vib}_{\rightarrow i_\alpha}$. By applying this to  the Hamiltonian~\eqref{vibron_tbm}, we get
 \begin{equation}
\label{current_op}
\begin{split}
I_{ \rightarrow i_{\alpha}}^{\rm vib}&=-\ii\sum_{\beta}\sum_{j_{\beta}>i_{\alpha}}{J}^*_{j_{\beta}i_{\alpha}}a_{i_{\alpha}}^{\dagger}a_{j_{\beta}}^{\phantom{\dagger}}+\text{H.c.},\\
I_{  i_{\alpha}\rightarrow}^{\rm vib}&=-\ii\sum_{\beta}\sum_{j_{\beta}>i_{\alpha}}{J}_{i_{\alpha}j_{\beta}}a_{j_{\beta}}^{\dagger}a_{i_{\sigma}}^{\phantom{\dagger}}+\text{H.c.},\\
\end{split}
\end{equation}
where we have used ${J}_{i_{\alpha}j_{\beta}}={J}_{j_{\beta}i_{\alpha}}$. In
the particular case of Eq.~\eqref{vibron_tbm}, the tunnelings are real.
However, we keep the above expression  general since it will be useful in
other sections below. In Figs.~\ref{bulk_cooling_2_ions}{\bf (b)}-{\bf (c)},
we let one of the Rabi frequencies vary in the range $\Omega_{{\rm L}^{\rm
sw}_{4_\tau}}\in[0.1\Gamma_{\tau},10\Gamma_{\tau}]$, which allows us to
modify the ratio $\Gamma_{\rm L}/\Gamma_{\rm R}$. As shown in these figures,
if the constraint $2\gamma_{\ell_{\tau}}\gg J_{i_{\alpha}j_{\beta}}$ is
fulfilled, there is an excellent agreement of both numerical solutions.

\subsubsection{ Thermal quantum wire: assessing Fourier's law}

Let us now consider a mesoscopic thermal quantum wire (TQW) with $N=20$ ions, which would have a length of $L\approx0.21\,$mm for the trap frequencies $(\omega_{\alpha x},\omega_{\alpha y},\omega_{\alpha z})/2\pi=(5,5,0.1)\hspace{0.2ex}$MHz. The
configuration of ions species is
\beq
\nonumber
\tau-\sigma-\cdots-\sigma-\cdots-\sigma-\tau,
\eeq
 where $\tau=$$^{24}{\rm Mg}^+$ and  $\sigma=$$^{25}{\rm Mg}^+$, and we choose the detunings $\Delta_{1_\tau}=-0.8\Gamma_{\tau}$, $\Delta_{N_\tau}=-0.6\Gamma_{\tau}$. We shall use this setup to test the validity of { Fourier's
 law}
 of thermal conduction. This law predicts the onset of a linear gradient in the number of carriers between the reservoirs $
 \langle n_{i_{\alpha}}\rangle_{\rm FL}=\bar{n}_{\rm L}+{i_{\alpha}(\bar{n}_{\rm R}-\bar{n}_{\rm L})}/{N}.
$

 In Fig.~\ref{fourier_law_app}, we represent the number of vibrons in the steady state of the TWQ if the laser-cooling Rabi frequencies are set to  $\Omega_{{\rm L}^{\rm sw}_{1_\tau}}=\Omega_{{\rm L}^{\rm
 sw}_{N_\tau}}=1.4\Gamma_{\tau}$.  These numerical simulations confirm the theoretical prediction~\eqref{occupation} to a good degree of accuracy. It is also clear from
 this figure that the number of vibrons  does not display a linear gradient, as predicted by Fourier's law, but is rather homogeneous. As mentioned in the main text, this apparent violation of Fourier's law is not a surprise, since  this  law applies to diffusive processes, whereas our vibron transport is ballistic. Let us now explore two possible mechanisms to introduce
diffusive dynamics in the problem.

\begin{figure}
\centering
\includegraphics[width=0.75\columnwidth]{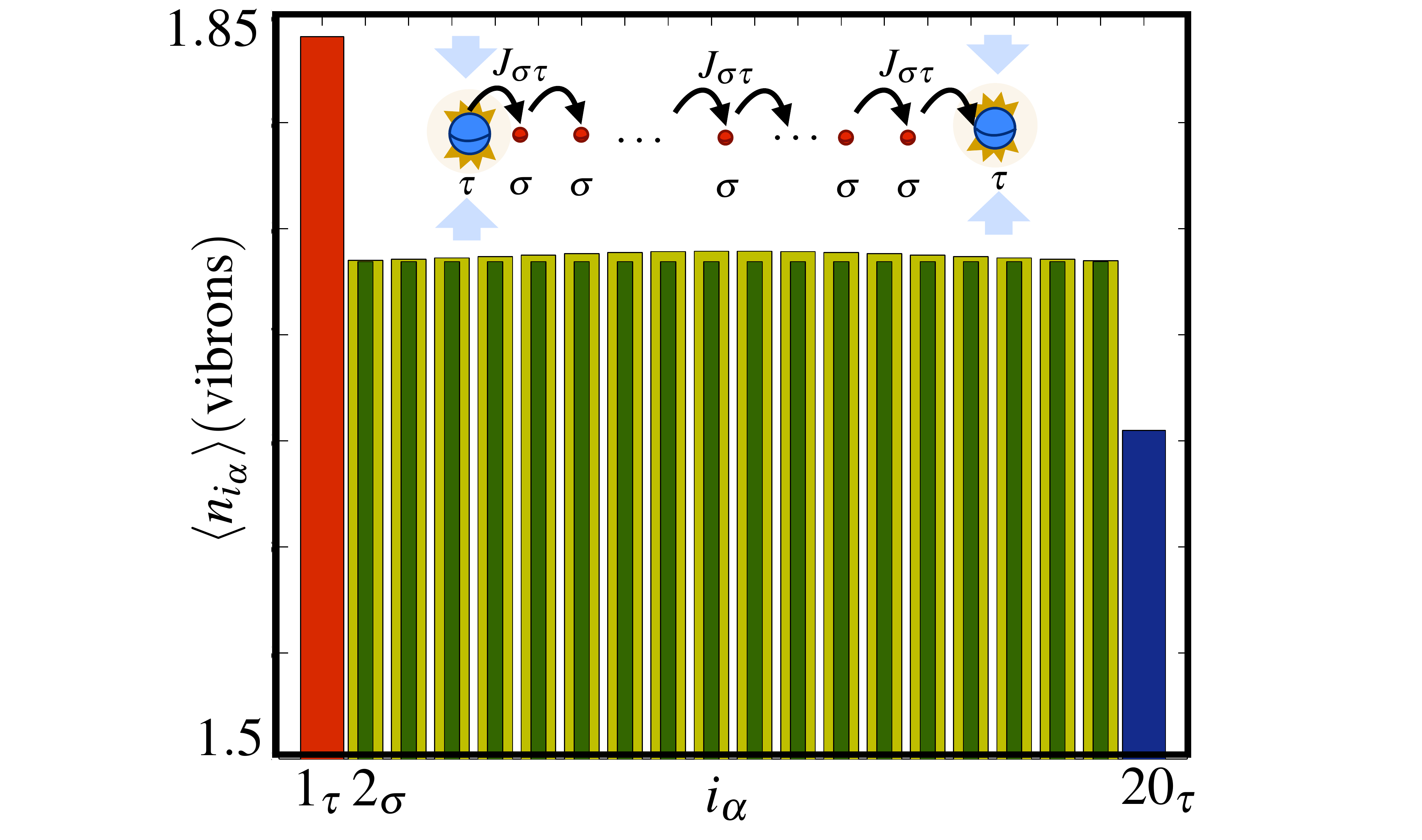}
\caption{ {\bf Thermal quantum wire:}  Steady-state number of vibrons $\langle n_{i_\alpha}\rangle$ along the ion chain (see the text for the particular parameters). The bulk of the TQW displays a homogeneous number of
vibrons, in  contrast to   Fourier's law. The  yellow (green) bars correspond to the numerical solution of Eqs.~\eqref{ode}-\eqref{edge_matrices} (Eqs.~\eqref{ode}-\eqref{bulk_matrices}).
}
\label{fourier_law_app}
\end{figure}

{\it i) Noise-induced  dephasing.--} A possible mechanism to introduce diffusion in the transport is to consider an engineered noise leading to dephasing in the vibron tunneling. This can be accomplished by injecting a
noisy signal in the trap electrodes~\cite{dephasing_noise_app}, leading to fluctuating trap frequencies that modify the on-site energies of the tight-binding Hamiltonian
\beq
\nonumber
H_{\rm vo}\to H_{\rm vo}+\delta H_{\rm vo}(t)=\sum_{i_\alpha}(\omega_{i_\alpha}+\delta\omega_{i_\alpha}(t))a^{\dagger}_{i_\alpha}a_{i_\alpha}.
\eeq
Here, we have considered that $\delta\omega_{i_\alpha}(t)$ is a zero-mean random Markov process that is stationary and Gaussian. Such process, usually known as the Ornstein-Uhlenbeck process~\cite{ou}, is typically characterised
by
a diffusion constant $c$, and a  correlation time $\tau_{\rm c}$, which we assume to be much shorter than the time-scales of interest $t\gg \tau_{\rm c}$. Moreover, we introduce a correlation length $\xi_{\rm c}$ in order to model
the extent of the noisy signal on the trap electrodes. The power spectrum of this noise is
\begin{equation}
\nonumber
S_{\delta\omega_{i_ \alpha},\delta\omega_{j_ \beta}}(\omega)={\rm Re}\left\{\int_0^{\infty}{\rm d}t\overline{\delta\omega_{i_ \alpha}(t)\delta\omega_{j_\beta}(0)}\ee^{+\ii\omega t}\right\},
\end{equation}
where the "bar" refers to the statistical average over the random process. In particular, the above three constants determine completely
the noise spectrum
\beq
\nonumber
S_{\delta\omega_{i_ \alpha},\delta\omega_{j_\beta}}(\omega)=\frac{\Gamma_{\rm d}}{1+(\omega\tau_{\rm c})^2}\ee^{-\frac{|z^0_{i_ \alpha}-z^0_{j_\beta}|}{\xi_{\rm c}}},
\eeq
where we have introduced the equilibrium positions  of the ions, and the dephasing rate $\Gamma_{\rm d}=c\tau_{\rm c}^2/2$.

By using a Born-Markov approximation to account for the fluctuating
trap frequencies, the master equation  becomes \beq
\nonumber
\dot{\mu}=\mathcal{L}_{\rm ddsv}(\mu)-\int_0^{\infty}{\rm
d}t'\overline{[\delta  H_{\rm vo}(t),[\delta H_{\rm vo}(t-t'),\mu]]}. \eeq
Using the above noise spectrum, the Liouvillian of the TQW gets the
additional contribution of a pure-dephasing super-operator $\mathcal{L}_{\rm
ddsv}\to\mathcal{L}_{\rm ddsv}+\mathcal{D}_{\rm d}$, where
\beq \label{dephasing}
\mathcal{D}_{\rm d}(\bullet)=\sum_{\alpha,\beta}\sum_{i_\alpha,
j_\beta}S_{\delta\omega_{i_\alpha},\delta\omega_{j_\beta}}(0)(n_{i_\alpha}\bullet
n_{j_\beta}-n_{j_\beta}n_{i_\alpha}\bullet)+{\rm H.c.}, \eeq such that the
dephasing rate only depends on the zero-frequency component of noise
spectrum. We also observe that $\xi_{\rm c}$ controls the collective effects
in the dephasing dynamics of the TQW: if $\xi_{\rm c}\to0$, we obtain a
purely local dephasing that introduces phase-breaking processes in the
vibron transport, whereas for $\xi_{\rm c}\to \infty$, the noise is purely
global, such that the tunneling dynamics is not affected, and remains  ballistic.

This collective dephasing modifies the system of differential equations~\eqref{ode}  for the two-point vibron correlators $C_{i_\alpha,j_\beta}=\langle a_{i_\alpha}^{\dagger}a_{j_\beta}\rangle$, which becomes
\begin{equation}
\label{ode_dephasing}
\frac{{\rm d}C}{{\rm d}t}=\ii[\mathbb{J},C]-(\mathbb{W}C+C\mathbb{W}^*)-\mathbb{D}C+\mathbb{K},
\end{equation}
where we have introduced the following matrix
\begin{equation}
\nonumber
\mathbb{D}=\sum_{\alpha,\beta}\sum_{i_\alpha,j_\beta}2\Gamma_{\rm d}\bigg(1-\ee^{-|z^0_{i_ \alpha}-z^0_{j_\beta}|/\xi_{\rm c}}\bigg)\ket{i_\alpha}\bra{j_\beta}.
\end{equation}
Here, $\{\ket{i_\alpha}\}_{i_\alpha=1}^N$ form an orthogonal basis of the $N$-dimensional  subspace of the two-point correlators.

\begin{figure}
\centering
\includegraphics[width=1\columnwidth]{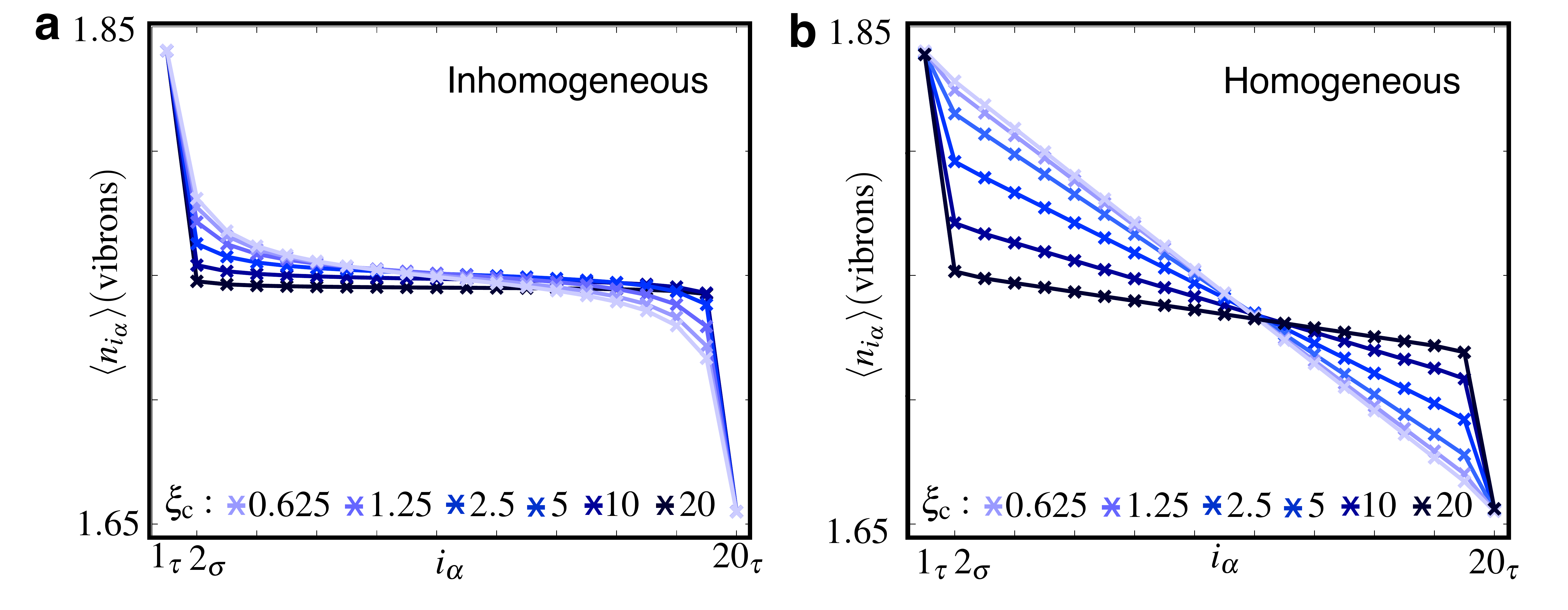}
\caption{ {\bf The dephasing route to Fourier's law:}  {\bf (a)} Steady-state number of vibrons $\langle n_{i_\alpha}\rangle$ for an inhomogeneous ion chain in a linear Paul trap. As the correlation length of the dephasing noise
$\xi_{\rm c}$ decreases (in units of the nearest-neighbour spacing at the centre of the chain), keeping $\Gamma_{\rm d}=10\gamma_{N_{\tau}}$, we observe a inhomogeneous distribution of vibrons across the chain. Far away from the
edges and close to the bulk of the chain, the distribution displays a linear gradient. {\bf (b)} Steady-state number of vibrons $\langle n_{i_\alpha}\rangle$ for an homogeneous chain in micro-fabricated ion trap array. As
$\xi_{\rm c}$ decreases and $\Gamma_{\rm d}=10\gamma_{N_{\tau}}$ is fixed, we observe a perfect linear gradient across the chain. Therefore, edge effects are less pronounced in this homogeneous scenario.}
\label{fourier_law_dephasing}
\end{figure}

In Fig.~\ref{fourier_law_dephasing}, we compute the steady state solution of the above system of differential equations~\eqref{ode_dephasing}. We consider the same experimental parameters as previously, and set the dephasing rate
to $\Gamma_{\rm d}=10\gamma_{N_\tau}$. As can be observed in Fig.~\ref{fourier_law_dephasing}{\bf (a)}, in the limit of large correlation lengths $\xi_{\rm c}\sim |\tilde{z}_{1_{\tau}}-\tilde{z}_{N_{\tau}}|$, the vibrons display
the same  homogeneous distribution that does not agree with Fourier's law (i.e. ballistic regime). As the correlation length is decreased, a linear gradient starts to develop at the bulk of the chain (diffusive regime). It is interesting
that
we have a single parameter to control the ballistic-diffusive crossover. However,  note that edge effects  mask the linear gradient. We have found that these edge effects are particularly strong for a linear Paul trap, since the equilibrium positions correspond to an inhomogeneous crystal. By modifying the dc trapping potentials, or by
considering micro-fabricated ion traps, it is possible to obtain a homogeneous ion crystal. In Fig.~\ref{fourier_law_dephasing}{\bf (b)}, we study numerically the distribution of vibrons in this regime. Our results show that edge effects are less pronounced, and a perfect linear gradient arises as predicted by Fourier's law.

{\it ii) Spin-assisted random disorder.--} In order to introduce another  diffusive mechanism for the transport of vibrons, we apply a spin-vibron coupling~\eqref{driving}. By  controlling the laser intensities, polarisations and
frequencies, we further impose that $\nu_\sigma=\Delta\omega_\sigma^+=0$, which leads to a static spin-vibron coupling
$
H_{\rm
sv}^{\sigma}=\sum_{i_\sigma}\half\Delta\omega^-_{\sigma}n_{i_ \sigma}\sigma_{i_ \sigma}^z,
$
whose strength $\Delta\omega_{\sigma}^-$ can be controlled at will.
The idea to mimic the effects of diagonal disorder is to use the spin degrees of freedom as a gadget to build a Liouvillian with random on-site energies. Here, the randomness is  inherited from the quantum superposition principle
in the spin degrees of freedom~\cite{paredes,Bermudez-NJP-2010_supp}.

Let us consider an initial pure state for the $\sigma$-spins of the TQW, namely $\mu_{\rm spin}(0)=\ketbra{{\Psi_0}}$. Without loss of generality, it can be expressed  as
$\ket{\Psi_0}=\sum_{\{s_{i_\sigma}\}}c_{\{s_{i_\sigma}\}}\ket{\{s_{i_\sigma}\}}$, where $\{s_{i_\sigma}\}=\{s_{2_\sigma},s_{3_\sigma},\cdots, s_{{N-1}_\sigma}\}$ is a particular spin configuration for the bulk $\sigma$-ions
$s_{i_\sigma}\in\{\uparrow_{\sigma},\downarrow_{\sigma}\}$. The reduced density matrix of the vibrons evolves in time according to
\beq
\nonumber
\mu_{\rm vib}(t)={\rm Tr}_{\rm spin}\{\ee^{\mathcal{L}_{\rm ddsv}(\{\sigma_{i_\sigma}^z\})t} \mu_{\rm spin}(0)\otimes \mu_{\rm vib}(0)\},
\eeq
where we have rewritten the spin-vibron Liouvillian~\eqref{ddsp} making explicit reference to its dependence on the spin operators $\mathcal{L}_{\rm ddsv}(\{\sigma_{i_\sigma}^z\})=\mathcal{L}_{\rm
ddsv}(\{\sigma_{2_\sigma}^z,\sigma_{3_\sigma}^z,\cdots,\sigma_{N-1_\sigma}^z\})$. From this expression,  the reduced density matrix evolves as
\beq
\nonumber
\mu_{\rm vib}(t)=\sum_{\{s_{i_\sigma}\}}p_{\{s_{i_\sigma}\}}\ee^{\mathcal{L}_{\rm ddsv}(\{s_{i_\sigma}\})t} \mu_{\rm vib}(0),
\eeq
which can be interpreted as an statistical average of the time-evolution under a stochastic Liouvillian. In particular, the Liouvillian $\mathcal{L}_{\rm ddsv}(\{s_{i_\sigma}\})$ depends on the binary variables $\{s_{i_\sigma}\}$,
which inherit their randomness from the quantum parallelism of the initial spin state. In fact, the associated probability distribution for the binary random variable  is $p_{\{s_{i_\sigma}\}}=|c_{\{s_{i_\sigma}\}}|^2$. Therefore,
we can formally write $\mu_{\rm vib}(t)=\overline{\mu_{\rm vib}(t)}$, where the "bar" refers to a statistical average over a random Liouvillian
\begin{equation}
\label{sdtb}
\mathcal{L}_{\rm ddsv}\to\mathcal{L}_{\rm sdtb}(\mu_{\rm vib})=-\ii[H_{\rm stb},\mu_{\rm vib}]+\sum_{\ell_\tau}\mathcal{D}^{\ell_\tau}_{\rm v}(\mu_{\rm vib}).
\end{equation}
Here, $\mathcal{D}^{\ell_\tau}_{\rm v}$ is the dissipator acting on the edge vibrons~\eqref{dissipation}, whereas the stochastic tight-binding Hamiltonian is
\begin{equation}
\nonumber
H_{\rm stb}\!\!=\!\!\sum_{\alpha,i_{\alpha}}\!\!\epsilon_{i_{\alpha}}a_{i_{\alpha}}^{\dagger}a_{i_{\alpha}}^{\phantom{\dagger}}\!+\hspace{-1ex}\sum_{\alpha,\beta}\sum_{ i_{\alpha}\neq
j_{\beta}}\!\!\!\big({J}_{i_{\alpha}j_{\beta}}a_{i_{\alpha}}^{\dagger}a_{j_{\beta}}^{\phantom{\dagger}}\!+\!\text{H.c.}\big).
\end{equation}
Here, the on-site energies of the bulk $\sigma$-ions are binary random variables sampling  $\epsilon_{i_{\sigma}}\in\{{\omega}_{i_ \sigma} -\half\Delta\omega^-_{{\sigma}},{\omega}_{i_ \sigma} +\half\Delta\omega^-_{{\sigma}}\}$. For an initial spin state $\ket{\Psi_0}=\otimes_{i_\sigma}(\ket{{\uparrow}_{i_\sigma}}+\ket{{\downarrow}_{i_\sigma}})/\sqrt{2}$, this diagonal disorder has a flat probability distribution
$p(\epsilon_{i_{\sigma}})=\half$.

 In order to study the steady state for the vibrons thermalizing under this disordered Liouvillian~\eqref{sdtb}, we can solve the system of differential equations for the two-point correlators~\eqref{ode}   for each realisation of
 the diagonal disorder
 \begin{equation}
\frac{{\rm d}C_{\{\epsilon_{i_\sigma}\}}}{{\rm d}t}=\ii[\mathbb{J},C_{\{\epsilon_{i_\sigma}\}}]-(\mathbb{W}C_{\{\epsilon_{i_\sigma}\}}+C_{\{\epsilon_{i_\sigma}\}}\mathbb{W}^*)+\mathbb{K}.
\end{equation}
 Then, we should average over the random variable according to the probability distribution $p(\epsilon_{i_{\sigma}})=\half$. Because of the disorder,
$
\mathbb{J}^{\rm edge}_{i_{\alpha}j_{\beta}}(\{\epsilon_{k_\sigma}\})=\epsilon_{i_\alpha}\delta_{i_{\alpha}j_{\beta}}+{J}_{i_{\alpha}j_{\beta}}
$ becomes stochastic.
After performing the statistical average $\overline{C}(t)$, we can reconstruct the vibron density of the disordered TQW.

We consider the same setup  as in Fig.~\ref{fourier_law_app} for the ordered TQW, namely a $N=20$ ion chain. Moreover, we use the same parameters introduced there. For the spin-induced disorder, we set
$\Delta\omega_{\sigma}^-=10\gamma_{N_\tau}$, which corresponds to a strong spin-vibron coupling. In Fig.~\ref{fourier_law_disorder}, we represent the distribution of vibrons along the TQW in the steady-state. In this case, the
predictions for both a homogeneous ion crystal (i.e. microtrap array), and an inhomogeneous one (i.e. linear Paul trap) coincide. As a consequence of the disorder-induced diffusion, the vibron layout is no longer homogeneous, but
rather displays a linear gradient far way from the edges of the chain.

Before closing this section, let us also comment on another interesting perspective for the TQW, namely the possibility of realising {\it noise-assisted quantum heat transport}. As demonstrated in~\cite{noise_assisted_transport},
the efficiency of transport in quantum networks including linear chains with disorder may be sometimes increased by the presence of local dephasing noise. In order to test this prediction in our current scenario, let us note first
that the presence of disorder~\eqref{sdtb} will partially  inhibit the heat transport. By switching on the local dephasing~\eqref{dephasing},  the interference leading to Anderson localization, or transport bottlenecks  due to energy
mismatches between neighboring sites, can be overcome thanks to the presence of noise, thus assisting the transport of heat. This can be probed by the current measurement described in a section below.

\begin{figure}
\centering
\includegraphics[width=.75\columnwidth]{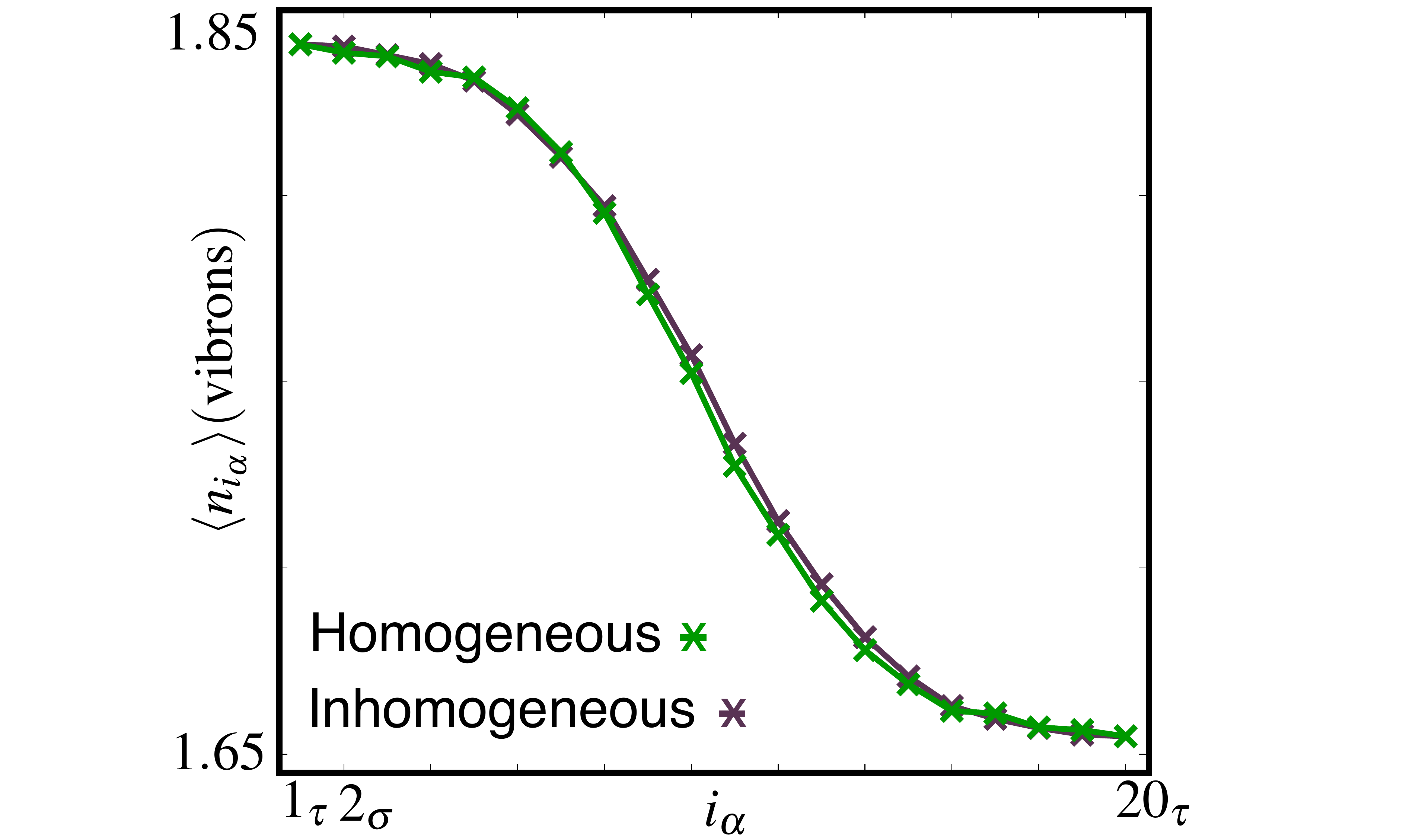}
\caption{ {\bf The disorder route to Fourier's law:}  Steady-state number of vibrons $\langle n_{i_\alpha}\rangle$ for an inhomogeneous ion chain in a linear Paul trap (purple crosses), and a  homogeneous ion chain in a microtrap
array (green crosses). We observe an inhomogeneous distribution of vibrons across the chain. Far away from the edges, and close to the bulk of the chain, the distribution displays a linear gradient. }
\label{fourier_law_disorder}
\end{figure}

\subsubsection{Thermal leads and single-spin heat switch}

We now consider a mesoscopic ion chain with $N$ ions. The configuration of ions species is
\beq
\nonumber
\tau-\sigma-\cdots-\sigma-\kappa-\sigma-\cdots-\sigma-\tau,
\eeq
where $\tau=$$^{24}{\rm Mg}^+$, $\sigma=$$^{25}{\rm Mg}^+$, and $\kappa=$$^{9}{\rm Be}^+$. The $\kappa$-ion plays the role of the thermal quantum dot (TQD), the $\tau$-ions act effective vibronic reservoirs, and the left/right
chain of $\sigma$-ions acts as a lead that connects the TQD to the reservoirs. We will start by discussing the conditions under which the $\sigma$-ions can be interpreted as effective thermal leads. Then, we will discuss how to
control the tunneling of vibrons across the $\kappa$-ion, which can be exploited to build a single-spin heat switch.

{\it i) Effective thermal leads.--} In order to devise  the leads, we apply a strong and static spin-vibron coupling~\eqref{driving} to the $\sigma$-spins, namely
$
H_{\rm
sv}=\sum_{i_\sigma}\half\Delta\omega_{\sigma}^-n_{i_ \sigma}\sigma_{i_ \sigma}^z.
$
In this case, we  consider the strong-driving regime
\begin{equation}
\label{condition_leads}
\tilde{J}_{i_{\alpha}j_{\beta}},\tilde{\Lambda}_{i_{\alpha}j_{\beta}}^{\pm}\ll
|\Delta\omega^-_{\sigma}|\ll\delta_{\ell_\tau},\gamma_{\ell_\tau},
\end{equation}
and the following initial state for the spins of the leads $
\ket{\psi_0}=\ket{{\downarrow_{\sigma}\cdots\downarrow_{\sigma}}}\ket{{\phi_{{p_\kappa}}}}\ket{{\uparrow_{\sigma}\cdots\uparrow_{\sigma}}},
$ where $\ket{{\phi_{{\kappa}}}}$ is an arbitrary spin state of the
$\kappa$-ion. In this regime, the spin-vibron
coupling provides a large and static shift of the vibron
on-site energies \beq \nonumber
{\omega}_{i_\sigma}\to\tilde{\omega}_{i_\sigma}={\omega}_{i_\sigma}-\half\Delta\omega^-_{\sigma}\theta(p_{\kappa}-i_{\sigma})+\half\Delta\omega^-_{\sigma}\theta(i_{\sigma}-p_{\kappa}),
\eeq where we have introduced the Heaviside step function $\theta(x)=1,$ if
$x>0$. Because of these shifts, the  thermalization of the bulk
ions described in Sec.~\ref{sec_diss_bulk} must be re-addressed. Assuming
that the separation of time-scales is valid, we can derive a similar master equation~\eqref{bulk_meq} for the
bulk ions. However, in the limit~\eqref{condition_leads} a rotating wave
approximation allows us to neglect all the tunneling processes that lead to
the thermalization between the two halves of the ion chain. This observation
allows use to  partition the master
equation into
\begin{equation}
\label{meq_leads}
\dot{\tilde{\mu}}_{{\rm bulk}}=\mathcal{L}_{\rm L}(\tilde{\mu}_{{\rm bulk}})+\mathcal{L}_{\rm L\kappa R}(\tilde{\mu}_{{\rm bulk}})+\mathcal{L}_{\rm R}(\tilde{\mu}_{{\rm bulk}}).
\end{equation}
Here, we have introduced the Liouvillian for each of the leads $\mathcal{L}_{\rm L/R}(\bullet)=-\ii[H_{\rm L/R},\bullet]+\mathcal{D}_{\rm L/R}(\bullet)$. For the left-most lead
\begin{equation}
\begin{split}
\nonumber
H_{\rm L}=\sum_{i_{\sigma},j_{\sigma}<p_{\kappa}}\hspace{-2ex}\tilde{J}_{i_\sigma,j_\sigma}a^{\dagger}_{i_{\sigma}}a^{\phantom{\dagger}}_{j_{\sigma}}\ee^{+\ii(\tilde{\omega}_{i_\sigma}-\tilde{\omega}_{j_\sigma})t}+{\rm H.c.}, \\
\end{split}
\end{equation}
where we have used the renormalized tunnelings of Eq.~\eqref{eff_J_app}. Additionally, the corresponding dissipators are
\begin{equation}
\nonumber
\begin{split}
\mathcal{D}_{\rm L}=\hspace{-1ex}\sum_{i_\sigma,j_\sigma<p_\kappa}\hspace{-1ex}\mathcal{D}[\tilde{\Lambda}^+_{i_\sigma,j_ \sigma},
a^{{\dagger}}_{i_ \sigma}(t),a^{\phantom{\dagger}}_{j_ \sigma}(t)]+\mathcal{D}[\tilde{\Lambda}^-_{i_ \sigma,j_ \sigma},a^{\phantom{\dagger}}_{i_ \sigma}(t),a^{{\dagger}}_{j_ \sigma}(t)],\\
\end{split}
\end{equation}
where we have used the dissipative couplings in Eq.~\eqref{eff_gamma_app}, the interaction picture operators $a_{i_ \sigma}(t)=a_{i_ \sigma}\ee^{-\ii\tilde{\omega}_{i_\sigma}t}$, and  the  generic super-operator~\eqref{generic_so}. Note that for the right-most lead, the expressions are equivalent, but we must sum over sites $i_{\sigma},j_{\sigma}>p_{\kappa}$.
The final part is the coupling of the leads to the TQD, which can be expressed as
\begin{equation}
\label{TQD_leads}
\mathcal{L}_{\rm L\kappa R}(\bullet)=-\ii\big[H_{\rm L\kappa R}(t), \bullet\big]+\mathcal{D}_{\kappa}(\bullet),
\end{equation}
where we have introduced the Hamiltonian
\beq
\label{LkR}
H_{\rm L\kappa R}(t)=\sum_{i_{\sigma}\neq p_{\kappa}}2\tilde{J}_{i_\sigma,p_\kappa}a^{\dagger}_{i_{\sigma}}a^{\phantom{\dagger}}_{p_{\kappa}}\ee^{+\ii(\tilde{\omega}_{i_\sigma}-\tilde{\omega}_{\ell_\kappa})t}+{\rm H.c.},
\eeq
and the dissipator due to the long-range tunneling between the reservoir and the TQD
\beq
\nonumber
\mathcal{D}_{\kappa}=\mathcal{D}[\tilde{\Lambda}^+_{p_\kappa,p_\kappa},
a^{{\dagger}}_{p_\kappa},a^{\phantom{\dagger}}_{p_\kappa}]+\mathcal{D}[\tilde{\Lambda}^-_{p_\kappa,p_\kappa},a^{\phantom{\dagger}}_{p_\kappa},a^{{\dagger}}_{p_\kappa}],
\eeq
From the master equation~\eqref{meq_leads}, we thus expect that the left/right half or the chain thermalizes individually with the left/right reservoir, such that the mean vibron number  is
$
\langle n_{i_\sigma}\rangle_{\rm ss}\approx\bar{n}_{\rm L}\theta(p_{\kappa}-i_{\sigma})+\bar{n}_{\rm R}\theta(i_{\sigma}-p_{\kappa}).
$
 Thus, the two chains of $\sigma$-ions serve as a   lead to connect the reservoirs to  the $\kappa$-ion,  modifying the local density of states seen by the TQD.

\begin{figure}
\centering
\includegraphics[width=1\columnwidth]{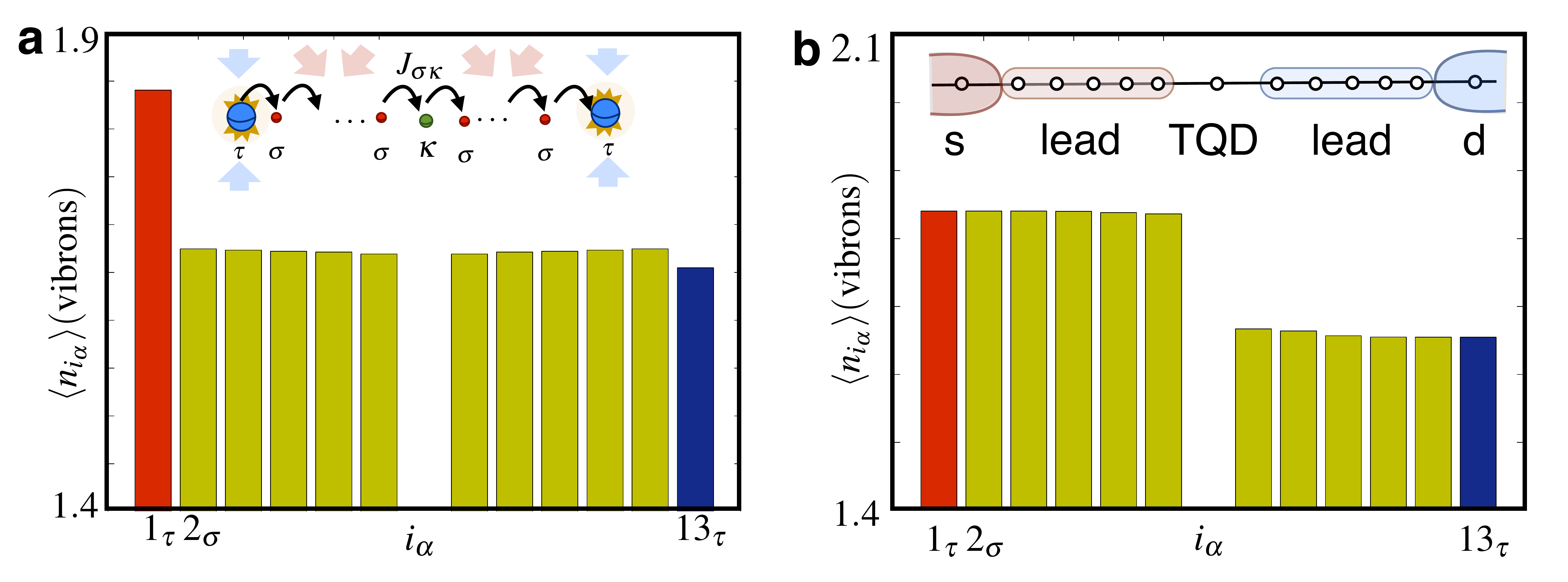}
\caption{ {\bf Thermal leads:}  {\bf (a)} Steady-state number of vibrons $\langle n_{i_\alpha}\rangle$ along the ion chain in the absence of the static spin-vibron coupling $\Delta\omega_{\sigma}^-=0$ (see the text for the
remaining parameters). The bulk of the chain yields a homogeneous number of vibrons. The  bars (yellow, red, and  blue)  correspond to the numerical solution of Eqs.~\eqref{ode_shift}. {\bf (b)} Same as
above but setting a strong spin-vibron coupling $\Delta\omega_{\sigma}^-=200J_{1_{\tau}2_\sigma}$. The number of vibrons displays a step-like function.
}
\label{thermal_contacts_fig}
\end{figure}

 In order to support this theoretical prediction, we
integrate numerically the system of differential equations for
the two-point correlators $C_{i_{\alpha}j_{\beta}}=\langle a_{i_{\alpha}}^{\dagger}a_{j_{\beta}}\rangle$, namely
\begin{equation}
\label{ode_shift}
\frac{{\rm d}C}{{\rm d}t}=\ii[\tilde{\mathbb{J}}^{\rm edge},C]-(\mathbb{W}^{\rm edge}C+C(\mathbb{W}^{\rm edge})^*)+\mathbb{K}^{\rm edge},
\end{equation}
where the matrices $\mathbb{W}^{\rm edge},\mathbb{K}^{\rm edge}$ have been
defined in Eq.~\eqref{edge_matrices}. Because of the on-site energy shifts,
we have to modify
$
\tilde{\mathbb{J}}^{\rm edge}_{i_{\alpha}j_{\beta}}=\tilde{\omega}_{i_\alpha}\delta_{i_{\alpha}j_{\beta}}+J_{i_{\alpha}j_{\beta}}$, $\alpha,\beta\in\{\sigma,\kappa,\tau\}.
$
In Fig.~\ref{thermal_contacts_fig}, we represent the mean value of vibrons in the steady state of a $N=13$ ion chain, where we recall that the chosen species are $\tau=$$^{24}{\rm Mg}^+$,  $\sigma=$$^{25}{\rm Mg}^+$, and
$\kappa=$$^{9}{\rm Be}^+$. We consider the following trap frequencies $(\omega_{\alpha x},\omega_{\alpha y},\omega_{\alpha z})/2\pi=(5,5,0.1)\hspace{0.2ex}$MHz, and the laser-cooling parameters
$\Delta_{1_\tau}=-0.8\Gamma_{\tau}$,
$\Delta_{13_\tau}=-0.6\Gamma_{\tau}$, and  $\Omega_{{\rm L}^{\rm sw}_{1_\tau}}=\Omega_{{\rm L}^{\rm sw}_{13_\tau}}=2.4\Gamma_{\tau}$, such that we expect each reservoir to thermalize to $\bar{n}_{1_{\tau}}=1.84$, and
$\bar{n}_{13_{\tau}}=1.65$. In Fig.~\ref{thermal_contacts_fig}{\bf (a)}, we represent our results in the absence of the on-site energy shifts $\Delta\omega_{\sigma}^-=0$. In analogy to the TQW [Fig.~\ref{fourier_law_app}{\bf
(a)}],
we recover the expected homogeneous mean number of vibrons along the whole bulk. In Fig.~\ref{thermal_contacts_fig}{\bf (b)}, we study the consequences of switching a very strong spin-vibron coupling
$\Delta\omega_{\sigma}^-=200J_{1_{\tau}2_{\sigma}}$. In this case, the left half of the chain thermalizes to the left reservoir $\langle n_{i_{\sigma}}\rangle_{\rm ss}\approx \bar{n}_{\rm L}$ for $i_{\sigma}<7_{\kappa}$, whereas
the right half thermalizes to $\langle n_{i_{\sigma}}\rangle_{\rm ss}\approx \bar{n}_{\rm R}$ for $i_{\sigma}>7_{\kappa}$. We can thus conclude that our prediction where each lead thermalizes to the neighbouring reservoir, describes
considerably well the actual steady state of the mixed ion chain.

Let us also note that, according to the coupling of the leads to the TQD described by $\mathcal{L}_{\rm L\kappa R}$~\eqref{TQD_leads}, the tunneling of vibrons across the TQD also becomes rapidly rotating in the regime of
strong couplings  $\Delta\omega_{\sigma}^{-}=200J_{1_{\tau}2_{\sigma}}$, such that the current through the TQD is inhibited. In fact, we find numerically that  the vibron current through the $\kappa$-ion is $\langle
I_{p_{\kappa}\rightarrow}^{\rm vib}\rangle_{\rm ss}\approx 1.6\cdot 10^{-4}{\rm vibrons}/s$. This must be contrasted to the case of $\Delta\omega_{\sigma}^-=0$, where $\langle I_{p_{\kappa}\rightarrow}^{\rm vib}\rangle_{\rm
ss}\approx 15{\rm vibrons}/s$. For experimental time-scales, we can consider that the  strong drivings  $\Delta\omega_{\sigma}^-$ suppress completely the vibron current through the TQD. Hence,  only the long-range tunnelings to
the
reservoirs influence the thermalization of the TQD  $\mathcal{L}_{\rm L\kappa R}\approx \mathcal{D}_{\kappa}$.

 {\it ii) Single-spin heat switch.--}
 We now describe a mechanism to switch on the vibron current across the TQD. We make use of the last ingredient in our toolbox~\eqref{ddsp}, a periodic spin-vibron coupling~\eqref{driving} applied to the $\kappa$-ions
$
H_{\rm
sv}^{\kappa}(t)=\half(\Delta\omega_{\kappa}^++\Delta\omega_{\kappa}^-\sigma_{p_ \kappa}^z)\cos(\nu_{\kappa}t-\varphi_{{\kappa}})n_{p_ \kappa}.
$
 According to  Sec.~\ref{sv_app}, and the explicit relations in Eqs.~\eqref{driving_equations_1}-\eqref{driving_equations_2}, we can achieve such a spin-vibron coupling by using a pair of laser beams with different frequencies. Moreover, by adjusting
 the
 laser intensities, detunings, polarizations, and phases, we impose
\beq
\label{switch_constraints}
\nu_{\kappa}=\half\Delta\omega^-_{\sigma},\hspace{1ex} \varphi_{\kappa}=0,\hspace{1ex}\Delta\omega_{\kappa}^-=r\Delta\omega_{\kappa}^+.
\eeq
 The idea is to use this periodic modulation to bridge the gradient of on-site energies between the two halves of the chain,  assisting in this way the tunneling through the TQD.  Moreover, we exploit the spin-dependent drivings, such that depending on the parameter $r$, we can  build a single-spin
 heat switch.

Let us supplement the Liouvillian~\eqref{TQD_leads} with the periodic spin-vibron coupling
$
H_{\rm L\kappa R}(t)\to \tilde{H}_{\rm L\kappa R}(t)= H_{\rm L\kappa R}(t)+H_{\rm
sv}^{\kappa}(t).
$
 In order to understand its effects,
we move into another interaction picture with respect to the periodic driving $a_{p_\kappa}\to U_{\rm sv}a_{p_\kappa}U_{\rm sv}^{\dagger}$, where $U_{\rm sv}={\rm exp}\{{\ii\int_0^t{\rm d}t'H^{\kappa}_{\rm sv}(t')}\}$. This leads
to $
a_{p_\kappa}\to a_{p_\kappa}\ee^{-\ii\zeta_{\kappa}(1+r\sigma_{p_\kappa}^z)\sin(\nu_{\kappa}t)}$, with $\zeta_{\kappa}=\frac{\Delta\omega_\kappa^+}{2\nu_{\kappa}},
$
which can be inserted in the the tunneling of vibrons between the TQD and the leads~\eqref{LkR}. By using the Jacobi-Anger expansion for the first-kind  Bessel functions $\mathfrak{J}_n(z)$,
together with the constraints~\eqref{switch_constraints}, it is possible to derive an effective Hamiltonian for the coupling of the leads to the TQD
\begin{widetext}
\beq
\label{pat}
H_{\rm L\kappa R}^{\rm PAT}\approx-\sum_{i_{\sigma}<p_\kappa}2\tilde{J}_{i_\sigma p_\kappa}\mathfrak{J}_{1}\big(\zeta_{\kappa}(1+r\sigma_{p_\kappa}^z) \big)a_{i_\sigma}^{\dagger}a_{p_\kappa}^{\phantom{\dagger}}+\sum_{i_{\sigma}>p_\kappa}2\tilde{J}_{i_\sigma p_\kappa}\mathfrak{J}_{1} \big(\zeta_{\kappa}(1+r\sigma_{p_\kappa}^z) \big)a_{i_\sigma}^{\dagger}a_{p_\kappa}^{\phantom{\dagger}}+{\rm H.c.}
\eeq
\end{widetext}
Here, we have considered that all species have the same trap frequencies, and  used a rotating wave approximation for
$\tilde{J}_{i_\sigma p_\kappa}\ll\half|\Delta\omega_{\sigma}^-|$. As announced previously, Eq.~\eqref{pat} shows that for the resonance condition $\nu_{\kappa}=\half\Delta\omega_{\sigma}^-$, the periodic
spin-vibron coupling is capable of assisting the tunneling of vibrons across the TQD. Moreover,  the spin-dependence of the effective tunneling  via $\mathfrak{J}_1 \big(\zeta_{\kappa}(1+r\sigma_{p_\kappa}^z) \big)$ can
be
exploited to build a single-spin heat switch. By setting $r=1$, we obtain $\mathfrak{J}_1 \big(\zeta_{\kappa}(1+\sigma_{p_\kappa}^z) \big)=\mathfrak{J}_1(2\zeta_{\kappa})\ketbra{{\uparrow_{\kappa}}}$, such
that the tunneling is only allowed if the $\kappa$-ion is in the spin-up state. Therefore, by controlling the $\kappa$-spin using microwave or laser radiation (i.e. $\pi$ pulses), it is possible to switch on/off the heat current.

In order to check these predictions numerically, we consider a simplified setup, namely  a $\sigma-\kappa-\sigma$ junction mimicking the connection of the thermal leads to the TQD. Rather than studying the steady state, we will
concentrate on the coherent dynamics to show that the tunneling can be switched on/off by controlling the spin state of the  the $\kappa$-ion. Let us define the parameters for this setup.  We consider the usual trap frequencies
$(\omega_{\alpha x},\omega_{\alpha y},\omega_{\alpha z})/2\pi=(5,5,0.1)\hspace{0.2ex}$MHz, and set the parameters of  static spin-vibron
coupling  for the $\sigma$-ions~\eqref{driving}
to $\Delta\omega^+_{\sigma}=-2(J_{2_\kappa 2_\kappa}-J_{1_\sigma 1_\sigma})$ and $\Delta\omega^-_{\sigma}=10^3J_{1_\sigma 2_\kappa}$. This provides an energy gradient that inhibits the tunneling across the TQD. The parameters of the periodic
spin-vibron coupling of the $\kappa$-ion~\eqref{driving}  are given by Eqs.~\eqref{switch_constraints}, where we set $r=1$. All these ingredients contribute to the dynamics given by  $H(t)=H_{\rm sv}^{\sigma}+H_{\rm sv}^{\kappa}(t)+H_{\rm tb}$,
which is solved numerically and compared to the theoretical predictions from $H_{\rm L\kappa R}^{\rm PAT}$~\eqref{pat}.

We consider the initial state $\rho(0)=\rho_{1_\sigma}\otimes\rho_{2_\kappa}\otimes\rho_{3_\sigma}$, where $\rho_{i_\alpha}=\ketbra{n_{i_\alpha}}\otimes\ketbra{s_{i_\alpha}}$ is determined by the vibrational Fock states
$n_{1_{\sigma}}=1$, $n_{2_{\kappa}}=0$, $n_{3_{\sigma}}=0$, and the spin states $s_{i_\alpha}\in\{\uparrow_{i_\alpha},\downarrow_{i_\alpha}\}$. We want to understand how the dynamics of  such an initial state for
$s_{1_\sigma}=\downarrow_{1_\sigma}$, $s_{3_\sigma}=\uparrow_{1_\sigma}$, is modified by changing $s_{2_\kappa}\in\{\uparrow_{2_\kappa},\downarrow_{2_\kappa}\}$. In Fig.~\ref{switch_fig}{\bf (a)}, we set
$s_{2_\kappa}=\uparrow_{2_\kappa}$, and observe how the vibron initially at the leftmost $\sigma$-ion tunnels through the TQD until it reaches the rightmost $\sigma$-ion. The agreement between both descriptions supports the validity of our derivation. Therefore, we expect that by interspersing
$\pi$ pulses that invert the $\kappa$-spin ${\ket{{\uparrow}_{2_\kappa}}}\leftrightarrow{\ket{{\downarrow}_{2_\kappa}}}$, we can switch on/off the vibron current. In Fig.~\ref{switch_fig}{\bf (b)}, we show that two consecutive
$\pi$ pulses allow us to switch off the vibron current momentarily, which thus confirms our prediction.

\begin{figure}
\centering
\includegraphics[width=1\columnwidth]{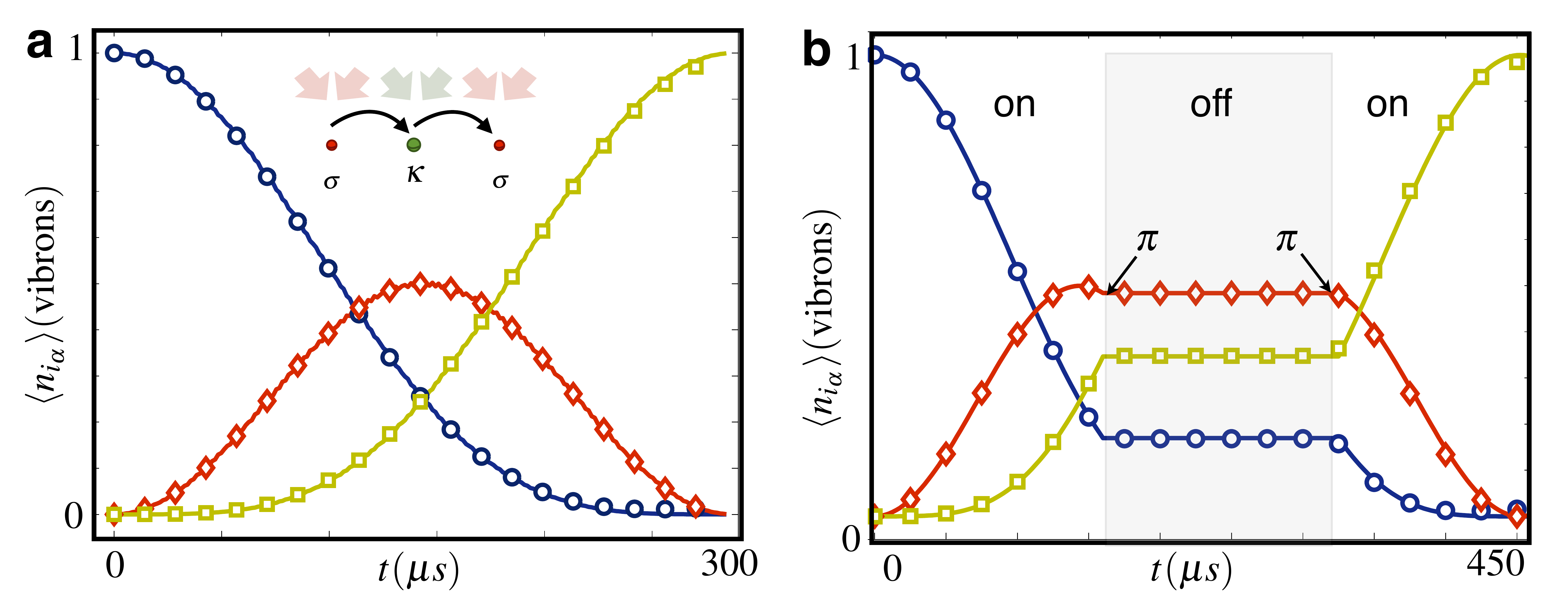}
\caption{ {\bf Single-spin heat switch:} {\bf (a)} Mean number of vibrons $\langle n_{i_\alpha}\rangle$ as a function of time in the regime of photon-assisted tunneling $s_{2_\kappa}=\uparrow_{2_\kappa}$ (see the text for the
remaining parameters).  The solid lines ($\langle n_{1_\sigma}\rangle$ blue, $\langle n_{2_\kappa}\rangle$ red, $\langle n_{3_\sigma}\rangle$ yellow) represent the exact solution of $H(t)$, while the open symbols ($\langle
n_{1_\sigma}\rangle$ circles, $\langle n_{2_\kappa}\rangle$ diamonds, $\langle n_{3_\sigma}\rangle$ squares) correspond to the effective photon-assisted-tunneling Hamiltonian $H_{\rm L\kappa R}^{\rm PAT}$. {\bf (b)} Mean number of
vibrons $\langle n_{i_\alpha}\rangle$ (same as in {\bf (a)}) as a function of time, where the $\kappa$-spin undergoes two consecutive $\pi$-pulses that switch off/on the current. Note that the $\pi$-pulses are synchronised with
the
period of the spin-vibron coupling $H_{\rm sv}^{\kappa}(t)$.  }
\label{switch_fig}
\end{figure}

\section{Spin-based measurements of heat transport}

The goal of this section is to present a detailed derivation, supported by numerical simulations, of the Ramsey probes for measuring  vibronic observables~\eqref{ramsey}. Then we particularise to the measurements of
the vibron number and the heat current.

\subsection{Ramsey measurement of  vibronic observables}

Let us start from the  bulk spin-vibron model in Eq.~\eqref{bulk_liouvillian}, and consider a generic spin-vibron coupling for the $\kappa$-spins
\begin{equation}
\label{generic_spin_vibron}
H_{\rm sv}^{\rm bulk}(t)\rightarrow \tilde{H}_{\rm
sv}^{O}=\sum_{i_{\kappa}}\half\lambda_OO_{i_ \kappa}\sigma_{i_ \kappa}^z,
\end{equation}
where the "tildes" refer to the interaction picture with respect to the spin and on-site vibron Hamiltonians $H_0=H_{\rm s}^{\sigma}+H_{\rm s}^{\kappa}+H_{\rm rvo}$. Here, we have introduced an arbitrary vibronic operator $O_{i_
\kappa}=O_{i_ \kappa}(\{a_{j_ \kappa}^{\phantom{\dagger}},a^{\dagger}_{j_{\kappa}}\})$, and the spin-vibron coupling $\lambda_0$. In the sections bellow, we will specify to measurements of the vibron numbers $O_{i_
\kappa}=n_{i_\kappa}$, and vibron currents $O_{i_\kappa}=I_{i_\kappa\rightarrow}^{\rm vib}$.

 Since we want to probe the steady state of the bulk ion chain, the above spin-vibron coupling should disturb minimally the dynamics of the vibrons. Therefore, we impose
 \begin{equation}
\label{condition_probe}
|\lambda_O|\ll\tilde{J}_{i_{\alpha}j_{\beta}},\tilde{\Lambda}_{i_{\alpha}j_{\beta}}^{\pm},
\end{equation}
which allows us to divide the bulk dissipative  model~\eqref{bulk_liouvillian} into
\begin{equation}
\label{L_ramsey}
\begin{split}
\tilde{\mathcal{L}}_{0}(\tilde{\mu}_{\rm bulk})&=-\ii[\tilde{H}_{\rm rvt},\tilde{\mu}_{\rm bulk}]+\tilde{\mathcal{D}}_{\rm bulk}(\tilde{\mu}_{\rm bulk}),\\
\tilde{\mathcal{L}}_{1}(\tilde{\mu}_{\rm bulk})&=-\ii[\tilde{H}_{\rm sv}^{O},\tilde{\mu}_{\rm bulk}],
\end{split}
\end{equation}
where $\tilde{H}_{\rm rvt}$ is tunneling part of the renormalized tight-binding model~\eqref{bulk_liouvillian}. The idea now is to project onto the steady-state of the bulk ions, which is given by ${\mathcal{L}}_{0}({\mu}_{\rm
bulk}^{\rm ss})=0$. We use again the projection-operator techniques in Eq.~\eqref{ad_el}, where the projector is now
$\mathcal{P}_{{\rm bulk}}\{\bullet\}= \mu^{\rm ss}_{\rm bulk}\otimes {\rm Tr}_{\sigma, {\rm spin}}\{{\rm Tr}_{\rm vib}\{\bullet\}\}$. This  yields an effective master equation for the $\kappa$-spins
\beq
\label{ramsey_spin_density}
\frac{{\rm d}\tilde{\mu}_{\kappa}^{\rm spin}}{{\rm d}t}=\mathcal{L}_{\rm Ramsey}(\tilde{\mu}_{\kappa}^{\rm spin})=-\ii[H_{\rm R},\tilde{\mu}_{\kappa}^{\rm spin}]+\mathcal{D}_{\rm R}(\tilde{\mu}_{\kappa}^{\rm spin}).
\eeq
Here, we have introduced a Hamiltonian that is responsible for the coherent part of the probe
$
H_{\rm R}=\sum_{i_ \kappa}\half\lambda_0\langle O_{i_{\kappa}}\rangle_{\rm ss}\sigma_{i_{\kappa}}^{z},
$
and   maps the information about the mean value of the vibronic operator $O_{i_ \kappa}$  onto the phase evolution of the spins. The vibronic fluctuations will be coded into the incoherent
part of the probe
\beq
\label{dephasing}
\mathcal{D}_{\rm R}(\bullet)=\sum_{i_{\kappa},j_{\kappa}}\fourth \lambda_{0}^2S_{O_{i_ \kappa}O_{j_ \kappa}}\hspace{-0.5ex}(0)(\sigma_{i_ \kappa}^z \bullet\sigma_{j_ \kappa}^z-\sigma_{j_ \kappa}^z\sigma_{i_ \kappa}^z
\bullet)+\text{H.c.}.
\eeq
Here, we have introduced the  spectral function of the  correlator between two vibronic observables
\beq
\label{fluc}
S_{O_{i_ \kappa}O_{j_ \kappa}}\hspace{-0.5ex}(\omega)=\int_0^{\infty}{\rm d}t\langle \tilde{O}_{i_ \kappa}(t)\tilde{O}_{j_ \kappa}(0)\rangle_{\rm ss}\ee^{+\ii\omega t},
\eeq
where  the operators $\tilde{O}_{i_ \kappa}={O}_{i_ \kappa}-\langle{O}_{i_ \kappa}\rangle_{\rm ss}$ quantify the fluctuations from the steady-state values, and we use
$
\langle \tilde{O}_{i_ \kappa}(t)\tilde{O}_{j_ \kappa}(0)\rangle_{\rm ss}={\rm Tr}\{\tilde{O}_{i_ \kappa}\ee^{\tilde{\mathcal{L}}_0t}\tilde{O}_{j_ \kappa}\mu^{\rm ss}_{\rm bulk}\}.
$
Therefore, the zero-frequency component of the  spectral function~\eqref{fluc} determines the dephasing of the probe~\eqref{dephasing}.

\begin{figure}
\centering
\includegraphics[width=.9\columnwidth]{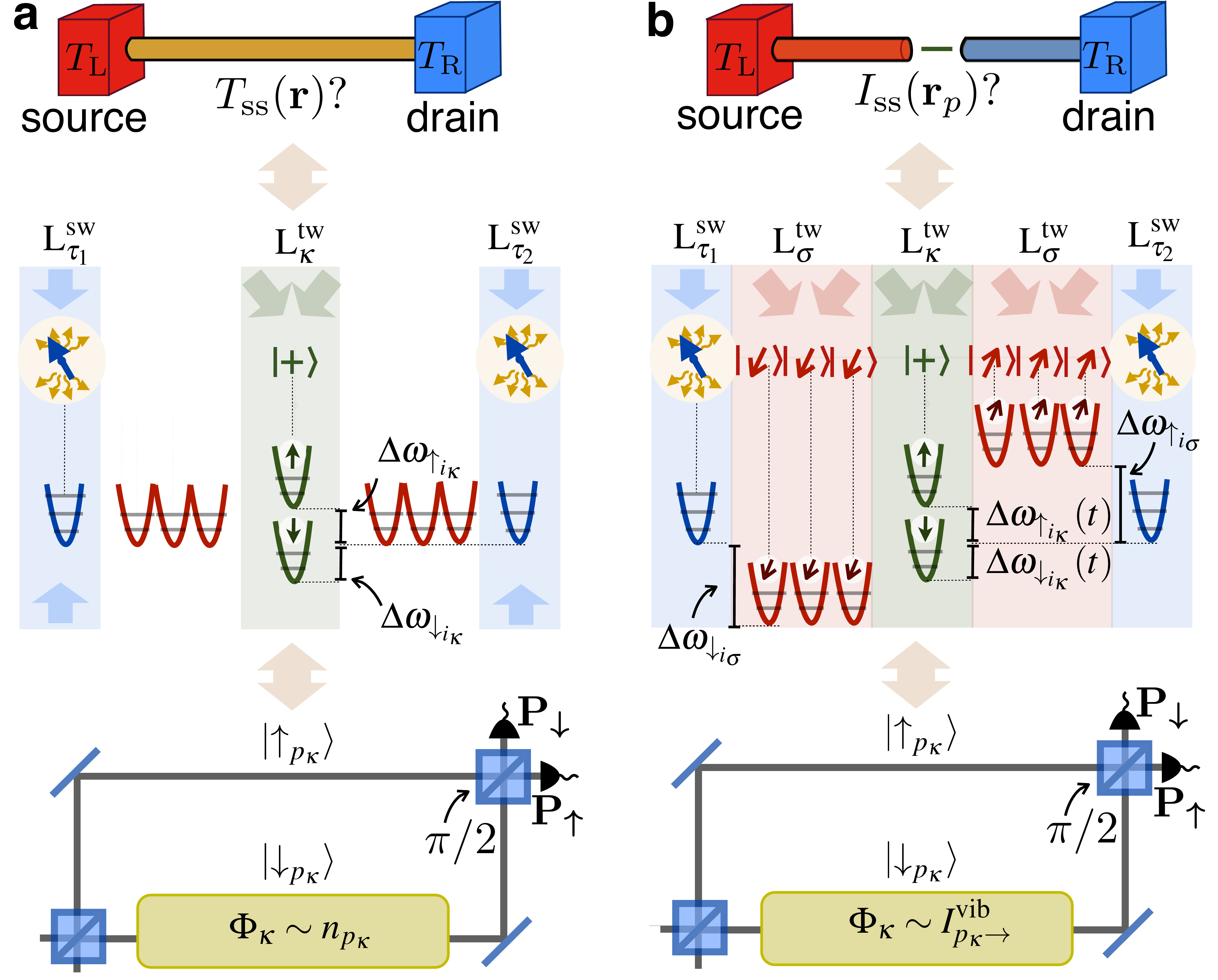}
\caption{ {\bf Spin-based measurements for heat transport:} {\bf (a)}  (upper panel) We consider a situation analogous to a thermal quantum wire (TQW) (i.e. a bar connected to two heat reservoirs with different temperatures). (mid panel) In the trapped-ion scheme [Fig.~\ref{fig_scheme}], we switch on the lasers ${\rm L}_{\kappa}^{\rm tw}$ for the $\kappa$-species leading to a static and weak
spin-vibron coupling~\eqref{driving}. If the $\kappa$-spins are initialised in a linear superposition $\ket{+_{i_{\kappa}}}=(\ket{{\uparrow_{i_\kappa}}}+\ket{{\downarrow_{i_\kappa}}})/\sqrt{2}$, the spin dynamics resembles a
Ramsey
interferometer capable of capturing the information about the mean vibron number and its fluctuations (lower panel). {\bf (b)} (upper panel) We consider a situation analogous to a thermal quantum dot (TQD) connected to two thermal
leads in equilibrium with two reservoirs held at different temperatures. (mid panel) In addition to the static spin-vibron coupling of the $\sigma$-ions of
({\bf
a}),  the lasers ${\rm L}_{\kappa}^{\rm tw}$ should induce now a  periodic and weak spin-vibron coupling~\eqref{driving}. In this case the driving is responsible for assisting the tunneling, but also for mapping the information
about the vibron current to the spin coherences in a Ramsey-type interferometer (lower panel).}
\label{ramsey_scheme}
\end{figure}

 We now describe in detail how the mean value and the fluctuations of the vibronic operator $O_{i_{\kappa}}$ can be measured in analogy to a Ramsey interferometer [Fig.~\ref{ramsey_scheme}{\bf (a)-(b)}]. Let us analyse the case
 where the probe is made of  a single $\kappa$-ion initialised by a $\pi/2$-pulse in  $\mu_{i_{\kappa}}^{\rm spin}(0)=\ketbra{+_{i_{\kappa}}}$, where
 $\ket{+_{i_{\kappa}}}=(\ket{{\uparrow_{i_\kappa}}}+\ket{{\downarrow_{i_\kappa}}})/\sqrt{2}$. Then, the bulk ions evolve under the Liouvillian $\mathcal{L}_0$~\eqref{L_ramsey}, such that their vibrons reach the steady state, while
 the  $\kappa$-spins evolve according to $\mathcal{L}_{\rm Ramsey}$~\eqref{ramsey_spin_density}, acquiring thus information about the vibron observable $O_{i_{\kappa}}$.  In order to recover this information, we perform another
 $\pi/2$-pulse, and measure the probability of observing  the $\kappa$-ion in the spin-down state $P_{\downarrow_{i_{\kappa}}}$. The second pulse, and the projective measurement, are equivalent to the measurement of the spin
 coherence $\langle \tilde{\sigma}_{i_{\kappa}}^x(t)\rangle$, where
 $\tilde{\sigma}_{i_{\kappa}}^x=\ket{{\uparrow_{i_\kappa}}}\bra{{\downarrow_{i_\kappa}}}\ee^{-\ii\omega_{0}^{\sigma}t}+\ket{{\downarrow_{i_\kappa}}}\bra{{\uparrow_{i_\kappa}}} \ee^{+\ii\omega_{0}^{\sigma}t} $, which according to
 Eq.~\eqref{ramsey_spin_density} evolves as
 \begin{equation}
 \langle\tilde{\sigma}_{i_{\kappa}}^x(t)\rangle=\cos(\lambda_{O}\langle O_{i_{\kappa}}\rangle_{\rm ss}t)\ee^{-\lambda_{O}^2{\rm Re}\{S_{{O}_{i_\kappa}{O}_{i_\kappa}}\hspace{-0.5ex}(0)\}t}.
 \end{equation}
Therefore, by measuring the spin populations as a function of time, we expect to get damped oscillations, the period of which gives us information about the mean number of vibrons, while their damping is proportional to the
vibron-number fluctuations in the steady-state. Let us note that the spin-population measurements can be performed through the state-dependent fluorescence of the trapped ion, a technique routinely used in many laboratories that
allow for accuracies reaching 100$\%$ for detection times in the millisecond range~\cite{haeffner_review}. Let us remark that, since we are interested in steady-state properties of the vibrons, this measurement scheme is not
sensitive to the time-resolution of the spin-state readout. Hence, this does not pose any limitation to the target accuracies reaching 100$\%$. Let us finally note that, according to Eq.~\eqref{fluc}, if the probe consists of several $\kappa$-ions, we will also have access to the two-point correlations  of distant ions.

\subsection{Particular applications: vibron number and current}

{\it i) Measurement of the vibron number.--} In order to tailor the coupling~\eqref{generic_spin_vibron} to probe the vibron number (i.e. $O_{i_\kappa}=n_{i_\kappa}$), we must resort to a weak and static  spin-vibron coupling~\eqref{driving}. According to
Eqs.~\eqref{driving_equations_1}-\eqref{driving_equations_2}, we can achieve such a spin-vibron coupling by using a pair of laser beams with equal frequencies, leading to $
H_{\rm
sv}^{\kappa}=\sum_{i_\kappa}\half\Delta\omega^-_{\kappa}n_{i_\kappa}\sigma_{i_\kappa}^z.
$
 In light of the notation used in
Eq.~\eqref{generic_spin_vibron}, we identify $O_{i_\kappa}=n_{i_\kappa}$ [Fig.~\ref{ramsey_scheme}{\bf (a)}], and $\lambda_0=\Delta\omega_{\kappa}^-$, which can be tuned to fulfil the required probe
condition~\eqref{condition_probe}. If we restrict to a single probing ion labeled by $p_{\kappa}$, according to Eqs.~\eqref{ramsey_spin_density}, the coherences evolve as follows
 \begin{equation}
 \label{coherences_density}
 \langle\tilde{\sigma}_{p_{\kappa}}^x(t)\rangle=\cos(\Delta\omega^-_{\kappa}\langle n_{p_{\kappa}}\rangle_{\rm ss}t)\ee^{-(\Delta\omega^-_{\kappa})^2{\rm Re}\{S_{n_{p_\kappa}n_{p_\kappa}}\hspace{-0.5ex}(0)\}t},
 \end{equation}
 which coincides with the description in the main text, and allows us to extract the vibron mean number and fluctuations.
\begin{figure}
\centering
\includegraphics[width=1\columnwidth]{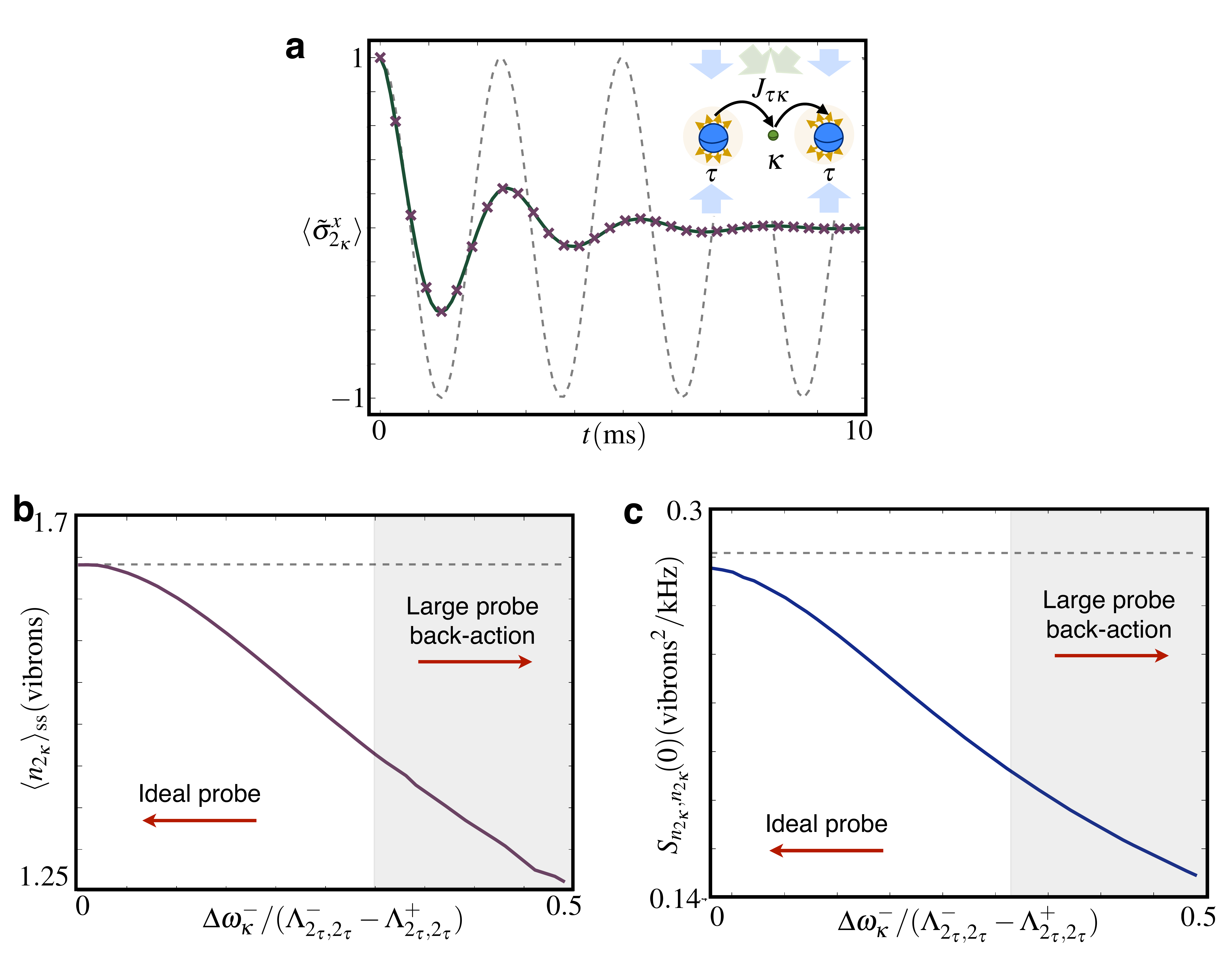}
\caption{ {\bf Ramsey measurement of the vibron number:}  {\bf (a)} Dynamics for the coherence of the probe spin $\langle\tilde{\sigma}_{2_\kappa}^x(t)\rangle$. The grey dashed line would
represent
the periodic oscillations in Eq.~\eqref{coherences_density}  in a noiseless scenario. However, because of the quantum noise $S_{n_{i_\kappa}n_{i_\kappa}}\hspace{-0.5ex}(0)$, such oscillations get damped as
shown
by the numerical solution (green solid line). The crosses correspond to  a numerical fit  $\langle\tilde{\sigma}_{2_\kappa}^x(t)\rangle=\cos(at){\rm exp}\{{-bt}\}$, with fitting parameters $a,b$, which allow us to recover the mean
value and fluctuations  via Eq.~\eqref{coherences_density}. {\bf (b)} Mean value of the vibron number obtained from the numerical fit $\langle n_{2_\kappa}\rangle_{\rm
ss}=a/\Delta\omega^-_{\kappa}$ (solid line). The dashed line represents the theoretical prediction~\eqref{s_o_predictions}. As expected, for $\Delta\omega^-_{\kappa}\to0$, the probe does not
disturb the  bulk vibrons, and we recover the predicted mean number of vibrons~\eqref{s_o_predictions} (dashed line). {\bf (c)} Quantum noise  of the vibron number obtained from the numerical fit $\langle
S_{n_{2_\kappa}n_{2_\kappa}}\hspace{-0.5ex}(0)\rangle_{\rm ss}=b/(\Delta\omega^-_{\kappa})^2$ (solid line). For $\Delta\omega_{\kappa}^-\to0$, we recover the prediction~\eqref{s_o_predictions} (dashed line).
}
\label{ramsey_vibron_number}
\end{figure}

To support our derivations, we analyse numerically the Ramsey measurement for the vibron number. Because of the introduction of the $\kappa$-spins, the dynamics of the system is no longer quadratic as in
 Sec.~\ref{mes_transport_chains}, which forbids finding a closed system of differential equations for the vibronic two-point correlators. Therefore, we have to obtain  numerically the time evolution of the complete density
 matrix
 $\mu_{\rm bulk}(t)$ given by Eq.~\eqref{L_ramsey}, and then calculate the observable  $\langle\tilde{\sigma}_{i_{\kappa}}^x(t)\rangle$. Because of the computational cost of this problem, let us simplify maximally the setup where
 the
 Ramsey measurement can be developed by considering a thermal quantum dot (TQD). However, in contrast to Sec.~\ref{single_ion_channel}, we will consider the arrangement $\tau-\kappa-\tau$, where $\tau=$$^{24}{\rm Mg}^+$
 and  $\kappa=$$^{9}{\rm Be}^+$. We use  the same parameters introduced in previous sections, but set the detunings $\Delta_{1_\tau}=-0.6\Gamma_{\tau}$, $\Delta_{3_\tau}=-0.5\Gamma_{\tau}$, and the Rabi frequencies
 $\Omega_{{\rm L}^{\rm sw}_{1_\tau}}=\Omega_{{\rm L}^{\rm sw}_{3_\tau}}=\Gamma_{\tau}$ for the laser cooling of the $\tau$-ions. The trap frequencies  are $(\omega_{\alpha x},\omega_{\alpha y},\omega_{\alpha z})/2\pi=(5,5,0.25)\hspace{0.2ex}$MHz, which lead to a  tunneling
 $J_{1_{\tau}2_{\kappa}}/2\pi=J_{3_{\tau}2_{\kappa}}/2\pi\approx35\hspace{0.2ex}$kHz.

 According to Eq.~\eqref{bulk_liouvillian},  the master equation of the $\kappa$-ion can be solved exactly, and we obtain the mean number of vibrons and the noise fluctuations by the quantum regression theorem
\begin{equation}
\label{s_o_predictions}
\bar{n}_{2_{\kappa}}=\frac{\Gamma_{\rm L} \bar{n}_{\rm L}+\Gamma_{\rm R} \bar{n}_{\rm R}}{\Gamma_{\rm L} + \Gamma_{\rm
R}},\hspace{1ex}S_{n_{2_\kappa},n_{2_\kappa}}\hspace{-0.5ex}(0)=\frac{\bar{n}_{2_{\kappa}}^2+\bar{n}_{2_{\kappa}}}{2(\tilde{\Lambda}^-_{2_{\kappa}2_{\kappa}}-\tilde{\Lambda}^+_{2_{\kappa}2_{\kappa}})}.
\end{equation}
Therefore, this particular TQD offers a neat playground to test the proposed measurement scheme.

 We now solve numerically the master equation~\eqref{bulk_liouvillian} considering the above realistic parameters, and compute the dynamics of the coherences $\langle\tilde{\sigma}_{2_\kappa}^x(t)\rangle$. In
 Fig.~\ref{ramsey_vibron_number}{\bf (a)}, we represent the numerical results for these coherences (solid line), and perform a numerical fit (crosses) to the expected behaviour in Eq.~\eqref{coherences_density}, which allows us to
 infer the mean value and the quantum noise of the vibron number $\langle n_{2_\kappa}\rangle_{\rm ss}$, $S_{n_{2_\kappa}n_{2_\kappa}}\hspace{-0.5ex}(0)$. In Figs.~\ref{ramsey_vibron_number}{\bf (b)-(c)}, we represent the results
 obtained from this numerical fit as a function of the probing strength $\Delta\omega_{\kappa}^-$. If the probe is too strong, there is an important back-action on the bulk vibrons, and the values obtained from the fit depart
 from the theoretical prediction~\eqref{s_o_predictions}. Conversely, for $\Delta\omega^-_{\kappa}\to 0$, the probe disturbs minimally the system, yielding an acceptable agreement with the theoretical
 predictions~\eqref{s_o_predictions}. It is important to emphasise that, although $\Delta\omega^-_{\kappa}\to 0$, the time  for the Ramsey measurement is fixed to $t=10\,$ms in the numerical simulations, which is a reasonable regime considering
 typical
 decoherence times in trapped-ion experiments. Hence, the limit $\Delta\omega_{\kappa}^-\to 0$ does not require to prohibitively large experimental times.

\begin{figure}
\centering
\includegraphics[width=1\columnwidth]{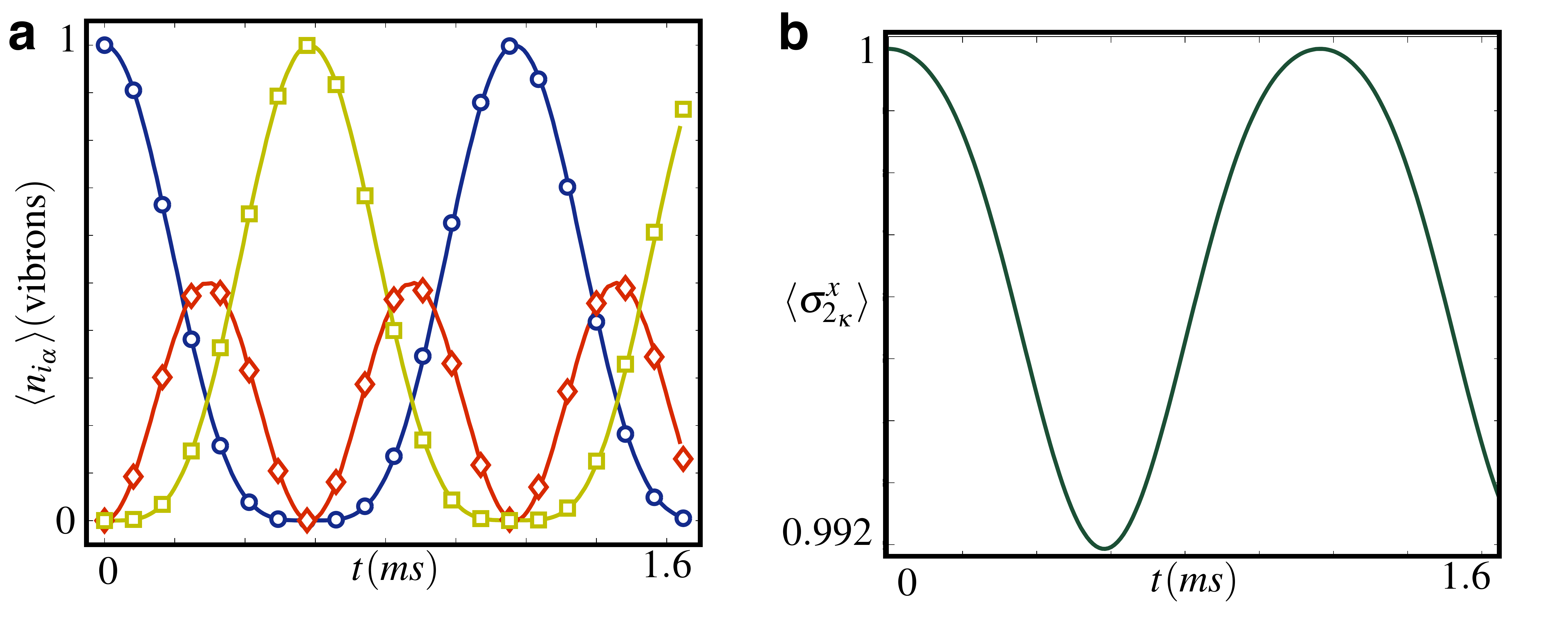}
\caption{ {\bf Effective tunneling for the current  Ramsey measurements:} {\bf (a)} Mean number of vibrons $\langle n_{i_\alpha}\rangle$ as a function of time in the regime of photon-assisted tunneling
$s_{2_\kappa}=\uparrow_{2_\kappa}$ (see the text for the remaining parameters).  The solid lines ($\langle n_{1_\sigma}\rangle$ blue, $\langle n_{2_\kappa}\rangle$ red, $\langle n_{3_\sigma}\rangle$ yellow) represent the exact
solution of $H(t)$, while the open symbols ($\langle n_{1_\sigma}\rangle$ circles, $\langle n_{2_\kappa}\rangle$ diamonds, $\langle n_{3_\sigma}\rangle$ squares) correspond to the effective photon-assisted-tunneling Hamiltonian
$H_{\rm L\kappa R}=H_{\rm L\kappa R}^{\rm Pat}+H_{\rm sv}^{I}$. {\bf (b)} Time evolution of the  spin coherence of the probe $\kappa$-ion, $\langle \sigma_{2_\kappa}^x(t)\rangle$, for an initial spin state
$\ket{\Psi_{2_\kappa}(0)}=\ket{+_{2_\kappa}}$. Because of the effective spin-current coupling~\eqref{pat_spinl}, the  the coherences display periodic oscillations that depend on the periodicity of the assisted tunneling.}
\label{ramsey_current_fig}
\end{figure}

{\it ii) Measurement of the vibron current.--} We now address a possible way of designing the coupling~\eqref{generic_spin_vibron} to probe the vibron current (i.e. $O_{i_\kappa}=I^{\rm vib}_{i_\kappa\rightarrow}$). In particular, we analyse the current through a TQD
 connected
 to the reservoirs by a couple of leads [Fig.~\ref{ramsey_scheme}{\bf (b)}]. Let us recall that this setup is described by the Liouvillian in Eq.~\eqref{meq_leads}, where the reservoirs correspond to the laser-cooled $\tau$-ions,
 the leads to the $\sigma$-chains, and the TQD to the $\kappa$-ion. As discussed below Eq.~\eqref{meq_leads}, the current through the TQD can be suppressed by means of a strong energy off-set between the two halves of the chain.
 Then, a periodic spin-vibron coupling~\eqref{driving} serves as gadget to switch on the current, such that the tunneling strengths depend on the particular spin state of the $\kappa$-ion (see Eq.~\eqref{pat}). In this section, we  make use of this spin-dependence to build a Ramsey probe for the vibron current.
  In particular, we exploit a bi-chromatic spin-vibron coupling
\begin{equation}
\nonumber
H_{\rm
sv}^{\kappa}(t)=\sum_{n=1,2}\half(\Delta\omega_{\kappa,n}^++\Delta\omega_{\kappa,n}^-\sigma_{p_ \kappa}^z)\cos(\nu_{\kappa,n}t-\varphi_{{\kappa,n}})n_{p_ \kappa}.
\end{equation}
 Moreover, by adjusting the laser parameters, we impose
\beq
\begin{split}
\nonumber
&\nu_{\kappa,1}=\half\Delta\omega^-_{\sigma},\hspace{1ex} \varphi_{\kappa,1}=\frac{\pi}{2},\hspace{1ex}\Delta\omega_{\kappa,1}^-=0,\hspace{1ex}\zeta_{\kappa,1}=\frac{\Delta\omega_{\kappa,1}^+}{2\nu_{\kappa,1}}=\pi ,\\
&\nu_{\kappa,2}=\half\Delta\omega^-_{\sigma},\hspace{1ex} \varphi_{\kappa,2}=0,\hspace{1ex}\Delta\omega_{\kappa,2}^+=0,\hspace{1ex}\zeta_{\kappa,2}=\frac{\Delta\omega_{\kappa,2}^-}{2\nu_{\kappa,2}}\ll1 ,\\
\end{split}
\eeq
where   $\Delta\omega^-_{\sigma}$ is the  off-set between the  halves of the chain.

In analogy to the derivation of Eq.~\eqref{pat},  to understand the effects of the bi-chromatic spin-vibron coupling,
we move into an interaction picture  with respect to the driving
$
a_{p_\kappa}\to a_{p_\kappa}=-a_{p_\kappa}\ee^{-\ii\zeta_{\kappa,1}\sin(\nu_{\kappa,1}t-\pi/2)}\ee^{-\ii\zeta_{\kappa,2}\sigma_{p_\kappa}^z\sin(\nu_{\kappa,2}t)}.
$
 By using the Jacobi-Anger expansion again,
together with the above parameter constraints, it is possible to derive an effective Hamiltonian for the coupling of the leads to the TQD
\begin{widetext}
\beq
\nonumber
H_{\rm L\kappa R}\approx\hspace{-1ex}\sum_{i_{\sigma}<p_\kappa,m\in\mathbb{Z}}\hspace{-1ex}-2\tilde{J}_{i_\sigma
p_\kappa}\mathfrak{J}_{-1-m}(\pi)\mathfrak{J}_{m}(\zeta_{\kappa,2}\sigma_{p_\kappa}^z)(\ii)^{-1-m}a_{i_\sigma}^{\dagger}a_{p_\kappa}^{\phantom{\dagger}}
+\hspace{-1ex}\sum_{i_{\sigma}>p_\kappa,m\in\mathbb{Z}}\hspace{-1ex}-2\tilde{J}_{i_\sigma
p_\kappa}\mathfrak{J}_{1-m}(\pi)\mathfrak{J}_{m}(\zeta_{\kappa,2}\sigma_{p_\kappa}^z)(\ii)^{1-m}a_{i_\sigma}^{\dagger}a_{p_\kappa}^{\phantom{\dagger}}
+{\rm H.c.},
\eeq
\end{widetext}
where  we have used a rotating wave approximation for
$\tilde{J}_{i_\sigma p_\kappa}\ll\half|\Delta\omega_{\sigma}^-|$. As announced previously, the expression above shows that for the resonance conditions $\nu_{\kappa,1}=\nu_{\kappa,2}=\half\Delta\omega_{\sigma}^-$, the bi-chromatic
spin-vibron coupling is capable of assisting the tunneling of vibrons across the TQD.

We will now exploit the spin-dependence of the effective tunneling strengths via the Bessel function $\mathfrak{J}_m(\zeta_{\kappa,2}\sigma_{p_\kappa}^z)$ to build a Ramsey probe of the vibron current. In particular, taking into
account that $\zeta_{\kappa,2}\ll1$, we can rewrite $H_{\rm L\kappa R}=H_{\rm L\kappa R}^{\rm PAT}+H_{\rm sv}^{I}$, where we have introduced the  Hamiltonian
\beq
\label{pat_spinless}
H_{\rm L\kappa R}^{\rm PAT}=\sum_{i_\sigma}(\tilde{J}^{\rm PAT}_{i_\sigma,p_\kappa}a_{i_\sigma}^{\dagger}a_{p_\kappa}^{\phantom{\dagger}}+{\rm H.c.}),\hspace{1ex}\tilde{J}^{\rm PAT}_{i_\sigma,p_\kappa}=-\ii2\tilde{J}_{i_\sigma
p_\kappa}\mathfrak{J}_1(\pi).
\eeq
This term describes a spin-independent tunneling of vibrons across the TQD, which will be responsible for setting a vibron current. The important feature of the assisted-tunneling strength is that it has become purely imaginary,
which becomes relevant in the definition of the current operators~\eqref{current_op}.
This turns out to be crucial to devise the Ramsey probe, since the remaining terms in the Hamiltonian can be written as
\beq
\label{pat_spinl}
H_{\rm sv}^{I}=\fourth\tilde{\lambda}_I (I_{p_\kappa\rightarrow}^{\rm vib}+I_{\rightarrow p_\kappa}^{\rm vib})\sigma_{p_\kappa}^z,
\eeq
where we have introduced the dimensionless coupling
$
\tilde{\lambda}_I=2\zeta_{\kappa,2}{(\mathfrak{J}_0(\pi)+ \mathfrak{J}_2(\pi))}/{\mathfrak{J}_1(\pi)}.
$
Remarkably enough, we can make the coupling of the Ramsey probe arbitrarily small by simply letting $\zeta_{\kappa,2}\to 0$, where we get an ideal Ramsey probe of the vibronic current. According to
Eq.~\eqref{ramsey_spin_density}, the $\kappa$-spin coherences evolve as
$ \langle\tilde{\sigma}_{p_{\kappa}}^x(t)\rangle=\cos(\tilde{\lambda}_{I}\langle I^{\rm vib}_{\rightarrow p_{\kappa}}\rangle_{\rm ss}t)\ee^{-\tilde{\lambda}_{I}^2{\rm Re}\{S_{I_{p_\kappa}I_{p_\kappa}}\hspace{-0.5ex}(0)\}t},
$
 where we have made use of the fact that $\langle I^{\rm vib}_{\rightarrow p_{\kappa}}\rangle_{\rm ss}=\langle I^{\rm vib}_{p_\kappa \rightarrow }\rangle_{\rm ss}$ in the steady state. We have also defined the zero-frequency
 intensity noise
$
 S_{I_{p_\kappa}I_{p_\kappa}}\hspace{-0.5ex}(0)=\int_0^{\infty}{\rm d}t\langle \tilde{I}_{p_ \kappa}(t)\tilde{I}_{p_ \kappa}(0)\rangle_{\rm ss},
$
 wherethe current fluctuations are given by
$ \tilde{I}_{p_ \kappa}=\half(I^{\rm vib}_{\rightarrow p_{\kappa}}+I^{\rm vib}_{p_{\kappa}\rightarrow })-\langle I^{\rm vib}_{ p_{\kappa}\rightarrow }\rangle_{\rm ss}.
$

In order to give supporting numerical evidence of this prediction, the minimal setup to explore would be a $\tau-\sigma-\kappa-\sigma-\tau$ chain. In analogy to the single-spin switch, computing the dynamics of
the spin coherences in this case becomes a non-linear problem that exceeds our numerical capabilities. Therefore, we cannot obtain the analogue of Fig.~\ref{ramsey_vibron_number} for the current operator. Instead, we  content
ourselves with showing that the  Hamiltonians in Eqs.~\eqref{pat_spinless}-\eqref{pat_spinl} describe the dynamics of a $\sigma-\kappa-\sigma$ setup in Fig.~\ref{ramsey_current_fig}.


\begin{references}
\bibitem{comment_cooling}
Note that laser cooling of mixed crystals has been shown for traveling waves~\cite{sympathetic_cooling_exp}, and standing-wave cooling may be achieved along the lines of Ref.~\cite{slowly_moving_standing_wave}. Possible
alternatives for the required strong cooling are  schemes based on the dynamical Stark-shift~\cite{ss_cooling}, EIT cooling~\cite{eit_cooling}, or pulsed sequences~\cite{fast_cooling}.
\bibitem{comment_drivings}
Note that similar couplings, but linear in the vibron operators, have been demonstrated experimentally~\cite{state_dep_forces_sigma_z}.
\bibitem{bulk_parameters}
{$\tilde{\Lambda}^{+}_{i_\alpha,j_\beta}=\sum_{\ell_\tau}\half\Gamma^{\ell_\tau}_{i_\alpha,j_\beta}\bar{n}_{\ell_\tau}$
and
$\tilde{\Lambda}^{-}_{i_\alpha,j_\beta}=\sum_{\ell_\tau}\half\Gamma^{\ell_\tau}_{i_\alpha,j_\beta}(\bar{n}_{\ell_\tau}+1)$ are the non-local heating/cooling strengths, expressed in terms of  the reservoir density of
states $ \rho_{\ell_\tau}(\epsilon) =
{\gamma_{\ell_\tau}}/{\pi([(\epsilon-\delta_{\ell_\tau})-\omega_{\ell_\tau}]^2+\gamma_{\ell_\tau}^2})$, with $\delta_{\ell_\tau}={\rm
Im}\{\Lambda^+_{\ell_\tau}-(\Lambda^-_{\ell_\tau})^*\}. $}
\bibitem{comment}
The parameter $\xi_{\rm c}$ may be controlled accurately in microtrap arrays, and to some extent in the more standard linear Paul traps.
\bibitem{comment_parameters}
The parameters of the bi-chromatic spin-vibron coupling are $\varphi_{\kappa,1}=\frac{\pi}{2}$, $\varphi_{\kappa,2}=0$, $\nu_{\kappa,1}=\nu_{\kappa,2}=\half\Delta\omega^-_{\sigma}$, and
$\Delta\omega_{\kappa,1}^-=\Delta\omega_{\kappa,2}^-=0$.
$\zeta_{\kappa,1}=\Delta\omega_{\kappa,1}^+/2\nu_{\kappa,1}=\pi$,
$\zeta_{\kappa,2}=\Delta\omega_{\kappa,2}^+/2\nu_{\kappa,2}\ll1$.
\end{references}
\end{document}